\documentclass[aps,prb,twocolumn,groupedaddress,10pt]{revtex4-2}

\usepackage{graphicx}
\usepackage{amssymb, amsmath}
\usepackage[utf8]{inputenc}
\usepackage{enumerate}
\usepackage{hyperref}

\newcommand{\ket}[1]{| #1 \rangle}
\newcommand{\bra}[1]{\langle #1 |}
\newcommand{\expval}[1]{\langle #1 \rangle}

\newcommand{\pdagger}{\phantom{\dagger}}
\DeclareMathOperator{\Tr}{Tr}

\newcommand{\bonnpi}{Physikalisches Institut, University of Bonn, Nussallee 12, 53115 Bonn, Germany}

\usepackage{xcolor}

\renewcommand\vec{\mathbf}

\begin{document}

\title{Majorana edge-modes in a spinful particle conserving model}

\author{Franco T. Lisandrini}
\affiliation{\bonnpi}
\author{Corinna Kollath}
\affiliation{\bonnpi}

\date{\today}

\begin{abstract}
We show the presence of Majorana edge modes in an interacting fermionic ladder with spin in a number conserved setting. The interchain single particle hopping is suppressed and only a pair hopping is present between the different chains of the ladder. Additionally, the hopping along the chains is spin imbalanced and a transverse magnetic field is applied breaking time-reversal invariance. We study the robustness of the topological phase with respect to an on-site interaction between the spin-up and spin-down fermions and the spin dependent imbalance of the hopping. 
The main result of the present work is that the topological phase survives for a finite region in the parameter space in the presence of interactions. 
The localized Majorana edge modes seems to be more stable in the case when the on-site interaction is an attraction. 

\end{abstract}

\maketitle

\section{Introduction}
Topologically protected edge states, such as Majorana edge modes, have attracted a lot of attention over the past \cite{Alicea2012, Beenakker2013, LutchynOreg2018, FlensbergStern2021}.
The interest in Majorana modes is motivated by curiosity to observe and understand these fundamental quasi-particles, but also by the key role such modes play in several quantum information protocols \cite{NayakDasSarma2008}.

One important model which covers Majorana zero modes was proposed by Kitaev \cite{Kitaev2001}. It consists of a single chain of spinless fermions with a pairing term that creates-annihilates pairs. 
As a consequence, the number of fermions in the system is not conserved. This minimal model has the advantage that many of its properties can be analytically determined. In particular, the features of the occurring Majorana modes could be explored in detail. 
Several experimental realizations of Majorana modes \cite{LutchynDasSarma2010, OregOppen2010, SatoFujimoto2009, JiangZoller2011}
have been proposed in the context of solid state and cold atomic gases relying on the coupling on superconducting/superfluid phases in order to realize a source and drain of pairs. 

Ten years after Kitaev's seminal work \cite{Kitaev2001} three different works on coupled chains in a particle number-conserving setting appeared almost at the same time \cite{ChengTu2011, FidkowskiFisher2011, SauDasSarma2011}. 
The motivation to study particle number-conserving settings is twofold, on one hand the conceptual question if it is possible to stabilize Majorana modes in a setup without a phase without long-range superconducting order and on the other hand whether the presence of a reservoir is required since several experimental realizations such as in cold atomic systems are easier using number conserved settings. Since then several particle-number conserving settings have been studied. For example, Kraus and collaborators\cite{KrausZoller2013} proposed that it was possible to couple two chains of spinless fermions interacting only by exchanging pairs to be able to ``simulate'' the pair creation term of the Kitaev chain. 
Due to the pair-hopping term, the particle number on each individual chain is not conserved but the parity is. They found with density matrix renormalization group method (DMRG) calculations that under certain parameters the ground state of the system supports Majorana edge modes. 
The signals they observe are (i) the double-degeneracy between both single-chain parity sectors, (ii) its entanglement spectrum is even-degenerated, (iii) the single particle correlations exponentially died in the bulk but had a finite revival on the other end, and (iv) all these properties are robust against static noise. 

 \begin{figure}
 \includegraphics[width=1.0\columnwidth]{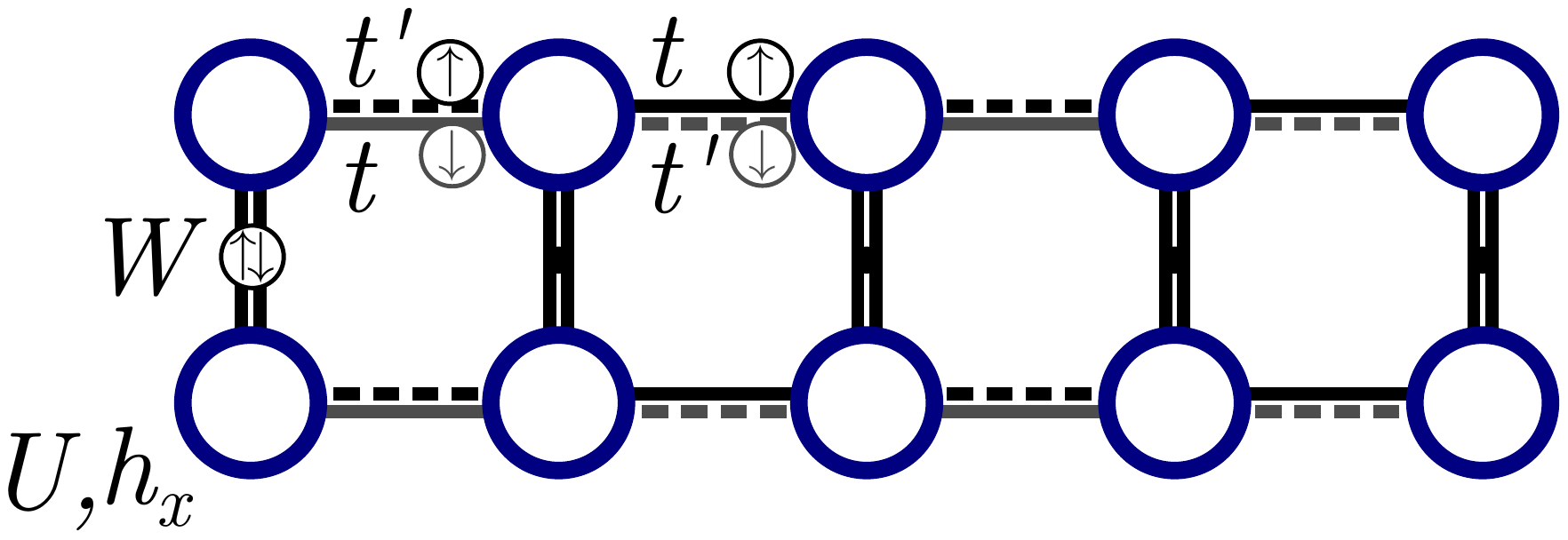}%
 \caption{ 
Sketch of the ladder structure with interacting spinful fermions. 
The interchain term is a pairing hopping with amplitude $W$. 
The on-site terms are an interaction $U$, that can be either attractive or repulsive, and a magnetic field in the $x$-direction $h_x$. 
 The single-particle intrachain hopping terms are different depending on the parity of the bond and the spin of the fermion: 
On odd bonds the hopping amplitude for the down-fermions is $t$ and for the up-fermions is $t'$; on even bonds it is the other way around, up-fermions hop with $t$ and down-fermions with $t'$. 
 \label{FIG_MODEL}}
 \end{figure}

A few years later some more work was done in this direction studying Majorana modes in variants of this model with spinless fermions \cite{OrtizBeenakker2014, IeminiDiehl2015, LangBuchler2015} and spinful models \cite{IeminiDalmonte2017}. 
In this work, we extend this previous studies by investigating a spinful Hubbard ladder with some additional terms as sketched in Fig.~\ref{FIG_MODEL}. Our motivation is to present alternative ways for the experimental realization in cold atom experiments where long range interactions between spinless fermions are very demanding to implement. In our proposal, the hopping terms of the spin up and spin down fermions along the chains are oppositely dimerized and an additional Coulomb interaction and magnetic field exist on each site, explicitly breaking the time-reversal symmetry. The only interchain term is a pair-hopping, and in contrast to the spinless case, this hopping only involves one site on each chain. The use of this pair hopping term which is confined to one site stems from recent findings on Floquet driven systems \cite{HuebnerSheikhan2022, Bauer2020} which allow one to enhance the pair hopping and at the same time suppress the single particle hopping in Hubbard systems. The same process can be used here in order to engineer a large pair hopping process compared to the single particle hopping. 

The dimerized tunnelling amplitudes on even and odd sites can be engineered using optical superlattice potentials. Such superlattice potentials are widely used in cold atomic gases. These superlattice potentials typically consist of two different laser beams with different wavelength and adjusting the phase between the laser beams the potential shape can be tuned over a wide range. In particular, lattices with a dimerized tunnelling amplitude can be engineered (see for example \cite{LohseBloch2016, WangXi2013, RomeroIsartGarciaRipoll2007, FoellingBloch2007, PeilPhillips2003}). 
The light potentials can be chosen to be spin-dependent, i.e.~that they couple differently to the different internal states of the atoms which represent the spin states. This is the case when the two counterpropagating laser beams have a linear polarization vectors with a relative angle between them \cite{JessenDeutsch1996, BrennenDeutsch1999, JakschZoller1999, BlochZwerger2008}. 

We explored the phase diagram of the system using a combination of analytical and numerical methods. 
Using Matrix Product States algorithms we computed the energy, entanglement spectrum and single particle correlations functions (among others observables). 
We were able to detect a finite region in the parameter space where 
the system exhibits simultaneously a vanishing energy difference between both single-chain parity sectors, an even degeneracy in the whole entanglement spectrum and finite edge-edge single particle correlations. 
Hence we identified the characteristic features of Majorana edge-modes in an extended region. 

Additionally we have use second order perturbation theory to obtain the effective model for strong interactions, both the strongly attractive and strongly repulsive limits. This allow us to gain some intuition on the topologically trivial phases surrounding the topological one.

In Sec.~\ref{SEC_Model} we introduce the model we are going to focus on in this work. 
We present also the connection with models from previous works, and the effective models for the strongly attractive and strongly repulsive limits. 
In Sec.~\ref{SEC_Methods} we discuss the details of the numerical implementation. 
In Sec.~\ref{SEC_TOP} we summarize our results focusing in the topological phase. 
The results are discussed in Sec.~\ref{SEC_RES}, including subsections focused on the energy difference, the entanglement spectrum, and the single particle correlations, and robustness of these features against static noise. 
The conclusions are presented Sec.~\ref{SEC_CONC}. 

\section{Model}
\label{SEC_Model}

In this paper we investigate a ladder of interacting spinful fermions with open boundary conditions, as the one depicted on Fig.~\ref{FIG_MODEL} and described by the following Hamiltonian
\begin{equation}
H =  H_{tt'} + H_{x} + H_{U} + H_{W},
\label{EQ_H}
\end{equation}
where 
\begin{align}
H_{tt'} =  &- \sum_{j=1}^{(L-1)/2}  \sum_{\alpha = a, b} ( 
 t \alpha^{\dagger}_{\downarrow 2j-1} \alpha^{\pdagger}_{\downarrow 2j} 
+ t' \alpha^{\dagger}_{\downarrow 2j} \alpha^{\pdagger}_{\downarrow 2j+1} + \text{h.c.} )  
\nonumber \\ \nonumber
 &-  \sum_{j=1}^{(L-1)/2}  \sum_{\alpha = a, b} ( 
 t' \alpha^{\dagger}_{\uparrow 2j-1} \alpha^{\pdagger}_{\uparrow 2j} 
+ t \alpha^{\dagger}_{\uparrow 2j} \alpha^{\pdagger}_{\uparrow 2j+1} + \text{h.c.} ), \\ \nonumber
H_{x} &= \sum_{j=1}^{L} \sum_{\alpha = a, b} ( 
h_x \alpha^{\dagger}_{\uparrow j} \alpha^{\pdagger}_{\downarrow j} + \text{h.c.} ), \\  \nonumber
H_{U} &= \sum_{j=1}^{L} \sum_{\alpha = a, b} 
U n^{}_{\alpha \uparrow j} n^{}_{\alpha \downarrow j}, \\
H_{W} &= \sum_{j=1}^{L} ( 
W a^{\dagger}_{\uparrow j} a^{\dagger}_{\downarrow j} b^{\pdagger}_{\downarrow j} b^{\pdagger}_{\uparrow j} + \text{h.c.} ) .
\end{align}
Here $L$ is the number of rungs of the ladder, $\alpha_{\sigma j}$ is the annihilation operator of a fermion with spin $\sigma$ on rung $j$ and on leg $\alpha = a, b$ of the ladder and $ n_{\alpha \sigma j} = \alpha^{\dagger}_{\sigma j} \alpha^{\pdagger}_{\sigma j} $ is the corresponding density operator. 
For simplicity, in Eq.~\ref{EQ_H} we chose $L$ to be odd. One can analogously write the Hamiltonian for a lattice with an even number of rungs, but will have to change the expression for $H_{tt'}$ accordingly. 

On each chain single-particle hopping terms exist. The amplitude of the single particle hopping is dimerized on even and odd bonds and is different for opposite spins. 
On odd (even) bonds the hopping amplitude for the down-fermions is $t$ ($t'$) and for the up-fermions $t'$ ($t$), respectively. The term proportional to $U$ is an on-site interaction term which can be either attractive ($U<0$) or repulsive ($U>0$). 
A transverse magnetic field is applied along the $x$-direction with amplitude $h_x$ inducing a spin flip. 
This term explicitly breaks the time-reversal symmetry, which is a fundamental requirement for the system to host Majorana modes \cite{Kitaev2001}. As a consequence, the total magnetization in the $z$-direction is not conserved.

To better understand the role of the spin-dependent hopping terms we can make a rotation into the direction of the magnetic field, if we do so, the spin-dependent hopping can be rewritten as normal hopping term for both spin projections (with amplitude $t_+ = (t+t')/2$) and a spin-flipping hopping reminiscent of a spin-orbit term. 
The amplitude of the spin-flipping hopping is $t_- = (t-t')/2$, and is zero when both hopping amplitudes in the unrotated Hamiltonian are the same. 
Therefore, the role of this term can be interpreted similarly to the spin-orbit coupling in other proposals of Majoranas-supporting systems in spinful settings \cite{LutchynDasSarma2010, OregOppen2010}.

All of the terms we have mentioned so far act on a single leg. 
The only interchain term is a pair hopping term between both chains which has an amplitude $W$. 
An immediate advantage of using spinful model is that the pair hopping terms now involve only one site on each leg, making a realization in cold atomic setups simpler. 
The Hamiltonian conserves the total number of fermions ($N$), but not separately the number of fermions on each individual chain ($N_a$, $N_b$). However, since by the pair hopping along the rungs we can only change the number of fermions in each chain in pairs, the parity ($P_a$, $P_b$) of each leg of the ladder is conserved. 
For the rest of this work we are going to consider an even number of particles will correspond to parity $P=0$ and an odd number of particles will correspond to $P=1$. 
Further, by the number conservation the parity ($P$) on the ladder structure is conserved. 
We choose to have an even parity in the whole ladder ($P=0$), then both chains have the same parity ($P_a=P_b$).

During the rest of the work $t$ will be the unit of energy, 
the pairing hopping will be $W=2.6t$ and the magnetic field will be fixed at $h_x = -t$. Typically, we will work at a fixed incommensurate density $n=N/2L= 0.32$, with systems that have $L = 25, \,50, \, 75, \,100, \,125, \,150$ rungs and $N = 16, \,32,\, 48, \,64, \,80, \,96$ fermions, respectively. In some situations we also show the results of a ladder with $L=13$ rungs and $N=8$ fermions, in order to have more systems lengths. In this situation, the density is $n=N/2L \simeq 0.308$.
We are going to explore the phase diagram varying $t'$ and $\pm U$ focusing on the regime where $t'$ is smaller than $t$.

\subsection{Connection to spinless fermionic ladders: the $(U=t'=0)$-limit}
\label{SEC_tU0}

 \begin{figure}
 \includegraphics[width=1.0\columnwidth]{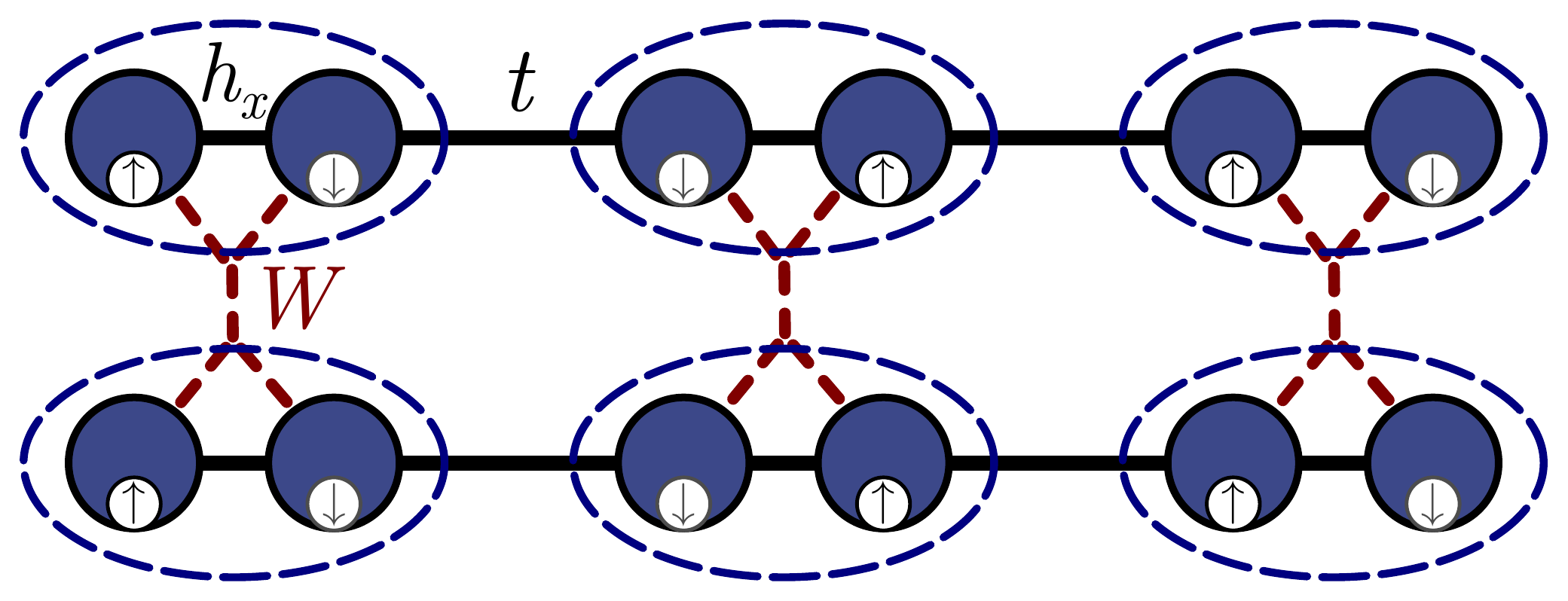}%
 \caption{ 
Sketch of the ladder structure when $U=t'=0$. 
For this special parameter set, the system can be mapped to a ladder of spinless fermions as the figure suggest.
 \label{FIG_MAP}}
 \end{figure}

An interesting limit of the considered model is the limit $U=t'=0$ (and $h_x = -t$). For this special parameter set, the system can be mapped to a ladder of spinless fermions as we will show in the following.

Taking $U=t'=0$ and $h_x = -t$, the Hamiltonian reduces to the form 
\begin{align}
&H = -t\sum_{j=1}^{(L-1)/2} \sum_{\alpha = a, b} 
( \alpha^{\dagger}_{\downarrow 2j-1} \alpha^{\pdagger}_{\downarrow 2j} 
+ \alpha^{\dagger}_{\uparrow 2j} \alpha^{\pdagger}_{\uparrow 2j+1} + \text{h.c.} ) \\ \nonumber
&-t \sum_{j=1}^{L} \sum_{\alpha = a, b} 
( \alpha^{\dagger}_{\uparrow j} \alpha^{\pdagger}_{\downarrow j} + \text{h.c.} ) 
+ W\sum_{j=1}^{L} ( a^{\dagger}_{\uparrow j} a^{\dagger}_{\downarrow j} 
b^{\pdagger}_{\downarrow j} b^{\pdagger}_{\uparrow j} + \text{h.c.} ), 
\end{align}
where we have explicitly written out the even and odd hopping terms. Again this expression corresponds to and odd number of rungs.

In order to map the system to a system of spinless fermions, we map the up- and down-operators to different sites doubling the number of sites in each chain as depicted in Fig.~\ref{FIG_MAP}. 
More precisely, the spin-up operator on an odd-site $2j-1$ will be mapped to a new site $4j-3$ and the spin-down operator on the same odd-site $2j-1$ to a new site $4j-2$, i.e.~
$$ \alpha_{\uparrow 2j-1}\rightarrow {\tilde \alpha}_{4j-3} \quad 
\alpha_{\downarrow 2j-1}\rightarrow {\tilde \alpha}_{ 4j-2}.$$ 
Here, the operators ${\tilde \alpha}_{j}$ are the spinless fermion operators. 
Similarly, we are going to map the spin-down operator on an even-site $2j$ to a site $4j-1$ and the spin-up operator on the same even-site $2j$ to a site $4j$, i.e.~
$$ \alpha_{\downarrow 2j}\rightarrow {\tilde \alpha}_{4j-1} \quad 
\alpha_{\uparrow 2j}\rightarrow {\tilde \alpha}_{4j}.$$ 
Using this mapping, we obtain a model of two chains of spinless fermions coupled by a pair-hopping term which is only applied on plaquettes which start with odd sites, i.e. 

\begin{align}
\label{EQ_KRAUS}
&H = -t\sum_{j=1}^{2L-1} \sum_{\alpha = a, b} 
( \tilde{\alpha}^{\dagger}_{j} \tilde{\alpha}^{\pdagger}_{j+1} + \text{h.c.} ) \\ \nonumber
&+ W\sum_{j=1}^{L} ( \tilde{a}^{\dagger}_{2j-1} \tilde{a}^{\dagger}_{2j} 
\tilde{b}^{\pdagger}_{2j} \tilde{b}^{\pdagger}_{2j-1} + \text{h.c.} ) .
\end{align}

The model with a pair-hopping term in every plaquette has been intensively studied before \cite{ChengTu2011, FidkowskiFisher2011, SauDasSarma2011, KrausZoller2013} and it is known to host pairs of Majoranas end-states in its ground state. 
We show in appendix \ref{SEC_ADIAB}, that this phase hosting Majorana modes seems to be connected adiabatically to the topological phase in the model of Eq.~\ref{EQ_H} which supports our claim that the observed modes are Majorana modes.

\subsection{Large-$U$ limits}
 \label{SEC_LARGEU}
 
In this section we discuss the limits of strong interactions, i.e.~ $U$ is the largest energy scale. 
We use 2nd order perturbation theory to map the system to an effective low energy model and study its properties first for the strongly repulsive limit and then for the strongly attractive limit. 

\subsubsection{Strongly repulsive interaction}
In the strongly repulsive case, the energy to create a pair is very high (of the order of the interaction energy $U$). Thus, in order to consider low energy features, we can restrict ourselves to the subspace with no pairs and only taking virtual excitations around this subspace into account. 
We use a 2nd order perturbation theory to derive an effective low energy Hamiltonian within this reduced subspace. The most important effect of this approximation is that the two legs of the ladder become effectively decoupled, since in the original Hamiltonian the two legs are only connected by pair-hopping terms which are of higher order. The pair hopping term only connects states with at least one pair to another state with one pair. This means that starting from the subspace of no pairs first the pair needs to be generated, followed by a pair hopping and a destruction of a pair. 
The approximate decoupling of the two legs of the ladder have certainly drastic consequences on the physical properties of the ladder and no Majorana modes are expected to survive in this limit. 

 \begin{figure}
 \includegraphics[width=1.0\columnwidth]{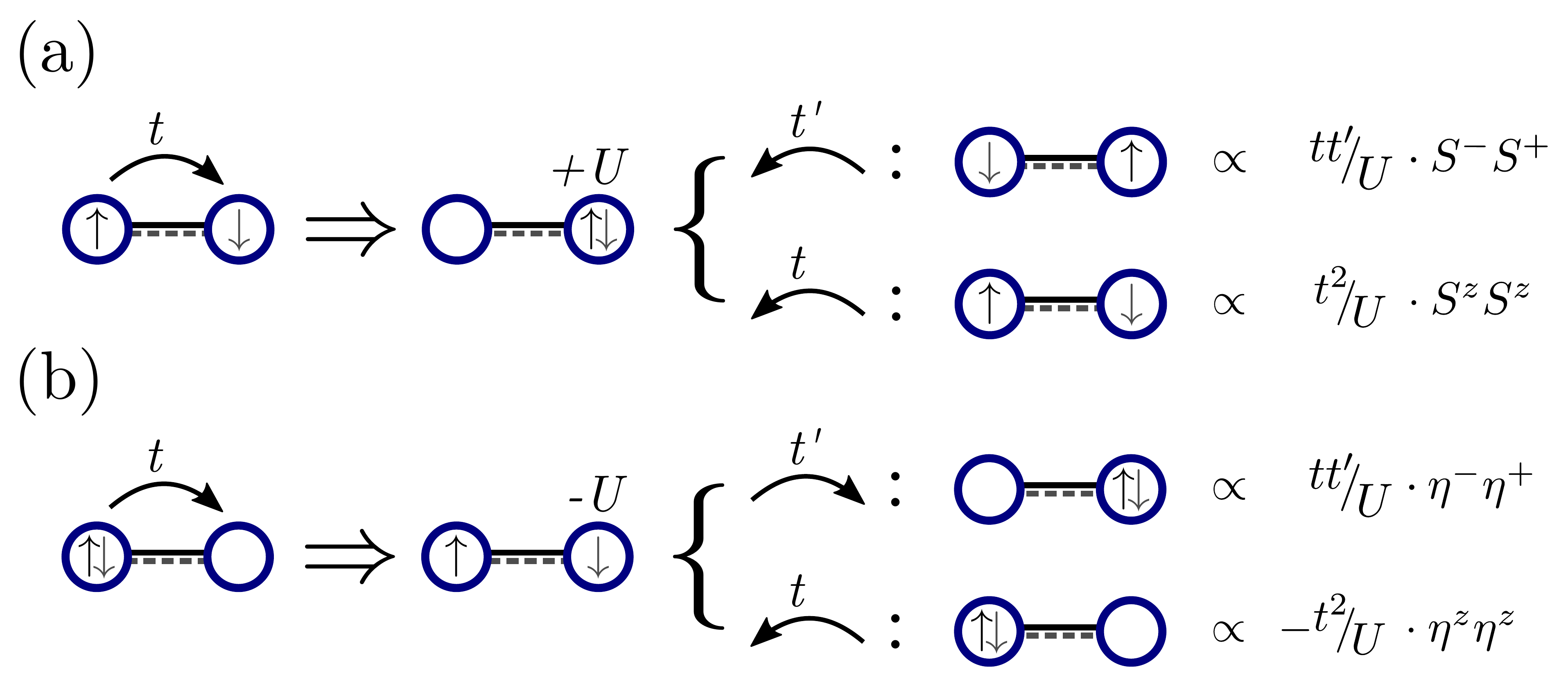}%
 \caption{ 
 Examples virtual processes that give rise to different terms in the effective Hamiltonian. 
 \label{FIG_ULIMS}}
 \end{figure}
 
Within the legs of the ladder, effective spin interactions are created as well known from the spin interactions arising in the high-$U$ limit of the Hubbard model. Here, due to the spin dependent hopping, the spin coupling is anisotropic along the different directions. The main effective processes are sketched in Fig.~\ref{FIG_ULIMS}. If two fermions with opposite spin occupy neighbouring sites, then one can virtually hop to the position of the other giving rise, effectively, to a spin interaction between neighbouring sites. If we have a fermion with spin up on an even site and a down-spin fermion on its right (see Fig.~\ref{FIG_ULIMS}), then the amplitude of the hopping will be $t$ and the system will have to raise its energy by $U$ (formation of a virtual pair). 
After this formation of a virtual pair, one of this two fermions has to hop back to the even site. 
If it is the down-spin fermion that hops, then the term of the effective Hamiltonian will be $\propto tt'/U S^- S^+$ where $\vec{S}=\left( S^x, S^y, S^z \right)$ is the spin-operator and $S^\pm = S^x \pm iS^y$ the ladder-operators. 
The definitions of the spin operators in terms of the fermionic operators are given by $S^{+}_{\alpha j} = \alpha^{\dagger}_{\uparrow j} \alpha^{\dagger}_{\downarrow j}$, $S^{-}_{\alpha j} = \alpha^{\dagger}_{\downarrow j} \alpha^{\dagger}_{\uparrow j}$ and $S^{z}_{\alpha j} = \left( n_{\alpha \uparrow j} + n_{\alpha \downarrow j} \right) /2 $.
In contrast, if it is the up-spin fermion that hops back, the amplitude will be $\propto t^2/U S^z S^z$. 
Similar terms arise from considering the other possible processes (i.e., starting on an odd site and/or with a spin down), combining all these terms we derive the following effective Hamiltonian
\begin{align}
\label{EQ_REP}
H_{sp} &=  \tilde{H}_{tt'} + \tilde{H}_{x} \\ \nonumber
&+ \sum_{j=1}^{L-1} 
J_z S^z_{\alpha, j} S^z_{\alpha, j+1} 
+ \sum_{j=1}^{L-1} 
\frac{J_{xy}}{2} ( S^+_{\alpha, j} S^-_{\alpha, j+1} + \text{h.c.} ),
\end{align}
where $J_z = 2(t^2 + {t'}^2)/U$ and $J_{xy} = 4tt'/U$. 
The first two terms ($\tilde{H}_{tt'}$ and $\tilde{H}_{x}$) correspond to the hopping and magnetic field terms in the original Hamiltonian restricted to the subspace with singly occupied sites. 

As previously mentioned, after this approximation we get two decoupled chains each of them representing an extended anisotropic $t-J$ model in a transverse magnetic field $h_x$ with $L$ sites. 
Let us note, that in the limit of low particle number, the first two terms are the dominating first order terms leading to decoupled fermionic chains subjected to a transverse magnetic field. For many of our findings in the regime of strongly repulsive interactions, this effective first order Hamiltonian is sufficient to explain the main effects. To have an even easier analytical handle on the state we assume that also the field $h_x$ is dominating, leading to an almost completely polarized state. Within such a completely polarized state we can treat the fermions as non-interacting and evaluate many of the quantities we are interested in, since the projector onto singly occupied sites is fulfilled. 

We will show that many features encountered in the regime of strong repulsion at the chosen parameters are explained by this polarized state.

\subsubsection{Strongly attractive interaction}
We will analogously proceed in the strongly attractive interaction limit since an energy of order $U$ is needed to break a pair. 
Considering that we start from an even number of fermions in each leg, we can assume that these are all paired in the lowest energy sector. 

We use a 2nd order perturbation theory to obtain the effective low energy Hamiltonian in the only-pairs subspace including effectively the processes to virtually break one pair and then rapidly form it again. In contrast to the repulsively interacting limit, for strong attraction the pair coupling between the two legs of the ladders can act in the low energy sector, since pairs are present. However, the magnetic field term is not acting on the subspace of only pairs and can be neglected in this situation. 
 
If we strictly restrict ourselves to the only-pairs subspace then to lowest order the rungs decouple. If only one pair is present on a certain rung, it delocalizes inside the rung and gains energy. Thus, at the considered low filling, the pairs try to distribute, such that they can delocalize on the rungs.
Considering also virtual pair breaking, allows the pair to interact along the chain direction as we can see in the sketch in Fig.~\ref{FIG_ULIMS}: 
If a pair on an even site has an empty neighbouring site, one of the fermions forming the pair can virtually hop there, breaking the pair. If the fermion that hops has its spin pointing up then the hopping amplitude is $t$. The resulting state is a virtual state with a broken pair. The breaking cost an energy of $|U|$. This virtual state can be resolved via two options: (a) either the fermion with down-spin also hops leading to an effective pair hopping or (b) the fermion with up-spin comes back leading to an effective interaction between neighbouring pairs. In the first case (a) the term on the effective Hamiltonian will be $\propto -tt'/|U| $ and in the second case (b) it will be $\propto t^2/|U|$. 

Putting together all the contributions of the different processes we obtain the effective low energy Hamiltonian 
\begin{align}
H_{pairs} &=  
 2W \sum_{j=1}^{L}
\frac{1}{2} (\eta^{+}_{a, j} \eta^{-}_{b, j} + \text{h.c.} ) \\
\nonumber
&+ J_z \sum_{j=1}^{L-1} \sum_{\alpha=a,b} 
\eta^{z}_{\alpha, j} \eta^{z}_{\alpha, j+1} \\ \nonumber
&- J_{xy} \sum_{j=1}^{L-1} \sum_{\alpha=a,b} 
\frac{1}{2} (\eta^{+}_{\alpha, j} \eta^{-}_{\alpha, j+1} + \text{h.c.} ),
\nonumber
\end{align} 
where $J_z = 2(t^2 + {t'}^2)/|U|$ and $J_{xy} = 4tt'/|U|$.
We defined the pair creation operator $\eta^{+}_{\alpha, j} = \alpha^{\dagger}_{\uparrow j} \alpha^{\dagger}_{\downarrow j}$, the pair annihilation operator $\eta^{-}_{\alpha j} = \alpha_{\downarrow j} \alpha_{\uparrow j}$ and $\eta^{z}_{\alpha j} = n_{\alpha \uparrow j} n_{\alpha \downarrow j} - 1/2 $ an operator related to the number of pairs.

Thus, the effective model we obtain in the strongly attractive interacting limit with only pairs consists of a pseudo-spin ladder describing the pairs where the inter-rung couplings are antiferromagnetic in the $z$-direction and ferromagnetic in the $xy$-plane. The intra-rung couplings are only in-plane and have an amplitude $2W$. 

Since, the intra-rung pair-hopping is a first order process, it is typically much larger than the inter-leg terms. This favours the formation of a rung singlet between the double occupied state and the empty state (or triplet, depending on the sign of $W$) that is separated from the others in-rung states by an energy $W$. 
For this reason at the low filling considered in this work, the pairs try to distribute on different rungs, where they will form a singlet. The remaining rungs will stay empty.

 \begin{figure}
 \includegraphics[width=1.0\columnwidth]{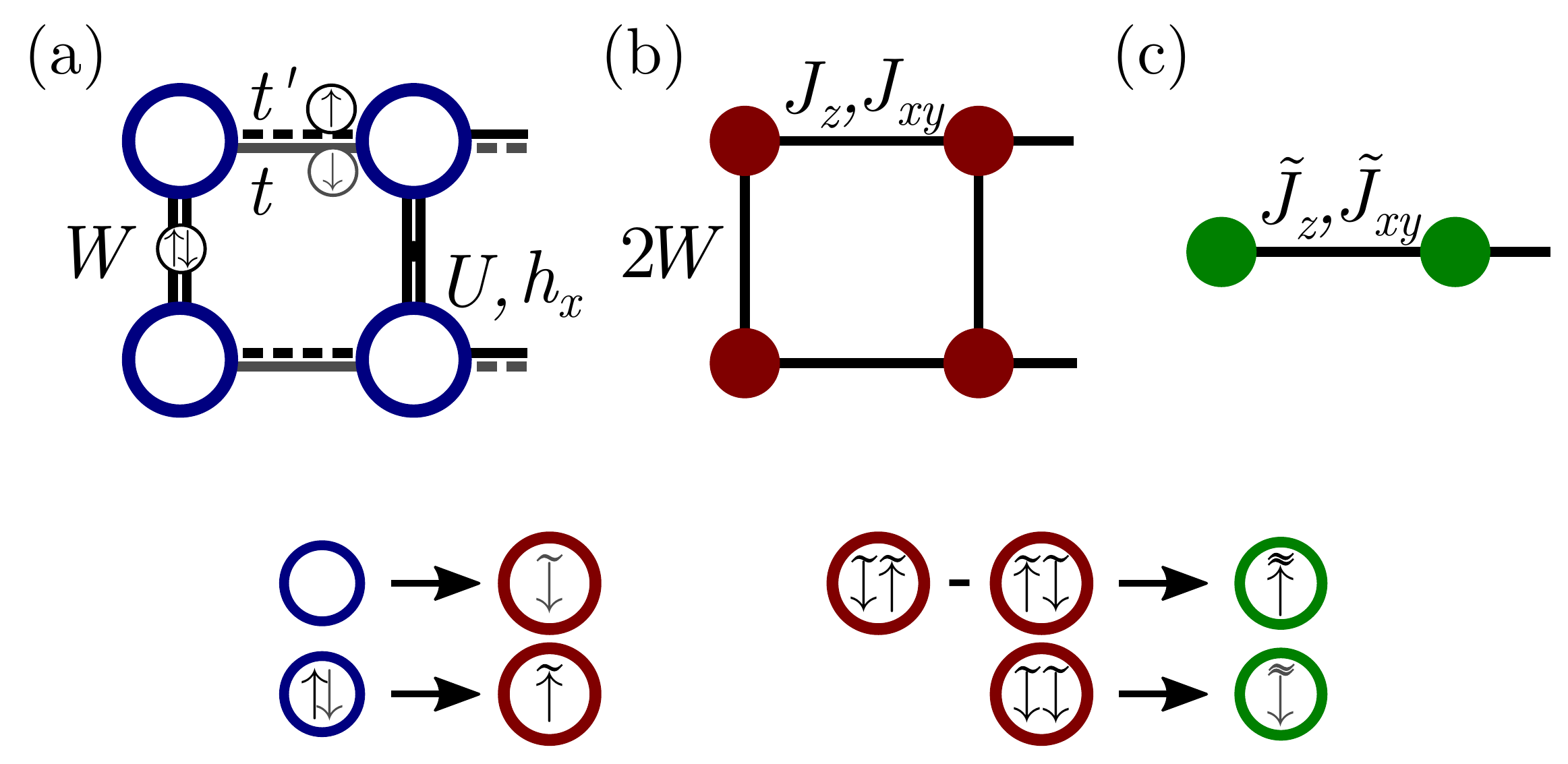}%
 \caption{ Sketch of the models we used to describe the strongly attractive regime. (a) The full model, (b) the spin ladder model describing how the pairs and empty sites interact, and (c) the spin chain model describing how the pair-singlets and the empty rungs interact (formed by a pair and an empty site in a single rung). 
 We show how the fermionic states are mapped to the spin ladder and how the in-rung states are mapped to the spin chain. 
 \label{FIG_SPCH}}
 \end{figure}

We can restrict ourselves to the subspace formed by this two in-rung states (the pair-singlet and the empty rung) and map this to a new pseudo spin states $\tilde{\uparrow}$ and $\tilde{\downarrow}$, effectively mapping each rung to a spin $1/2$. 
In this representation the system is mapped to a single spin chain of length $L$, and the magnetization will be fixed by the number of pairs. The interactions will be related to the ones we derived before by 2nd order perturbation theory by $\tilde{J}_z = J_z$ and $\tilde{J}_{xy} = 2J_{xy}$.
The effective Hamiltonian corresponding to the single spin chain is 
\begin{align}
H_{spin} &=  
 \tilde{J}_z \sum_{j=1}^{L-1} 
\tilde{S}^{z}_{j} \tilde{S}^{z}_{j+1}\\ \nonumber
&- \tilde{J}_{xy} \sum_{j=1}^{L-1} 
\frac{1}{2} (\tilde{S}^{+}_{j} \tilde{S}^{-}_{j+1} + \text{h.c.} ) \\ \nonumber
&+ \frac{\tilde{J}_z}{2} \left( \tilde{S}^{z}_{1} + \tilde{S}^{z}_{L} \right).
\end{align} 
As a consequence of the mapping a magnetic field of value $\tilde{J}_z/2$ appears only on sites $1$ and $L$. 
The new spin operators are obtained from the previous ones by
\begin{align}
\eta^{z}_{\alpha,j} &\to \frac{1}{2} \left( \tilde{S}^{z}_j - \frac{1}{2} \right), \nonumber \\
\eta^{+}_{\alpha,j} &\to \pm \frac{1}{\sqrt{2}} \tilde{S}^{+}_j  .
\end{align}
The plus-minus sign depends on the chain index $\alpha = a, b$, but since the ladder operators in the Hamiltonian always come in pairs this sign does not affect the effective Hamiltonian (but it should be taken into account if we want to measure, i.e.~ the correlations $xy$ between different legs).
We summarize the mapping realized on this section on Fig.~\ref{FIG_SPCH}.

\section{Method}
\label{SEC_Methods}

We use approximate methods and matrix product state (MPS) algorithms to obtain the phase diagram of the model. 
The MPS methods are variational methods based on matrix product states and are numerical exact methods \cite{White1992, White1993, VidalKitaev2003, DaleyVidal2004, Vidal2004, WhiteFeiguin2004, Schollwoeck2011}. 

To devise an MPS representation, we choose to snake the ladder as a one-dimensional system such that a general MPS is represented by
\begin{align}
\ket{\psi} = \sum_{\sigma_{a,1},\dots,\sigma_{b,L}} & 
A^{\sigma_{a,1}} 
A^{\sigma_{b,1}} \dots 
A^{\sigma_{a,L}}
A^{\sigma_{b,L}}
\ket{\sigma_{a,1} \dots \sigma_{b,L}}, 
\end{align}
using two matrices per rung ($a$ and $b$). 
In principle, the representation of any state by an MPS is exact. 
However, to make the approach feasible and efficient, the matrix dimensions are truncated. 
Here we use a so-called two-site truncation scheme, where always the matrix elements of two matrices are optimized at the same time. 
To do so it relies on making a singular value decomposition of the $A$ matrices, then keeping only the $m$ states corresponding to the largest singular values $s_i$ and truncating the rest. We will call $m$ the maximal bond dimension.
If the state is normalized then the sum of the squares of the singular values should be equal to 1 ($\sum_i s_i^2 = 1$), but after the compression a small amount of weight is lost. Hence, we define the discarded weight or truncation error as $ \epsilon = 1 - \sum_{i=1}^{m} s_i^2$. Since the algorithm is standard by now, we refer for details for example to the review \cite{Schollwoeck2011}. 

Given the ground state, if we bisect the system at a given bond into two subsystems $A$ and $B$, we are able to trace the degrees of freedom of one half and define the reduce density matrix of this bisection as
\begin{equation}
\rho_A = \Tr_{B} \ket{\psi}\bra{\psi} 
 = \sum_{i} \Lambda_i \ket{i}_A {}_A\bra{i},
\end{equation}
where $\left\lbrace \ket{i}_A \right\rbrace$ is an orthonormal bases of the subsystem $A$ obtained after a Schmidt decomposition of the state $\ket{\psi}$.
The eigenvalues $\Lambda_i$ of the reduced density matrix are the squares of the singular values that arise from the before mentioned compression at a given bond, meaning $s_i^2 = \Lambda_i$. 
This eigenvalues form the entanglement spectrum given by $-\ln \Lambda_i$. 
In order to measure the entanglement between subsystems in the closed system, one can use the von Neumann entropy of a subsystem A defined by
\begin{equation}
S(\rho_A) = -\Tr( \rho_A \ln \rho_A ) = - \sum_{i} \Lambda_i \ln \Lambda_i.
\end{equation}
This quantity is the von Neumann entropy of the reduced density matrix $\rho_A$ and it is a measure of the entanglement of the subsystem A with the remainder of the system. 
In the rest of this paper we are going to refer to the entanglement spectrum of the subsystem which arises by a bisection at the central bond of the system within the MPS representation. For an even number of rungs, this cuts the system between two rungs and for an odd number of rungs, this cuts the central rung of the system.

It is important to mention that finding the ground state within the topological phase can be a challenge. This is due to the nearly degenerate topological edge modes in this model. Care has to be taken not to be trapped in a meta-stable state. Here we use conserved quantities such as the parity on each chain, where we found this problem to be typically much less pronounced. To ensure the convergence of the ground state we used different initial states, noise during the MPS first few sweep and carefully checked the usual convergence parameters. The simulations on this work were made with the DMRG algorithm using the Itensor libraries \cite{FishmanStoudenmire2020}.

\section{Characterization of topological properties and summary of results}
\label{SEC_TOP}

 \begin{figure*}
 \includegraphics[width=0.65\columnwidth]{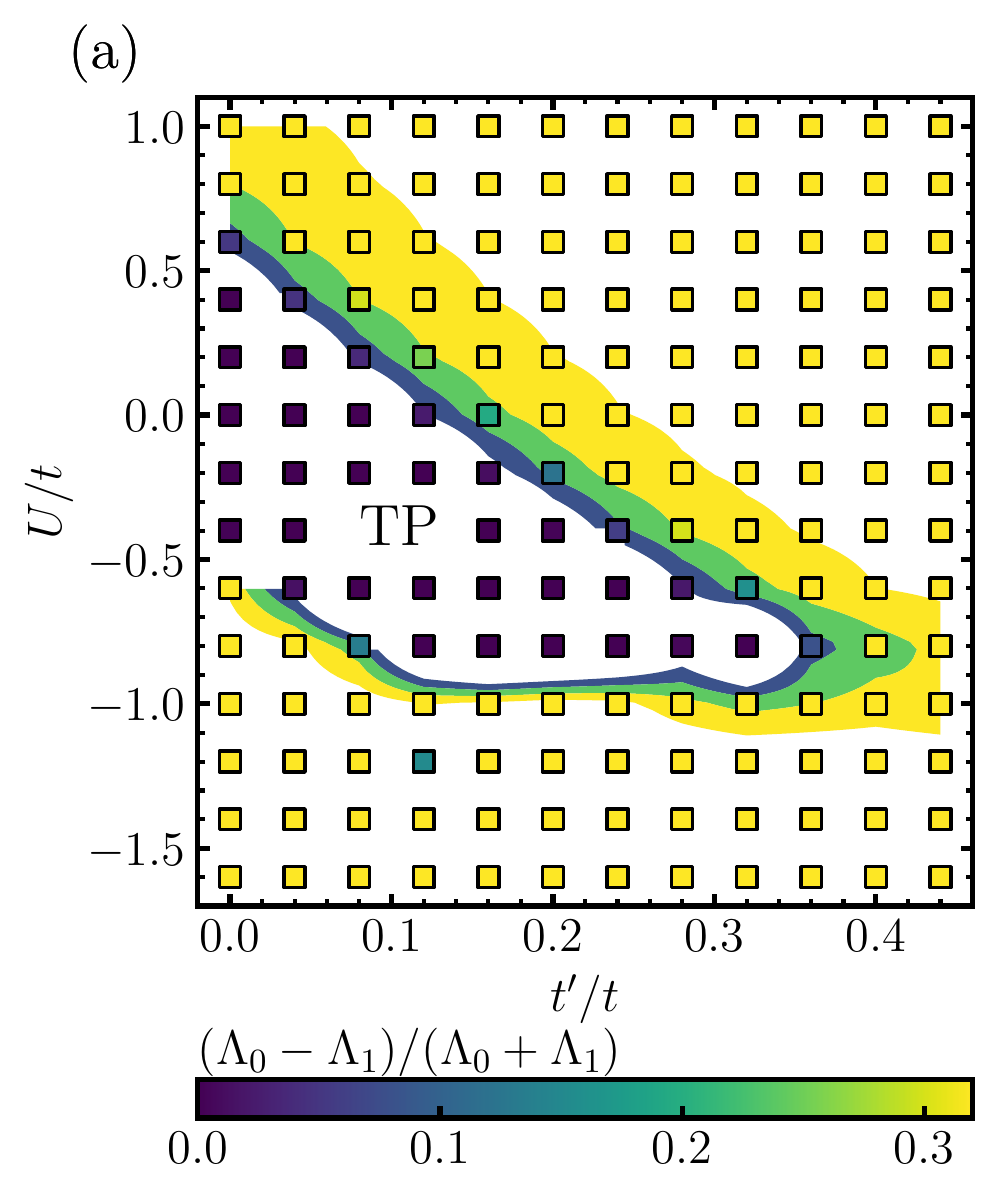}%
 \includegraphics[width=0.65\columnwidth]{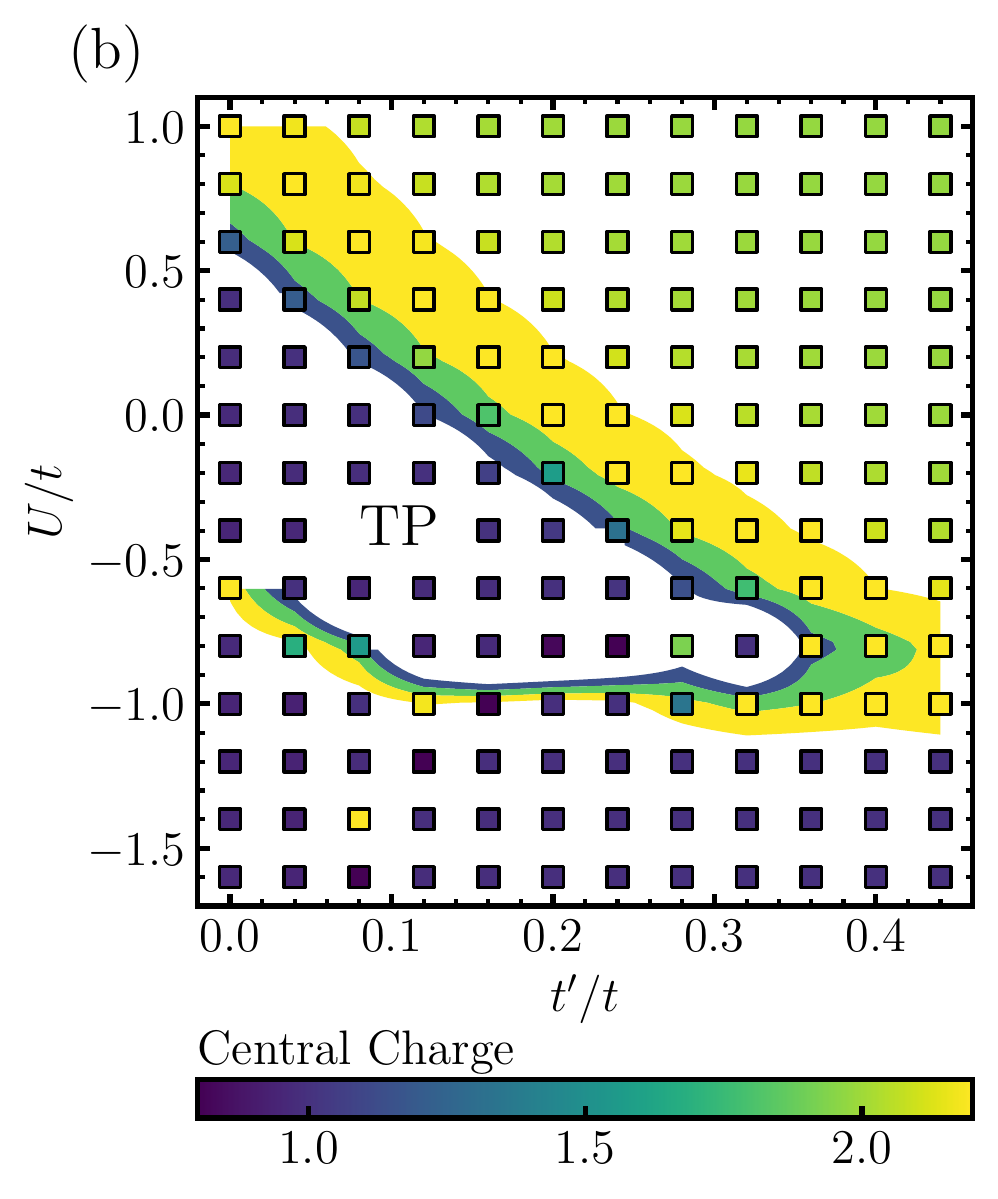}%
 \includegraphics[width=0.65\columnwidth]{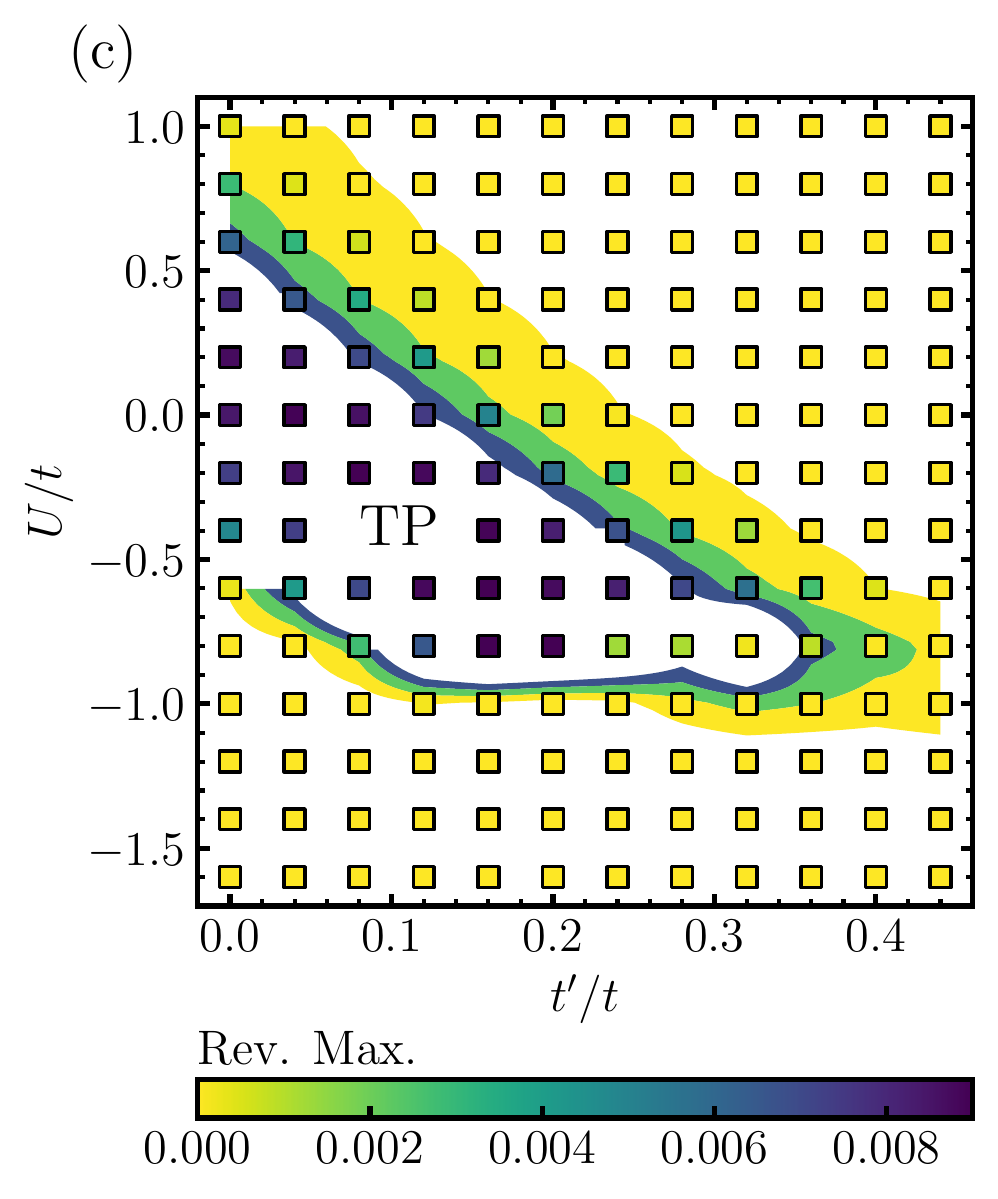}%
 \includegraphics[width=0.10\columnwidth]{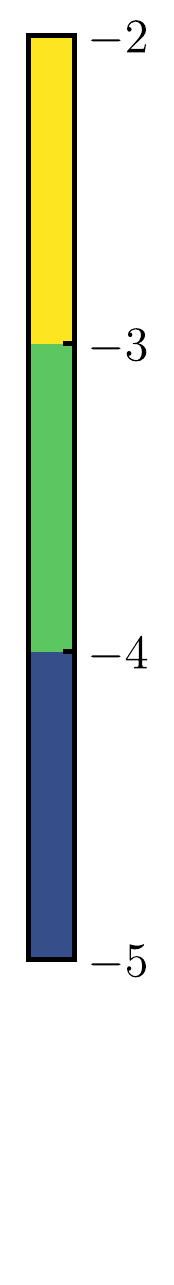}%
 \caption{Ground state phase diagram of the model $H$ in Eq.~\ref{EQ_H} for a system with $L=100$ rungs, concentrating on the topologically phase in the central area (marked with `TP').
 The color plots in the background of all panels show the same quantity, the logarithm of the energy difference between the two parity sectors [$ \log(\Delta_{Pa}) $]. Hence they correspond to the criterion (i). The corresponding color bar is shown on the right. In panel (a), the coloured symbols correspond to criterion (ii), since the difference of the first two eigenvalues $(\Lambda_0 - \Lambda_1)/(\Lambda_0 + \Lambda_1)$ of the reduced density matrix is shown. The symbols correspond in panel (b) the central charge, the topological region has a central charge $c=1$. In panel (c) the symbols show the difference between the maximum value of the correlations in the opposite end of the ladder and the maximum in the middle of the ladder. 
 \label{FIG_SUM}}
 \end{figure*}
 
In order to determine whether a system presents a topologically non-trivial phase supporting Majoranas as end-states, we rely on the detection of the following four specific features \cite{KrausZoller2013}: 

\begin{enumerate}[(i)]
\item The lowest lying states in each parity sector ($P_a = 0, 1$) are quasi-degenerate with an energy difference [$ \Delta_{P_a} = E(P_a = 1)-E(P_a = 0) $] that decays exponentially with the system size.
\item The entire entanglement spectrum has an even degeneracy.
\item The single particle correlations on each leg 
($ A^{\uparrow \uparrow}_{1j} = \expval{ \alpha^{\dagger}_{\uparrow 1} \alpha^{\pdagger}_{\uparrow j} } $) decay exponentially into the bulk, but have a revival on the opposite edge.
\item All these properties have to be robust against local perturbations that preserve the parity symmetry.
\end{enumerate}

If all these properties are fulfilled this will indicate the presence of a topological non-trivial phase with Majorana end-states.

\paragraph*{Summary of results:}
In the remainder of this section we want to summarize the results which we found concerning the topological nature of phases in the model under consideration. In Fig.~\ref{FIG_SUM} we show a phase diagram concentrating on the existence of the topological phase (TP). 
 
 We start the study of the system at $t'=U=0$ and $h_x=-t$ where the system is a variant of the spinless model considered in Ref.~\cite{KrausZoller2013}. 
However, instead of $W=1.8t$ for which the topological phase was investigated, we use the larger value of $W=2.6t$ which in preliminary runs seemed to be more favouring the topological phase in our model. We hand-wavingly explain this enhanced pair hopping that it corrects for the fact that we have only every second pair hopping and therefore, the actual amplitude of the pair hopping needs to be larger.

Exploring further the parameter space by switching on $t'$ and $U$, we find that this phase extents for a finite regime of parameters preserving the same general features. This can be seen in Fig.~\ref{FIG_SUM}. In panel (a) we plot contour lines of the logarithm of the energy difference, $\log(\Delta_{Pa})$, and the difference between the first two eigenvalues $(\Lambda_0 - \Lambda_1)/(\Lambda_0 + \Lambda_1)$ of the reduced density matrix with coloured squares. 
Both quantities are expected to be exponentially close to zero in the topological phase corresponding to criteria (i) and (ii) outlined above. We can see that both approximately vanish for a finite region around $U=t'=0$. In particular, for $U=t'=0$ the energy difference $\Delta_{Pa}$ is smaller than $10^{-12}$. We will analyse this behaviour in the following sections concentrating on the criterion (i) in Section \ref{SEC_GAP} and criterion (ii) in Section \ref{SEC_ENT}. In particular the scaling with system size will be discussed in more detail.
In Fig.~\ref{FIG_SUM}b we show the behaviour of the central charge which is extracted from the von Neumann entropy. Typically, the central charge rises at phase boundaries when a gap closing occurs. 
Within the topological phase, we expect the central charge to take the value $c=1$, since we have a gapless sector present in this phase \cite{KrausZoller2013}. We see that a maximum of the central charge surrounds the topological phase. Additionally, we see that there is a large value occurring below the topological phase separating two phases from each other.
In Fig.~\ref{FIG_SUM}c we show the maximum of the single particle correlation function for a distance of almost the system size minus the maximum value for a distance close to the middle of the system. 
The single particle correlation function in the topological phase following criterion (iii) should show a revival signalling the Majorana mode. In panel (c) we see a maximum occurring within the topological phase. Outside the topological phase, the revival is clearly missing. We discuss this behavior in Section \ref{SEC_SPC}.

On top of this, we have checked that all these properties are robust against static noise and we show the results in Section \ref{SEC_NOI}. 
Thus, we conclude, that we could identify in the interacting fermionic model, Majorana edge modes occurring in an extended topologically non-trivial phase marked by `TP' in Fig.~\ref{FIG_SUM}.

\section{Results}
\label{SEC_RES}

\subsection{Criterion (i): Behaviour of the gap $\Delta_{Pa}$}
\label{SEC_GAP}

 \begin{figure}
 \includegraphics[width=1.0\columnwidth]{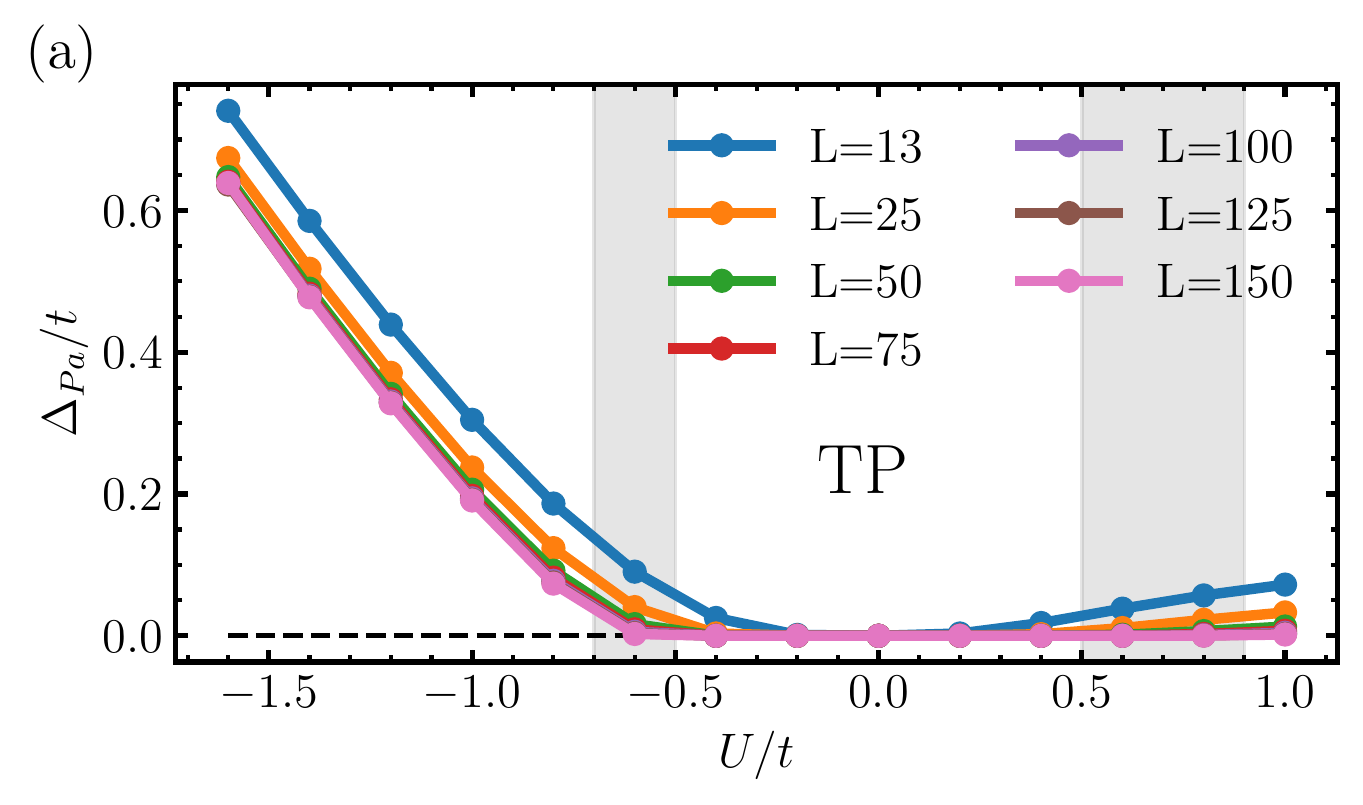}\\
 \includegraphics[width=1.0\columnwidth]{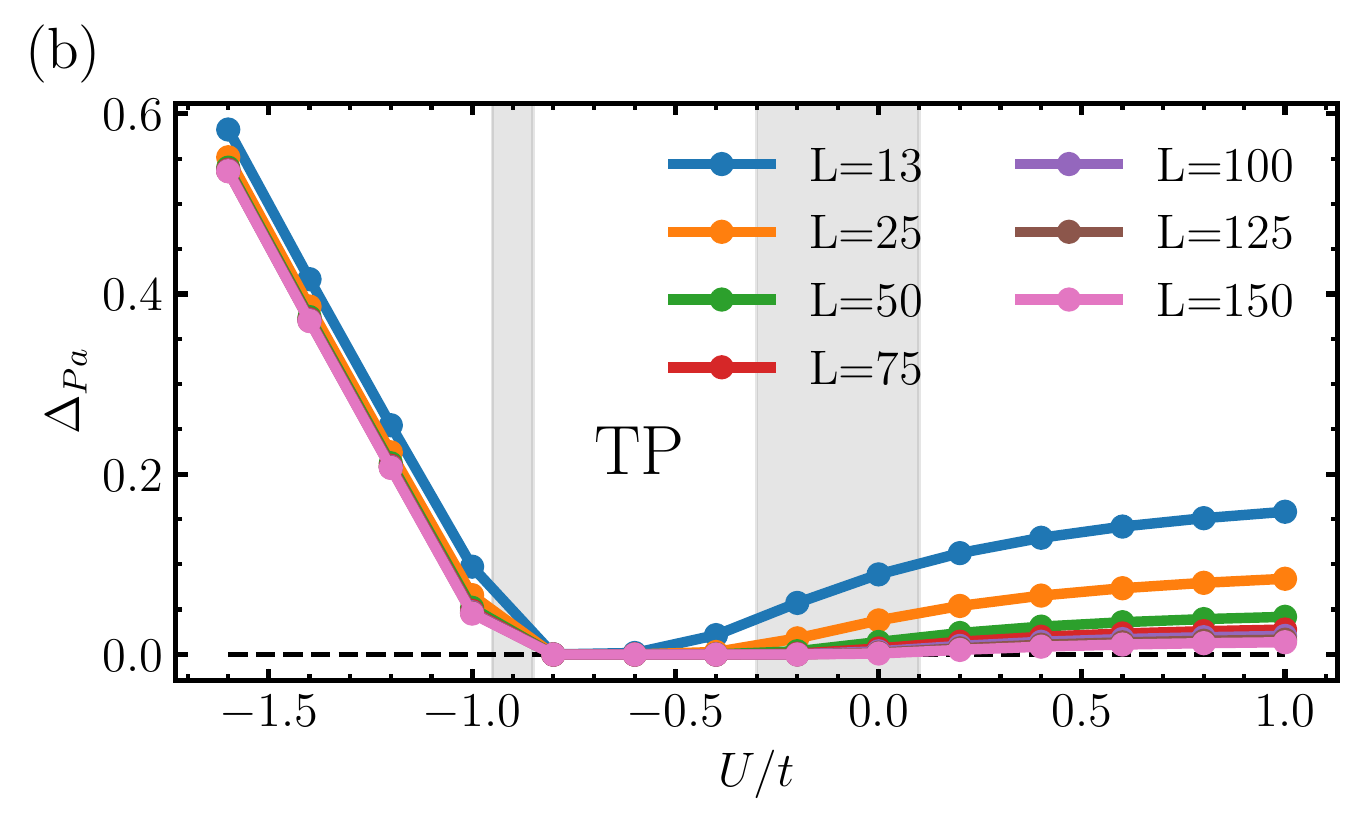}
 \caption{ 
 Cuts for the gap $\Delta_{Pa}$ between both parity sectors. We identify the region TP with the topologically non-trivial phase. 
 We show cuts at $t'=0$ and $t'=0.2t$ on panels (a) and (b), respectively. 
 The gray regions are the same as the ones plotted in Figs.~\ref{FIG_ENTR_tp00} and \ref{FIG_ENTR_tp20}.
 \label{FIG_GAP}}
 \end{figure}

As defined in criterion (i), the topological non-trivial phase possesses a vanishing energy difference $\Delta_{Pa}$ between even and odd parity sectors of chain $a$, namely $ \Delta_{Pa} = E(P_a = 1) - E(P_a = 0) $. 
The closing of the gap $\Delta_{Pa}$ with the system size is exponentially fast. The MPS calculations with different system sizes are done conserving the single chain parity. The gap is evaluated calculating the energy for both parity sectors ($P_a= 0, 1$), separately, and taking the difference. 

The overview of the behaviour of the gap was given in Fig.~\ref{FIG_SUM} for a fixed system size of $L=100$ rungs. We associated the region in which the values of the gap $\Delta_{Pa}/t$ lies below $10^{-5}$ with the topologically non-trivial phase. 
In this section we will explore in more detail the finite system scaling in order to show the exponential closing of the gap $\Delta_{Pa}$ in the topological phase. 

In Fig.~\ref{FIG_GAP} we show the gap $\Delta_{Pa}$ for different system sizes and cuts through the parameter space shown in the phase diagram Fig.~\ref{FIG_SUM}. 

In panel (a) we show the cut for $t'=0$ versus the interaction strength $U$ for different lengths of the ladder. For the non-interacting case at $U=0$, the gap is indistinguishable from zero within the symbol size for all system sizes shown. 
Around $U=0$, the gap remains zero for a finite range of $U$. Outside of this region the value of the gap becomes finite. 
The general behaviour is not symmetric with respect to $U$. While for a repulsive interaction ($U>0$) the energy difference smoothly increases, for an attractive one ($U<0$) the rise is a lot faster. 

A similar behaviour can be seen for a cut at $t'=0.2t$ on panel (b), where we can clearly distinguish the same three regions. The size of the central region in which the energy difference is close to zero, becomes smaller at finite $t'$ than it was at $t'=0$ for each system size. Additionally, the region is displaced down to more attractive interactions around $U\simeq -0.5t$. In particular, the non-interacting system ($U=0$) shows only for large system sizes the closing of the gap. 
In general for repulsive $U$ the gap changes considerably with the system size, whereas at attractive $U$ the value of the gap is much more stable with the system size.
 
In the following we analyse this further by means of a finite size scaling of the three different regions. It is worth mentioning that the system with $L=13$ rungs has been included on the fittings since we consider that the possible deviation arising from the small difference in density is negligible.

\begin{figure}
 \includegraphics[width=1.0\columnwidth]{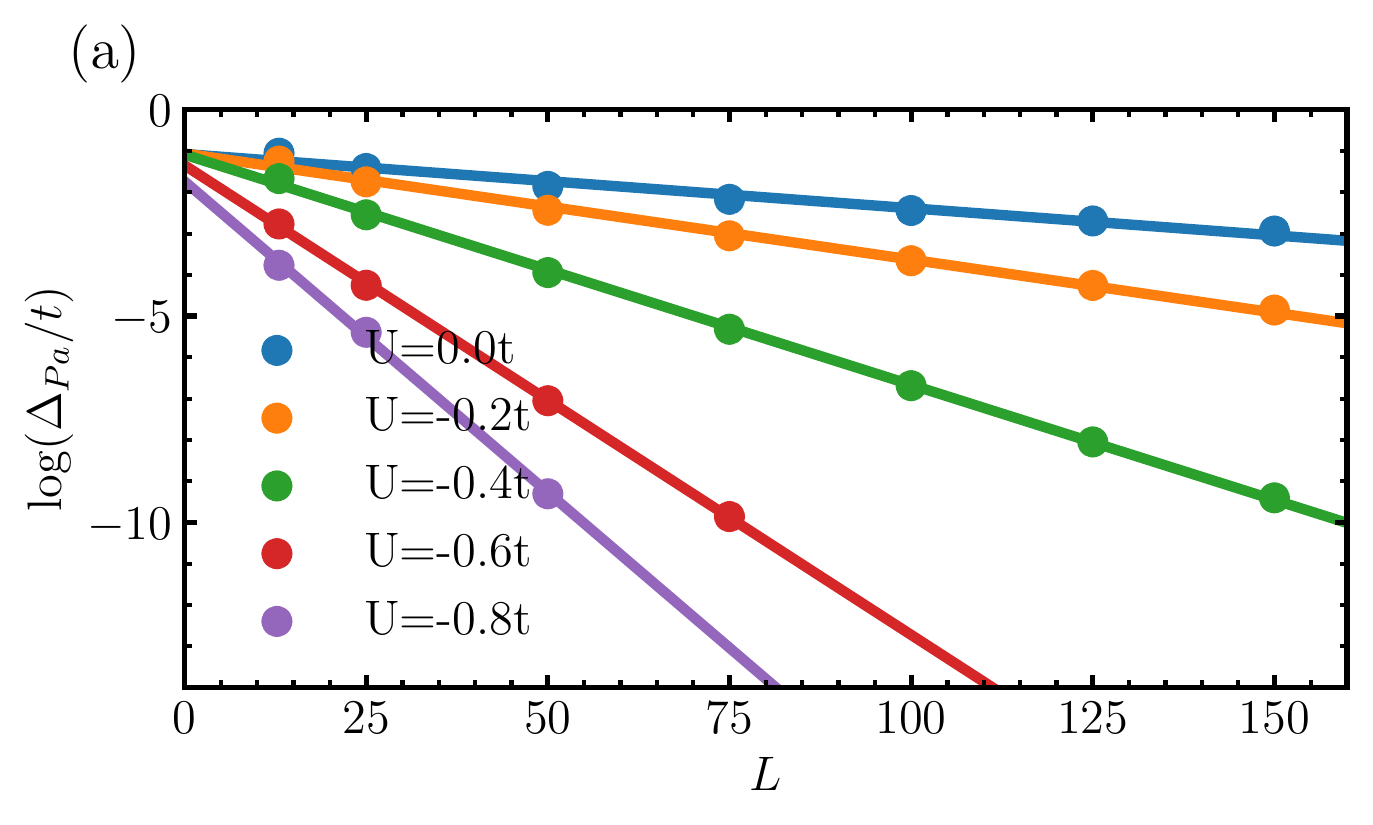}\\
 \includegraphics[width=1.0\columnwidth]{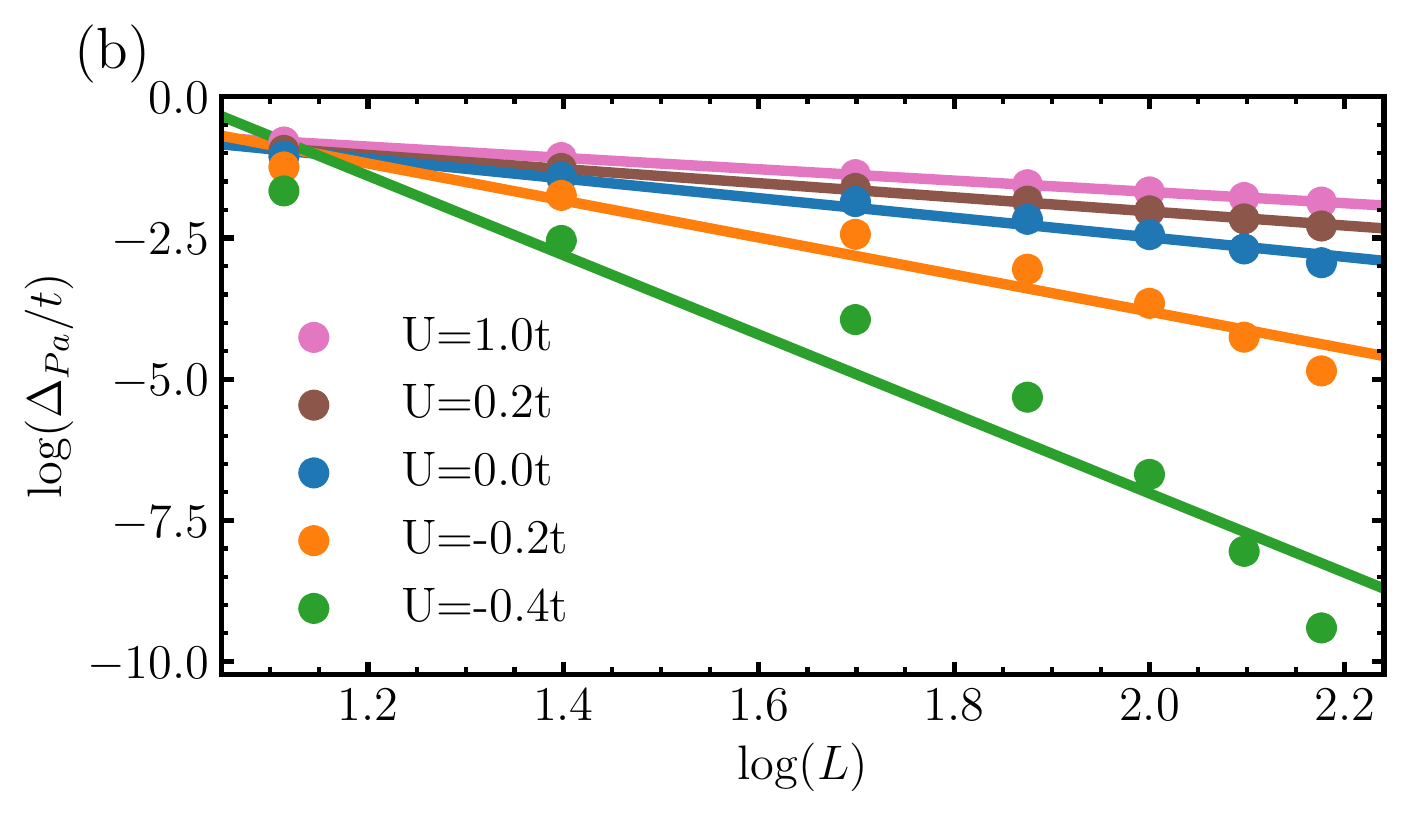}\\
 \includegraphics[width=1.0\columnwidth]{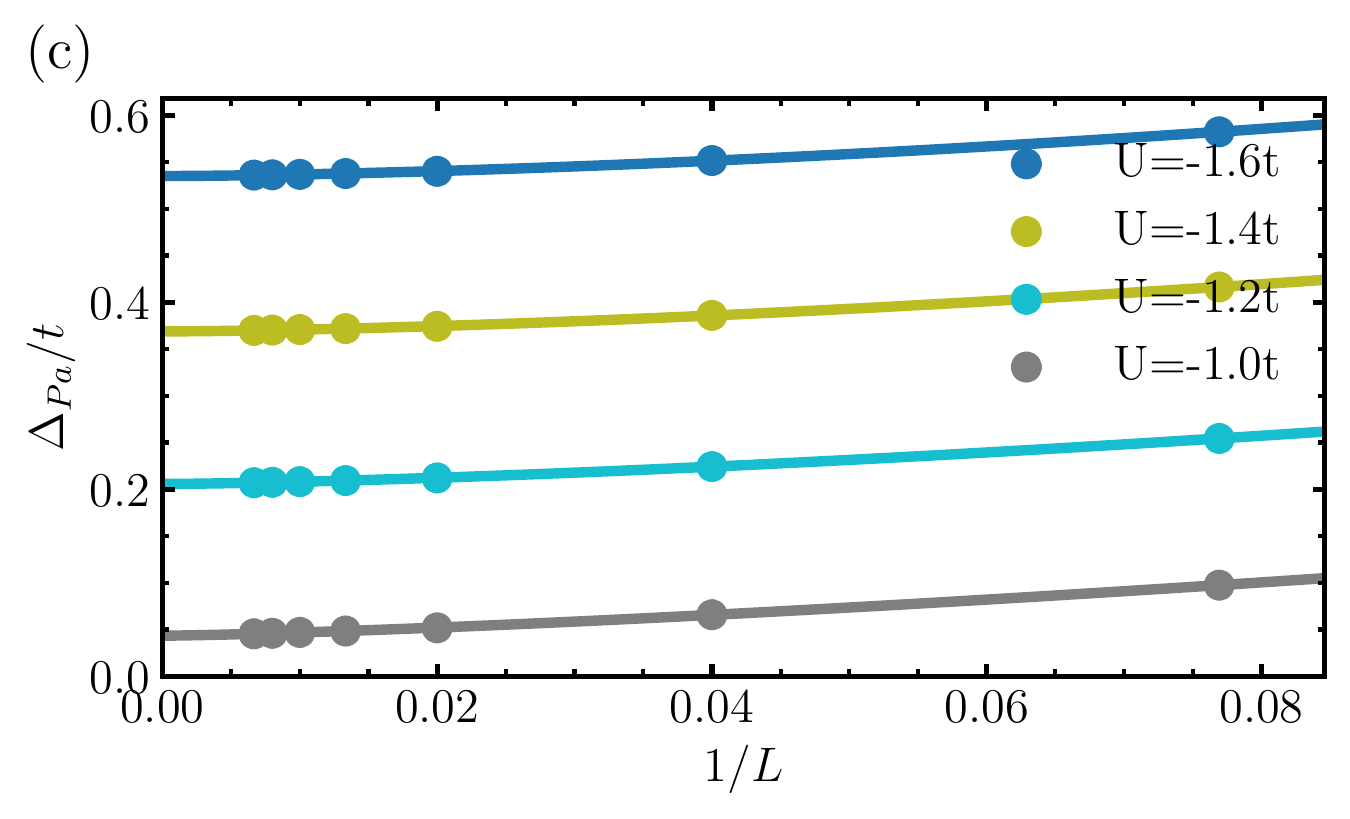}%
 \caption{ 
 Scaling of the energy difference between parity sectors.
 Here we present the results for $t'=0.2t$ on the three different regions. Panel (a) corresponds to points near the topological phase. 
 Panel (b) corresponds to the region above the topological phase (larger $U$). Panel (c) corresponds to the region below the topological phase (smaller $U$). The lines in panels (a) and (b) to linear fittings, in panel (c) correspond to the fitting of a function $a+b(1/L)^\alpha$. Every value below $10^{-10}$ was set to zero. 
 \label{FIG_GAPSCAL}}
 \end{figure}

In the topological region, the criterion (i) predicts an exponential scaling of the gap with the system size. Here we focus on $t'=0.2t$. 
Our results in this phase, e.g. $U=-0.6t$ in panel (a) of Fig.~\ref{FIG_GAPSCAL}, are in good agreement with an exponential scaling with system size. We verified that this scaling is stable in the region where we see a vanishing gap in Fig.~\ref{FIG_GAP}b. Also for values of $U=-0.2t, -0.4t, -0.8t$ in panel (a) of Fig.~\ref{FIG_GAPSCAL} the exponential scaling with system size seems to work very well. In panel (b) we show for comparison the log-log plot where a straight line would signal an algebraic scaling. For values of $U= -0.2t, -0.4t, -0.6t, -0.8t$ the exponential scaling seems to be more appropriate than the algebraic. 
This is supported by the coefficient of determination $R^2$ of the corresponding fits shown in Table \ref{TAB_R2}. Whereas the exponential fits work very well ($R^2$ is close to one) for interaction strength in the topological phase, i.e.~ $U=-0.2t, -0.4t, -0.6t, -0.8t$, the algebraic fits are not as good. In contrast for values of the interaction $U=0$ and larger, the algebraic fits work better than the exponential fits. 
Thus, the scaling of the gap (and the behaviour of $R^2$) would suggest that the upper phase boundary of the topological phase lies around $U=-0.1t$. 

\begin{table}[]
\begin{tabular}{|l|l|l|}
\hline
 $U/t$  & lin-log (exp) & log-log (pow)  \\ \hline
 $1.0$  &    -     & $0.99992$  \\ \hline
 $0.2$  &    -     & $0.9971$  \\ \hline
 $0.0$  & $0.971$  & $0.979$  \\ \hline
 $-0.2$ & $0.996$  & $0.938$  \\ \hline
 $-0.4$ & $0.9993$  & $0.919$  \\ \hline
 $-0.6$ & $0.9997$ & -  \\ \hline
 $-0.8$ & $0.9988$ & -  \\ \hline
\end{tabular}
 \caption{ 
 Coefficient of determination $R^2$ for the linear fitting ($Ax+B$) for $\log(\Delta_{Pa}/t)$ vs $L$ (lin-log, linear dependence corresponds to an exponential) $\log(\Delta_{Pa}/t)$ vs $\log(L)$ (log-log, linear dependence corresponds to a power law). This results correspond to $t'=0.2t$.
}
\label{TAB_R2}
\end{table}

The regions above and below the topological phase clearly show a different dependence on system size. The results for different system sizes in the region of attractive interaction outside of the topological phase ($U \lesssim -0.9t$) are almost converged to a finite value and the scaling with the system length is rather stable (see Fig.~\ref{FIG_GAPSCAL}c). The value of the gap tends approximately as $L^{-1.5}$ to a constant value as shown in panel (c) of Fig.~\ref{FIG_GAPSCAL}. 	
 
Let us explain this behaviour considering the strongly attractive $U$ limit. 
For strongly attractive $U$ the system wants to pair all fermions (see Section \ref{SEC_LARGEU}). 
For the even sector ($P_a = 0$) this is possible, but in the odd sector ($P_a = 1$) both legs are forced to have an odd number of fermions and therefore one pair has to be broken with respect to the even sector. 
If we only consider first order in the attractive case for $P_a=0$ ($P_a=1$) the fermions will form $\frac{N}{2}$ ($\frac{N}{2}-1$) isolated singlet states between a pair and an empty on different legs.
In this limit the energy will come only from $H_W$ and $H_U$ and it will be $E(P_a=0) = (U-W)\frac{N}{2}$ and $E(P_a=1) = (U-W)(\frac{N}{2}-1)$.
The energy of breaking a pair is proportional to $\Delta_{Pa}=W-U$, since a pair is broken ($-U$) and also it cannot be delocalized over a rung ($W$). 

\begin{figure}
 \includegraphics[width=1.0\columnwidth]{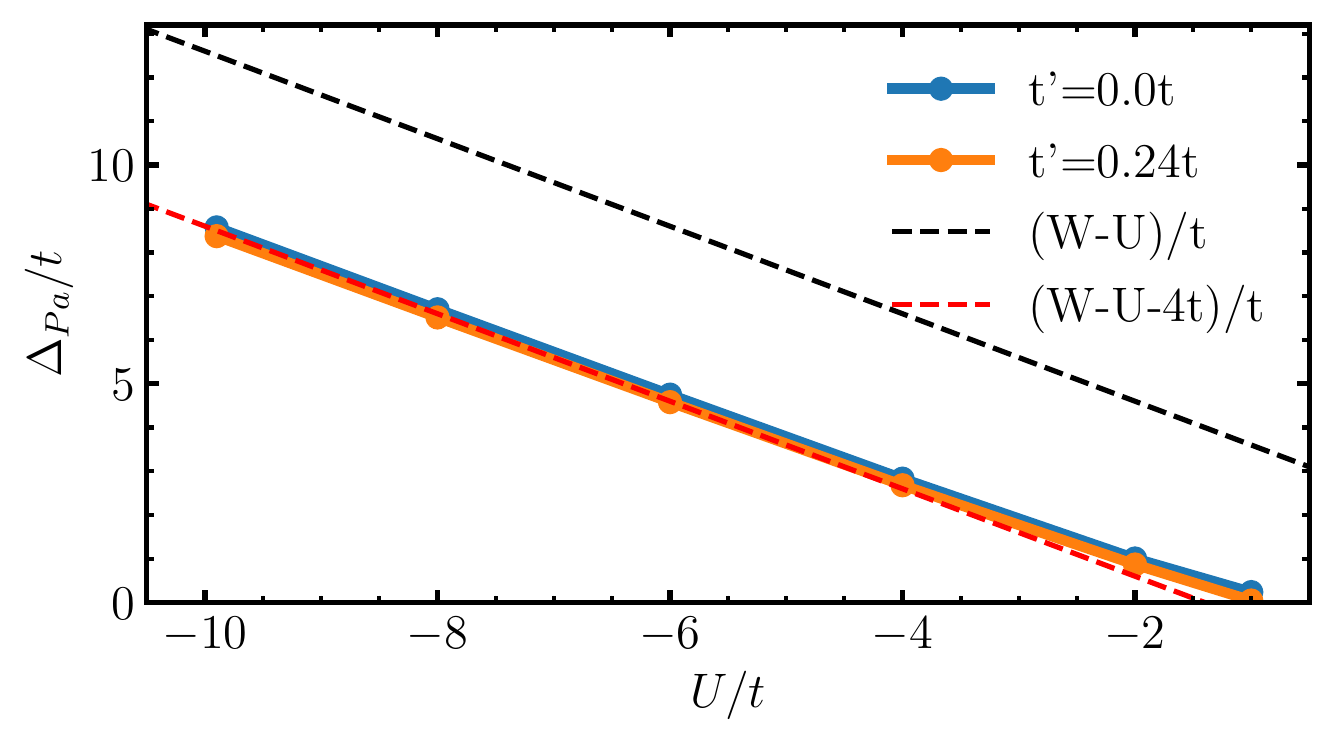}%
 \caption{ 
 Gap $\Delta_{Pa}$ between parity sectors as a function of $U/t$ in the strongly attractive limit for a ladder with $L=25$ rungs. 
 The energy difference agrees with $W-U-4t$ (red dotted line).
 \label{FIG_GAPvsU}}
 \end{figure}

In Fig.~\ref{FIG_GAPvsU} we see that the behaviour of the gap with the interaction strength is approximately linear supporting this statement. Even more, the results agree well with $\Delta_{Pa}= W-U-4t$ signalling the expected behaviour with the interaction and the pair hopping. The shift $-4t$ is introduced by hand, since we expect that the two unpaired fermions can delocalize within each chain giving an energy of the order of a combination of $t$ and $t'$. 

In contrast, the region above the topological phase ($U > - 0.1t$ for $t'=0.2t$) shows a very strong dependence on the system size (even though not as strong as the exponential closing of the gap in the topological region).
In particular, the energy difference $\Delta_{P_a}$ seems to vanish algebraically in the thermodynamic limit. 
In panel (b) of Fig.~\ref{FIG_GAPSCAL} the straight lines correspond to the fitting of a function $\log(\Delta_{Pa}/t)=a+b\log(L)$.
In particular for $U=t$ the scaling goes approximately as $1/L$ since the fitted exponent has a value $b=-1.0037$. 
But as we decrease $U$ and we get closer to the topological phase this exponent grows until we are not able to distinguish it from an exponential.

Such a scaling with $1/L$ can be justified from the strongly repulsive regime. 
For large enough $U$, the single particle part of the Hamiltonian $\tilde{H}_{t t'}+\tilde{H}_x$ (see Eq.~\ref{EQ_REP} on Section \ref{SEC_LARGEU}) governs the behaviour of the system. This corresponds to two decoupled chains of fermions. 
The low energy properties of these chains at typical fillings can be described by a Luttinger liquid. 
One characteristics of a Luttinger liquid is the linear dispersion relation. 
Since changing from the even sector to the odd one corresponds to transferring a fermion from one chain to the other, this only changes the filling in each Luttinger liquid. 
Due the linear dispersion the scaling of the resulting gap with the system size is $1/L$.

This argument is further supported by direct calculations assuming large interaction and sufficiently large magnetic field, such that the system is almost fully polarized along the field direction. 
In this limit the system constituents are free fermions, and changing from the even sector to the odd one corresponds to transferring an fermion from one chain to the other. 
This only changes the filling in each chain. 
 
To determine the exact upper boundaries of the topological phase is difficult, since we need to distinguish between the $1/L$ scaling of the gap and the exponential scaling. 
Therefore, we cannot give a detailed upper phase boundary of the topological region. 
Nevertheless, from the scaling behaviour we estimate that the topological phase survives up to a value around $U^{U}_{C} \simeq -0.1t$ for $t'=0.2t$ as shown in the different panels of Fig.~\ref{FIG_GAPSCAL}.

\subsection{Criterion (ii): signatures in the entanglement spectrum}
\label{SEC_ENT}

 \begin{figure}
 \includegraphics[width=1.0\columnwidth]{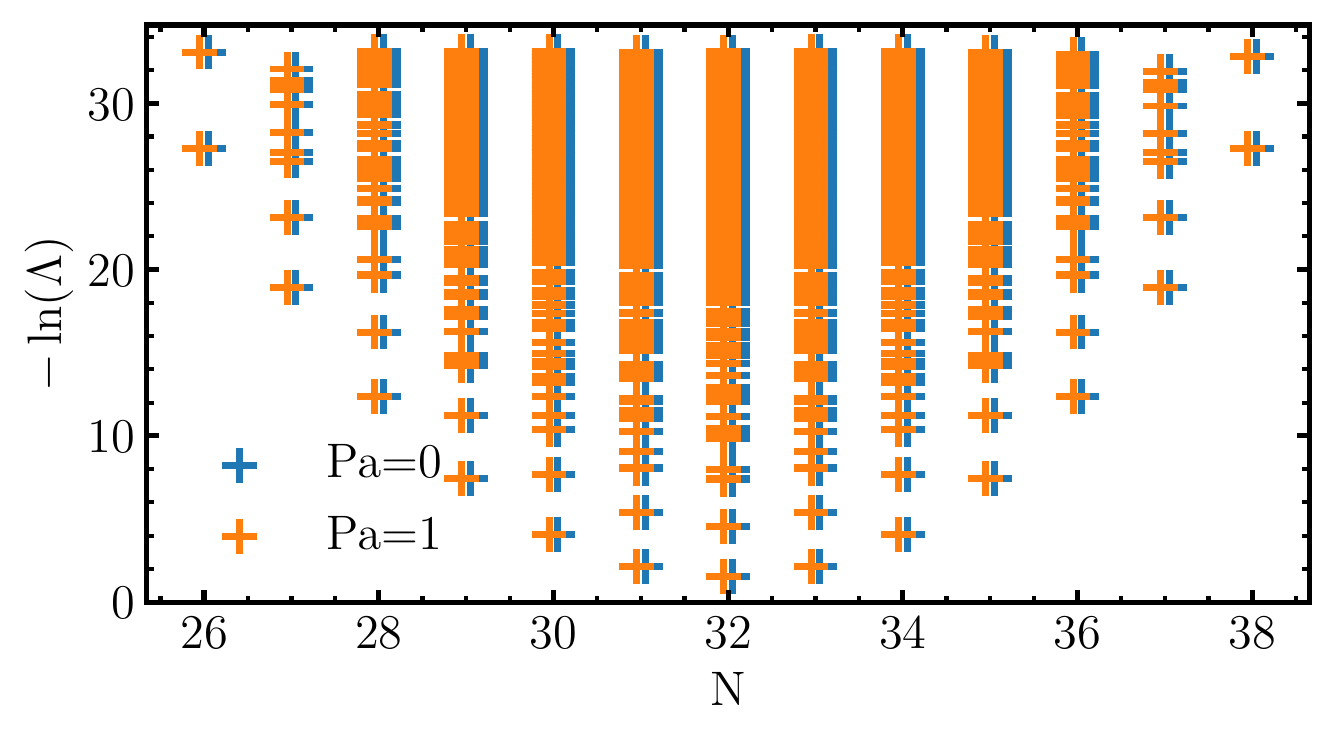}%
 \caption{ 
Entanglement spectrum in each sector corresponding to the particle number ($N$) in one half of the system. The eigenvalues are additionally labelled with the parity sector ($P_a$) to which they belong. The results correspond to a lattice with $L=100$ rungs at $t'=0.08t$ and $U=-0.4t$ and show the expected double degeneracy. 
 \label{FIG_PARAB}}
 \end{figure}

In this section we study the entanglement properties of the system concentrating on the entanglement spectrum and central charge, which is extracted from the behaviour of the von Neumann entropy as a function of the bisected bond (see appendix \ref{SEC_CCFIT}). 
As listed above, one of the features that characterizes a topological phase is a double degeneracy on its entire entanglement spectrum [feature (ii)]\cite{TurnerBerg2011}. 
In section \ref{SEC_TOP} in Fig.~\ref{FIG_SUM} we gave a set of summary plots depicting different quantities that allows us to describe the phase diagram of the model. 
In panel (a) we show the behaviour of the degeneracy of the lowest eigenvalues of the density matrix $(\Lambda_1-\Lambda_0) / (\Lambda_1+\Lambda_0)$. 
The topological phase is clearly distinguished by the vanishing difference of the lowest eigenvalues (dark color). The degeneracy is broken in the surrounding phases for the shown system size ($L=100$ rungs).
 
 \begin{figure}
 \includegraphics[width=1.0\columnwidth]{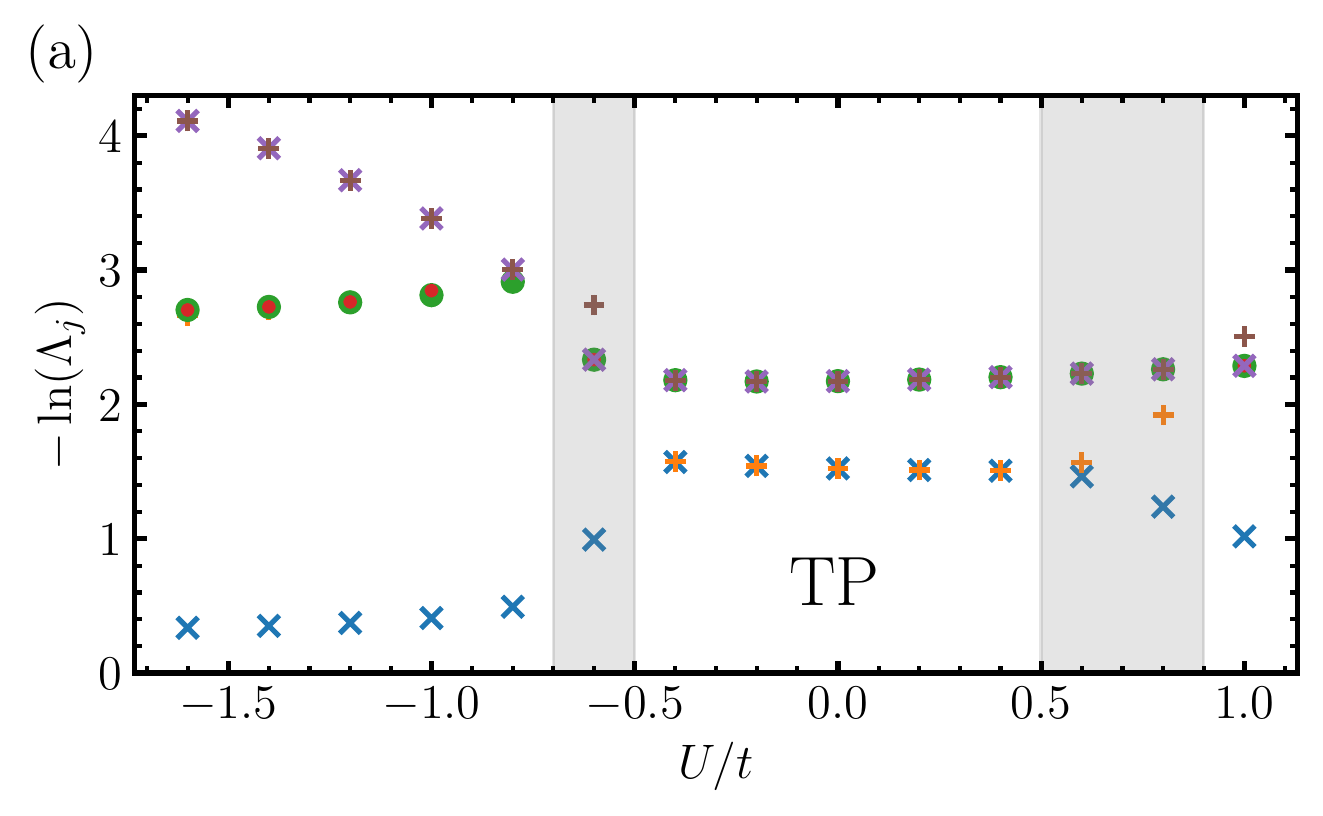} \\
 \includegraphics[width=1.0\columnwidth]{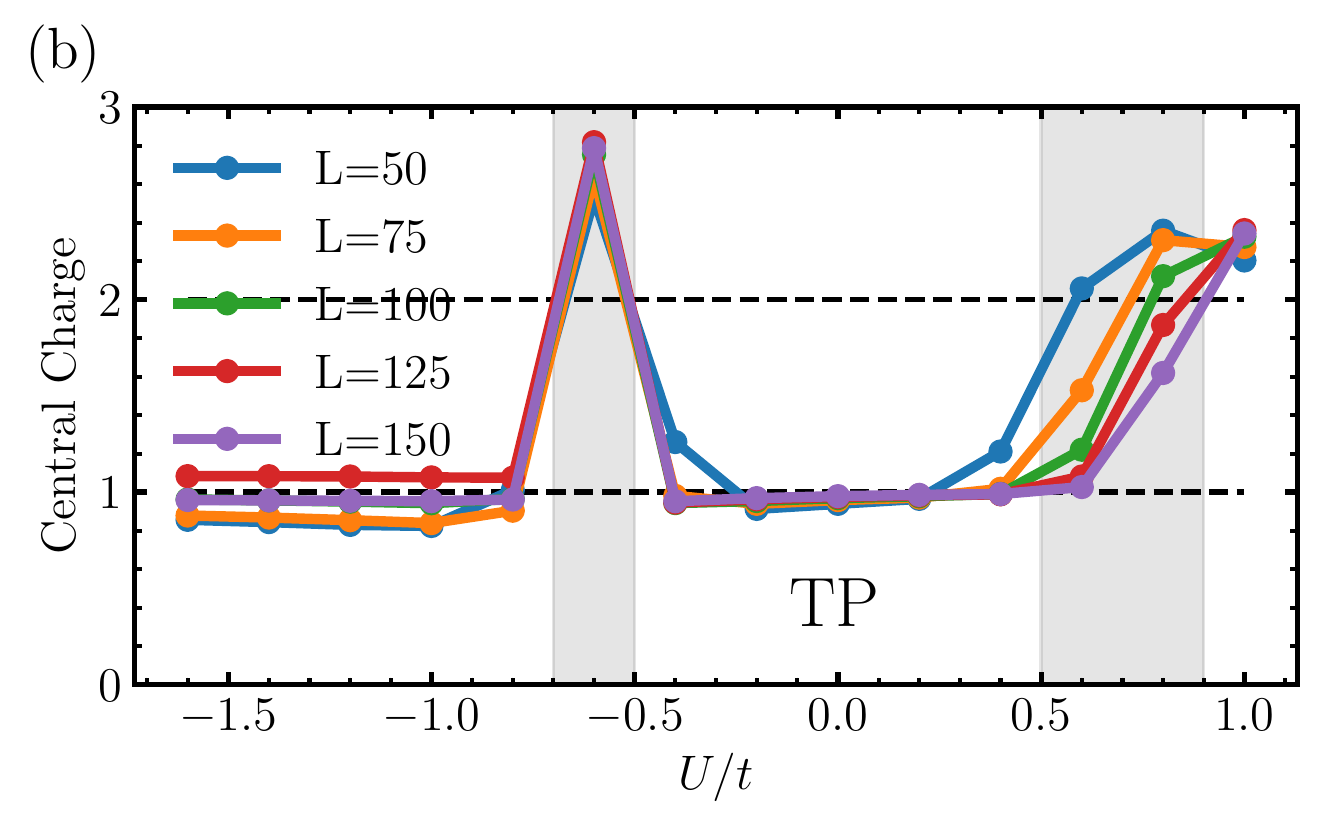}%
 \caption{ 
 (a) The entanglement spectrum and (b) central charge as a function of the interaction $U$ for $t'=0$. 
 The topological phase is characterized by a double degeneracy on the entanglement spectrum, here for a region approximately in between $ -0.5t < U < 0.5t $.
 After that the even degeneracy is broken and the system is in a topologically trivial state on both directions. 
 The reorganization in the entanglement spectrum (gray region) coincide with the regions where the central charge peaks. 
 \label{FIG_ENTR_tp00}}
 \end{figure}

 \begin{figure}
 \includegraphics[width=1.0\columnwidth]{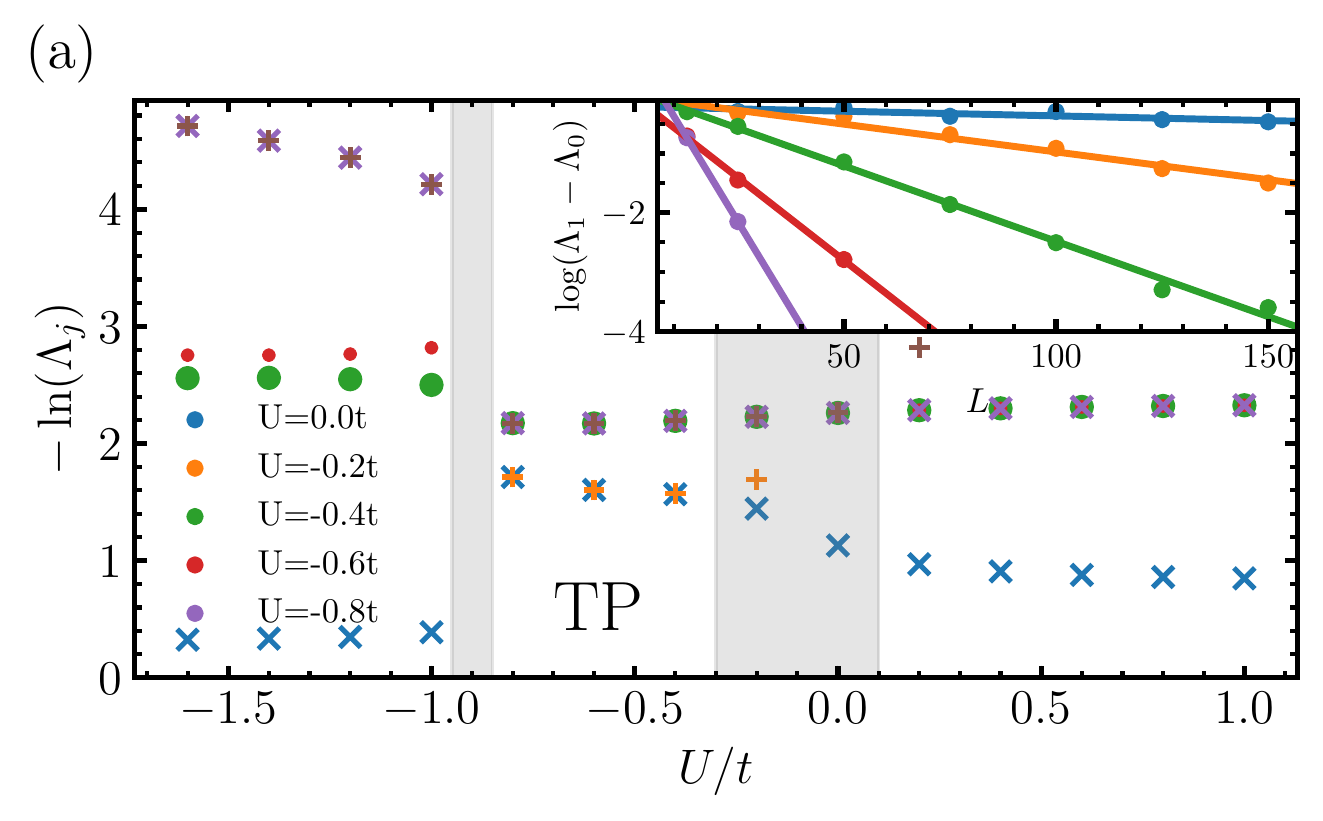} \\
 \includegraphics[width=1.0\columnwidth]{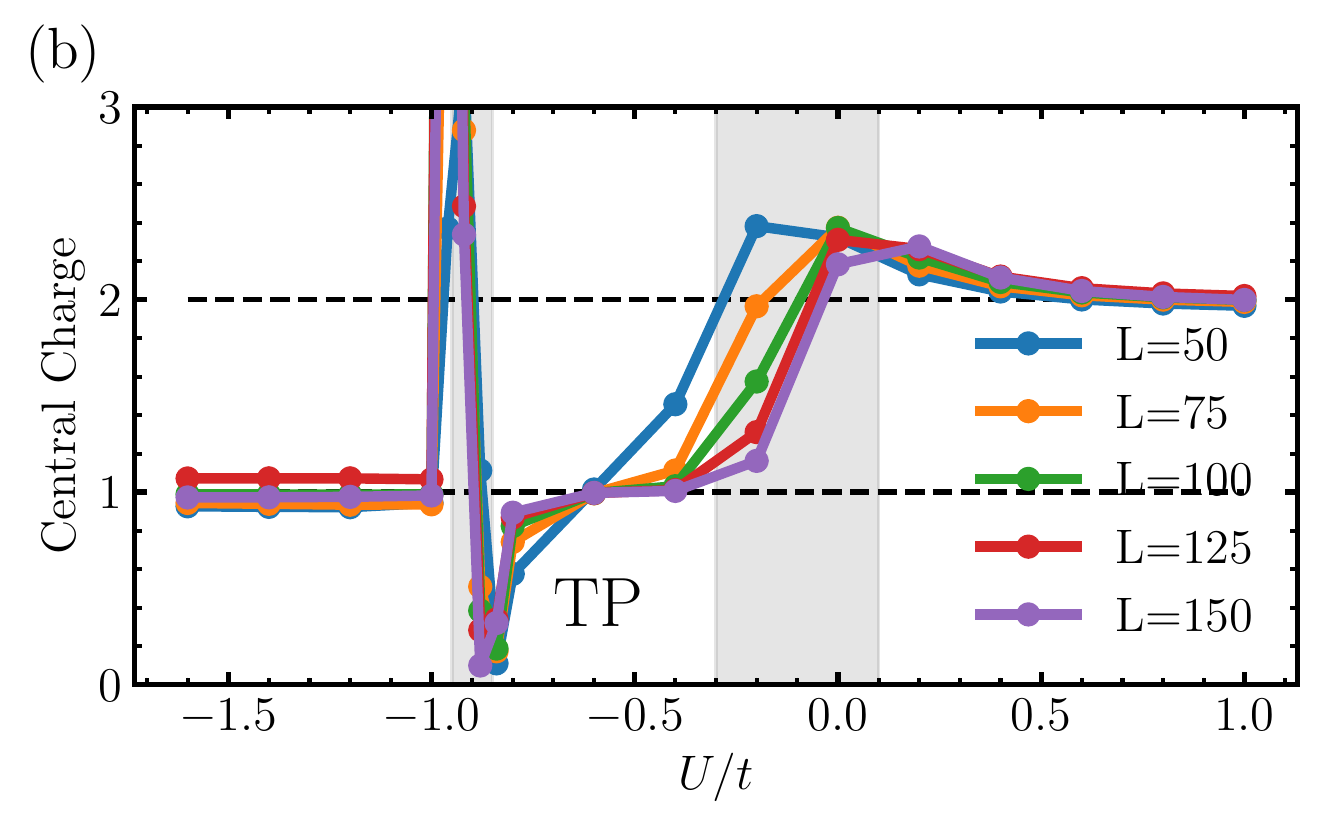}%
 \caption{ 
 (a) The entanglement spectrum and (b) central charge as a function of the interaction $U$ for $t'=0.2t$. 
 The even degeneracy for $L=100$ survives for a region approximately in between $ -0.9t < U < -0.3t $, a bit narrower than before.
 The reorganization of the spectrum (gray region) coincide with the regions where the central charge has its maximum.
 The inset show the exponential scaling of the difference between the two biggest eigenvalues inside the topological phase. 
 \label{FIG_ENTR_tp20}}
 \end{figure}
 
For one point inside the topological phase we plot in Fig.~\ref{FIG_PARAB} the entanglement spectrum as a function of the particle number and parity sector of a half of the system with $L=100$ rungs. 
Here one can appreciate the degeneracy between parity sectors and the parabola envelope expected for the lowest eigenvalues \cite{Lauchli2013, Laflorencie2016}.

In order to analyse feature (ii), the degeneracy of the entanglement spectrum, more in detail we present in Figs.~\ref{FIG_ENTR_tp00} and \ref{FIG_ENTR_tp20} the lower part of the entanglement spectrum and the central charge versus the interaction $U$ for $t'=0$ and $t'=0.2t$. 
The central charge is plotted for different system sizes of $L=25$, $L=50$, $L=75$ and $L=100$ rungs. The entanglement spectrum only for $L=100$ to avoid an overcrowded plot. 
As for the behaviour of the energy difference between the different parity sectors, we again see three different regions arising.

For the central region (TP), the entanglement spectrum seems degenerate as expected in the topological phase. 
This is even more evident in the inset in panel (a) of Fig.~\ref{FIG_ENTR_tp20} where we show the scaling of the difference between the lowest two eigenvalues with system size. This difference goes to zero exponentially and already for system sizes above 50 it is very close to zero. 

The breakdown of the topological phase can be appreciated on the degeneracy of the entanglement spectrum [see both panels (a) of Figs.~\ref{FIG_ENTR_tp00} and \ref{FIG_ENTR_tp20}]. Above and below the degenerate region, the degeneracy is already broken for the lowest eigenvalue at the shown system sizes.
For $t'=0$, the topological phase survives for a region around $U=0$ approximately in between $ -0.5t < U < 0.5t $ (Fig.~\ref{FIG_ENTR_tp00}), but when $t'=0.2t$ this region shrinks and displaces to occupy a region approximately in between $ -0.9t < U < -0.3t $ (Fig.~\ref{FIG_ENTR_tp20}). 
On both cases we have indicated with gray the region where the eigenvalues (of the $L=100$ ladder) cross with each other and reorganize the entanglement spectrum. The shaded regions in Figs.~\ref{FIG_ENTR_tp00} and \ref{FIG_ENTR_tp20} are also plotted in Fig.~\ref{FIG_GAP} and later in Fig.~\ref{FIG_REVMAXCUTS}. 

If we compare in more detail the upper and lower boundary of the topological phase, we notice that at the lower boundary all the plotted eigenvalues rearrange. None of the degeneracies stays unchanged. 
This is in strong contrast with the upper boundary where the breaking of the even degeneracy occurs in the lower branch but the next branch remains unaltered. 
On top of these differences, one is able to notice that for $t'=0$ (Fig.~\ref{FIG_ENTR_tp00}) the breaking of the topological phase at the lower boundary is less abrupt than for $t'=0.2t$ (Fig.~\ref{FIG_ENTR_tp20}). 
For the first case $t'=0$ we can distinguish that for $t'=-0.6t$ the degeneracy of the lower branch is broken but the 4-fold degeneracy of the second branch still approximately survives. 
For the second case $t'=0.2t$ below $U=-0.9t$ the degeneracy on both branches is broken. 

We show in Fig.~\ref{FIG_SUM}b the central charge extracted from the von Neumann entropy at each point of the phase diagram. 
We obtain a central charge close to $c=1$ inside the topological phase. This hints that the gapless symmetric sector present in the non-interacting fermionic model \cite{KrausZoller2013} is surviving in the present setup.
More in detail, we have pointed out in Sec.~\ref{SEC_tU0} the close connection between our model at $t'=U=0$ (a point inside the topological phase) and the model in Kraus et al.\cite{KrausZoller2013}. There they study the topological phase by means of bosonization and find that the bosonized Hamiltonian can be split into two decoupled sectors, the symmetric and the antisymmetric. 
The antisymmetric sector is described by the sine-Gordon Hamiltonian and by using re-fermionization it corresponds to the continuum limit of the Kitaev chain (gapped). 
In contrast, the symmetric sector of the system is described by a Tomonaga-Luttinger liquid which has a central charge of $c=1$.
This seems to be the same case in our model, such that the central charge in the topological phase $c=1$ identifies this gapless symmetric sector. 

The topological phase is surrounded by peaks of the central charge which signal a change of the phase. This is even better seen in the cuts which are presented in Fig.~\ref{FIG_ENTR_tp00} and Fig.~\ref{FIG_ENTR_tp20}. In the topological phase the central charge approaches $c=1$ with increasing system size. For $t'=0$ a stable and relatively wide plateau can be seen at large system length $L=100$. For $t'=0.2$ only for the largest system sizes $L=100$ a few points approach the value of $c=1$ due to the smaller region. 

Also in the central charge the different phase transitions manifest. For the cuts at $t'=0.2t$, the transition to the attractive-$U$ phase is a lot more abrupt than for $t'=0$. For $t'=0$ we can see a peak arising in between both phases hinting towards a closing a gap. 
For $t'=0.2t$ the transition is particularly sharp and a small dip can be appreciated prior to the breaking of the topological phase ($U \simeq -0.8t$). 

For the attractive-$U$ phase below the topological phase we also obtain a central charge of $c=1$ indicating a gapless phase. 
In Sec.~\ref{SEC_LARGEU} we have discussed that the strongly attractive limit of our model can be mapped into a XXZ spin chain considering only the singlets formed on each rung. 
The effective spin model for the singlet states is gapless at the considered magnetization (corresponding to an specific filling in the original ladder) for the effective parameters $\tilde{J}_{z} > \tilde{J}_{xy}$ of the spin chain. 
Since we have a gapless mode the central charge will be $c=1$ in agreement with the calculations. 
We note that despite of being able to detect the phase boundaries in most of the cases by a rise in the central charge, the topological phase and the attractive phase have the same central charge and we have to distinguish them by other quantities.

In Sec.~\ref{SEC_LARGEU} we have obtained the strongly repulsive limit of our model. There the number of pairs is negligible and since the only connection between the legs is the pairing hopping both legs decouple. 
Both chains will independently behave as a Luttinger liquid with a central charge of $c=1$. 
We expect a central charge of $c=2$ for the repulsive phase above the topological phase which is in agreement with our results (Fig.~\ref{FIG_SUM}).

\subsection{Criterion (iii): Edge to edge correlations}
\label{SEC_SPC}

In this section we are going to have a closer look at the single particle correlation functions $ A^{\uparrow \uparrow}_{1j} = \expval{ a^{\dagger}_{\uparrow 1} a^{\pdagger}_{\uparrow j} } $. Criterion (iii) for the topological phase tells us that single particle correlation functions on each leg of the ladder should decay exponentially into the bulk, but then show a revival at the opposite edge signalling the Majorana edge modes. 

 \begin{figure}
 \includegraphics[width=1.0\columnwidth]{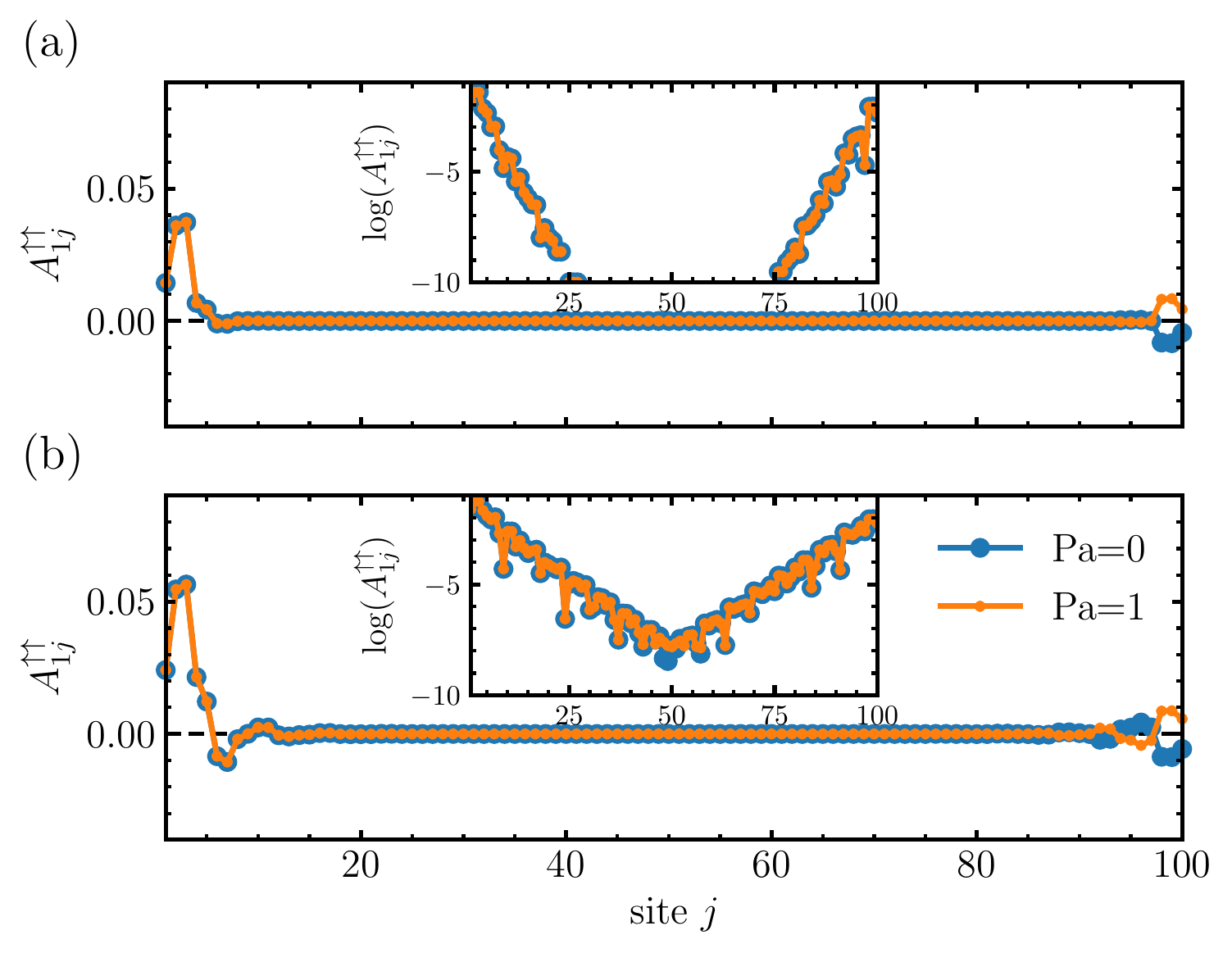}%
 \caption{ 
 Single particle correlations $ A^{\uparrow \uparrow}_{1j}$ for one leg of a ladder with $L=100$ rungs and for both parity sectors ($P_a = 0, 1$). 
 These results correspond to (a) $t'=U=0$ and (b) $t'=0.2t$, $U=-0.6t$. 
 We can see that the correlations are exponentially suppressed into the bulk and re-emerge on the other end of the chain. 
 \label{FIG_CORR_TOP}}
 \end{figure}

In Figs.~\ref{FIG_CORR_TOP}, \ref{FIG_CORR_T1}, \ref{FIG_CORR_TFREE}, \ref{FIG_CORR_T2}, and \ref{FIG_CORR_T3} we show the results for the single particle correlations for one leg of a ladder with $L=100$ rungs and for calculations in both parity sectors.

In Fig.~\ref{FIG_CORR_TOP} we show the correlations for two different points inside the topological phase. 
In panel (a) we show the $U=t'=0$ case, which as we discussed can be connected to the spinless ladder model in Eq.~\ref{EQ_KRAUS}. In panel (b) a more general case $t'=0.2t$ and $U=-0.6t$. 
The single particle correlations decay exponentially with distance, if the second site where the correlation is taken lies in the bulk (see inset). However, as expected for the presence of Majorana edge modes, if the second site approaches the other end of the chain, the single particle correlations re-emerge also exponentially to a finite edge-edge correlation. Further, changing the parity sector shows that the revival of the correlation changes sign. 
This typical behaviour of the single particle correlations in the topological phase is found in the entire topologically non-trivial phase.

This is summarized in panel (c) of Fig.~\ref{FIG_SUM} where we present a color plot of the maximum of the revival of the single particle correlations at the opposite edge of the ladder minus the maximum of the correlations in the middle of the ladder. Clearly, the amplitude of the revival is maximal in the topological phase. Outside of this phase the maximum takes values close to zero or negative (not visible since we saturate the scale at zero for better visibility). 
 
In the following we address how this behaviour is changing, when leaving the topological phase. 
From the study of the energy difference between both parities and the entanglement properties we have concluded that the breakdown of the topological phase can have different characteristics: 
if we destroy the phase by adding an attraction (limit $U\to-\infty$) we see a sharp change in the previous studied quantities which does not have any strong dependence on the system size. In contrast, if we destroy the topological phase by increasing the repulsion $U$ or increasing $t'$ the quantities smoothly change and strongly depend on the size of the system, making it difficult to determine a precise limit for the phase. 
To better understand how the topological phase is going over to the other phases we study the single particle correlations near these limits. 

\begin{figure}
 \includegraphics[width=1.0\columnwidth]{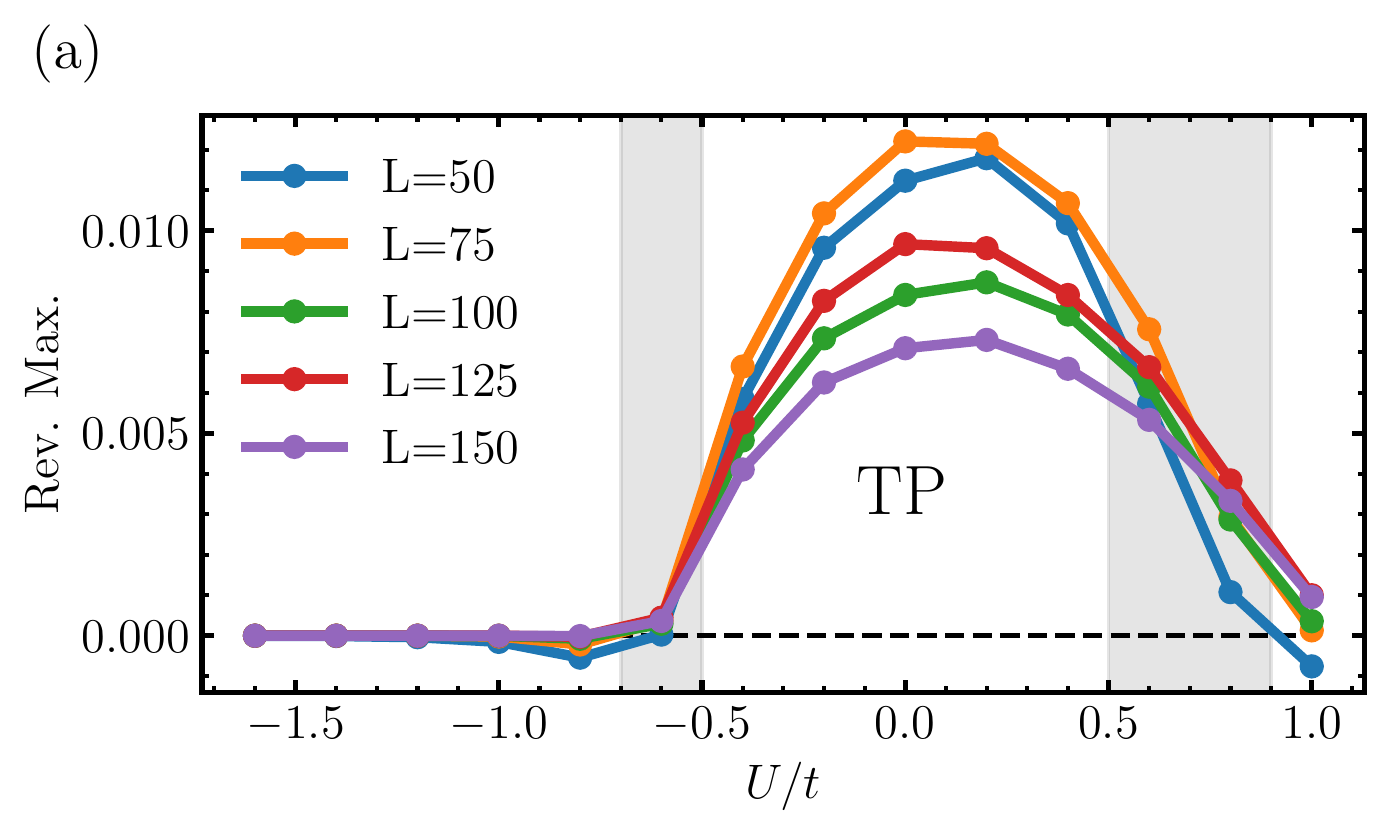}\\
 \includegraphics[width=1.0\columnwidth]{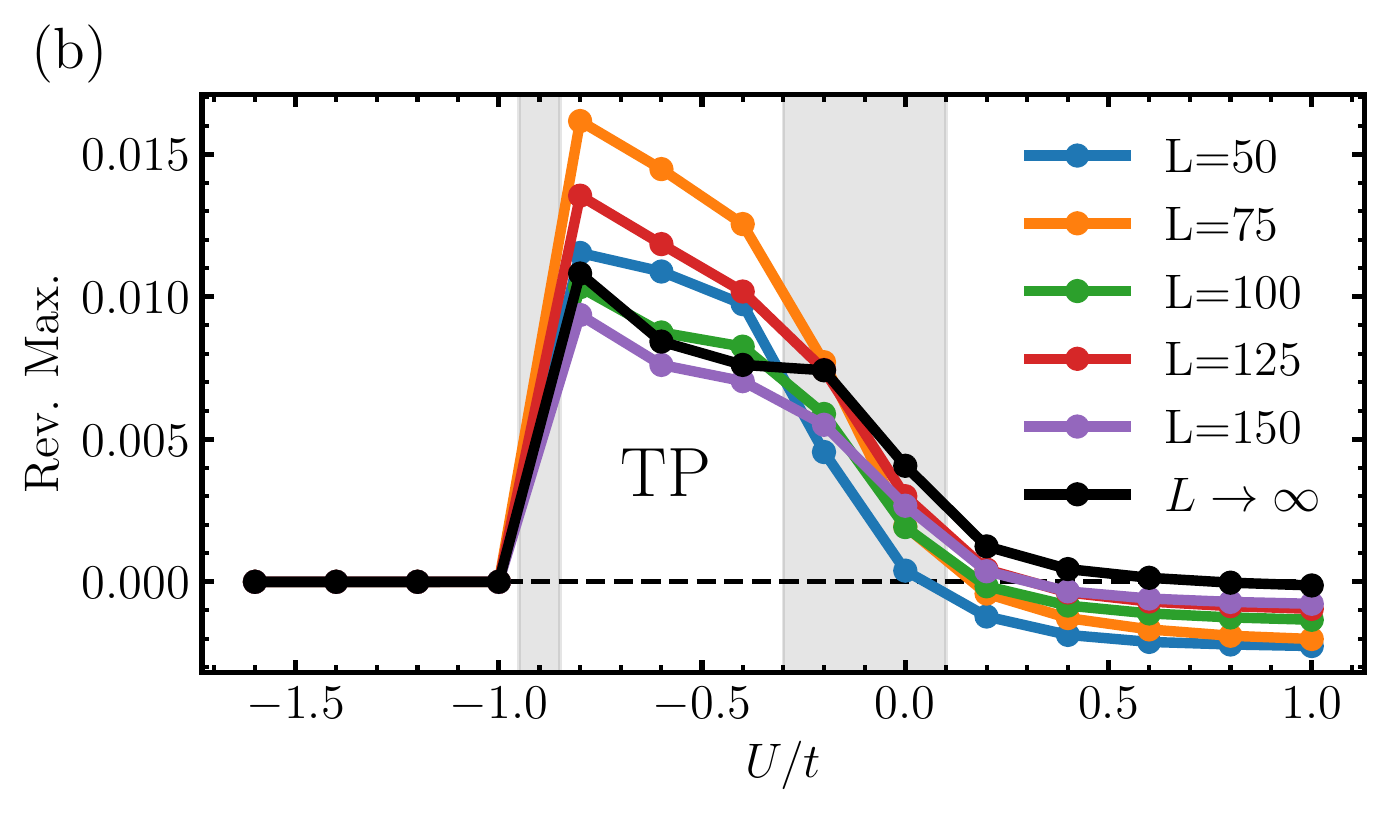}%
 \caption{ 
 Cuts for the maximal revival of the single particle correlation functions. 
 We present cuts at (a) $t'=0$ and (b) $t'=0.2t$. 
 In panel (b) we also include an extrapolation to the thermodynamic limit. 
 The gray regions are the same as the ones plotted in Figs.~\ref{FIG_ENTR_tp00} and \ref{FIG_ENTR_tp20}.
 \label{FIG_REVMAXCUTS}}
 \end{figure}

\begin{figure}
 \includegraphics[width=1.0\columnwidth]{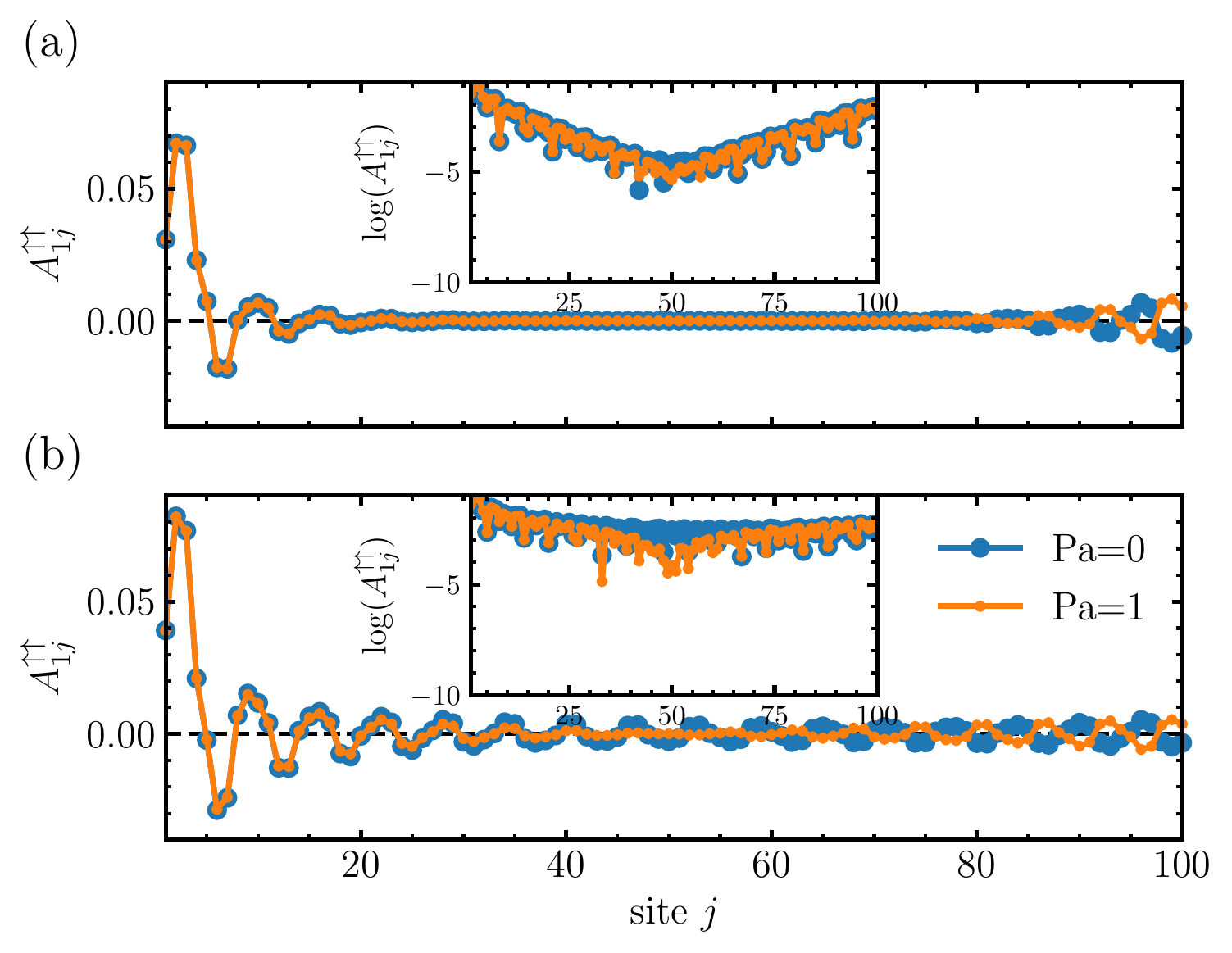}%
 \caption{ 
 Single particle correlations $ A^{\uparrow \uparrow}_{1j} $ for one leg of a ladder with $L=100$ rungs and for both parity sectors ($P_a = 0, 1$). 
 These results correspond to $t'=0.2t$ for (a) $U=-0.4t$ and (b) $U=0$. 
 We can see that the Majoranas spread into the bulk until they are not localized any more and the system goes to a trivial state with non-vanishing correlations. 
 \label{FIG_CORR_T1}}
 \end{figure}

\begin{figure}
 \includegraphics[width=1.0\columnwidth]{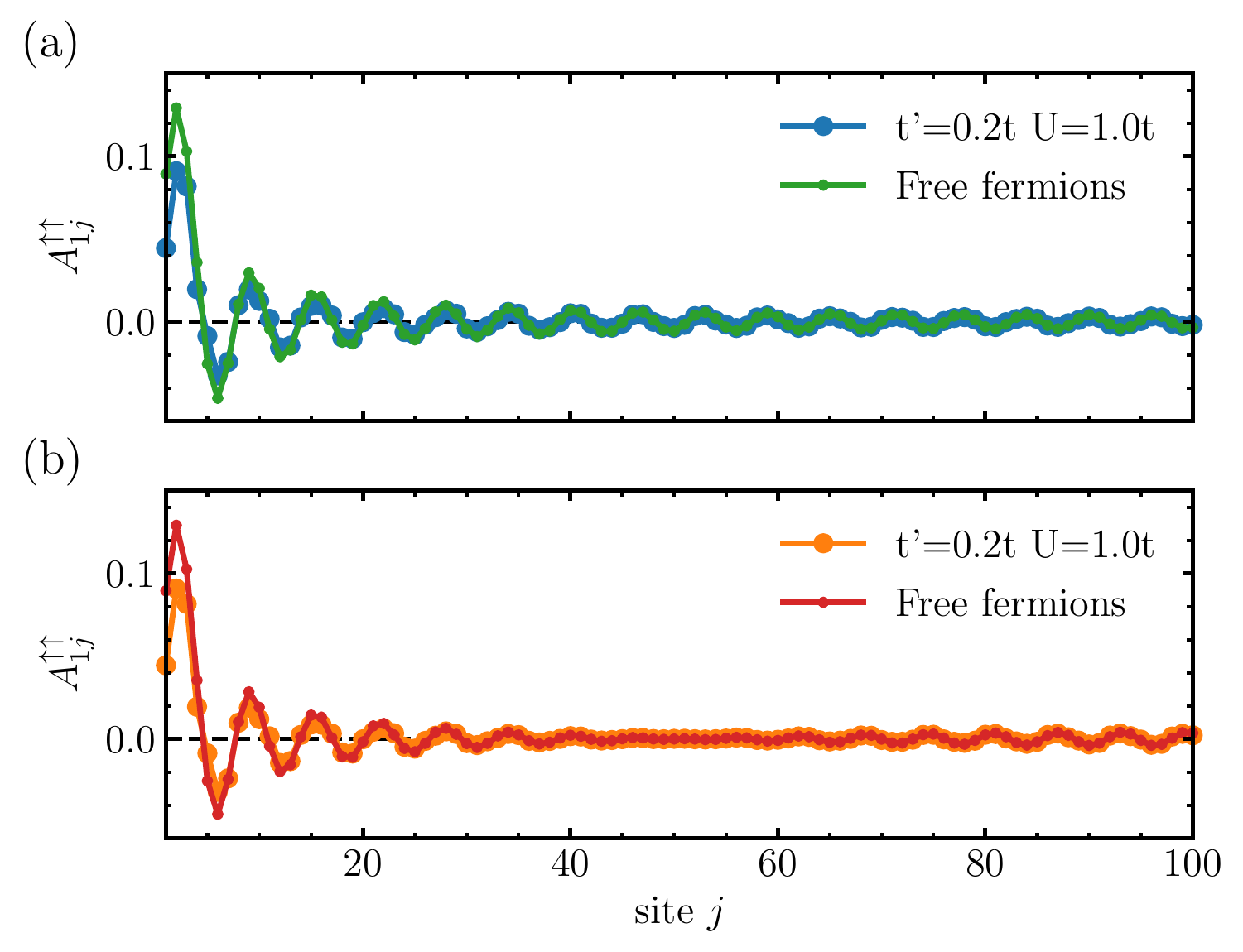}%
 \caption{ 
 Single particle correlations $ A^{\uparrow \uparrow}_{1j} $ for both parity sectors: (a) $P_a = 0$ and (b) $P_a = 1$. 
 Here we compare the results corresponding to the correlations in one leg of a ladder with $L=100$ rungs at $t'=0.2t$ for $U=t$ with the results of the effective model of free fermions in the strongly repulsive and fully-polarized limit. 
 We can see a good agreement between the two in both cases ($t'=0.2$ and $U=t$). 
 The magnetization on the $x$-direction is (given by $M_x = \sum_{j \alpha} \expval{\alpha^{\dagger}_{\uparrow j} \alpha^{}_{\downarrow j} + \text{h.c.}} $) $M_x \simeq 62.24 $, close to its saturated value $M_x^{sat} = 64$ (the number of fermions).
 \label{FIG_CORR_TFREE}}
 \end{figure}

\begin{figure}
 \includegraphics[width=1.0\columnwidth]{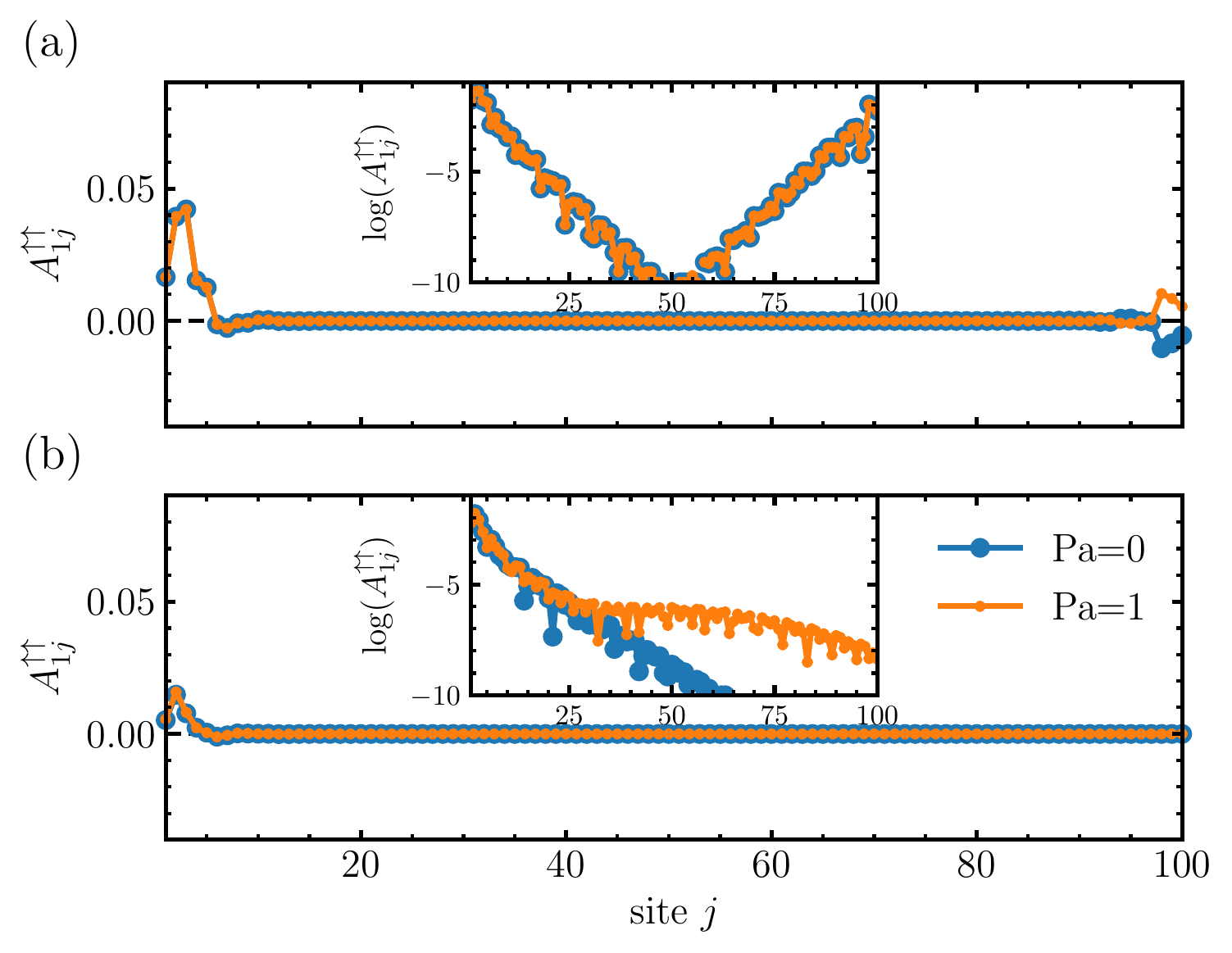}%
 \caption{ 
 Single particle correlations $ A^{\uparrow \uparrow}_{1j} $ on one leg with $L=100$ and for both parity sectors ($P_a = 0, 1$). 
 These results correspond to $t'=0.2t$ for (a) $U=-0.8t$ and (b) $U=-t$. 
 Even though the Majoranas do not spread into the bulk we can see that as soon as the interaction term goes below $U_C \simeq -0.9t$ we completely loose the correlations between different edges of the chain and even the correlations on the same edge get largely reduced. 
 \label{FIG_CORR_T2}}
 \end{figure}

\begin{figure}
 \includegraphics[width=1.0\columnwidth]{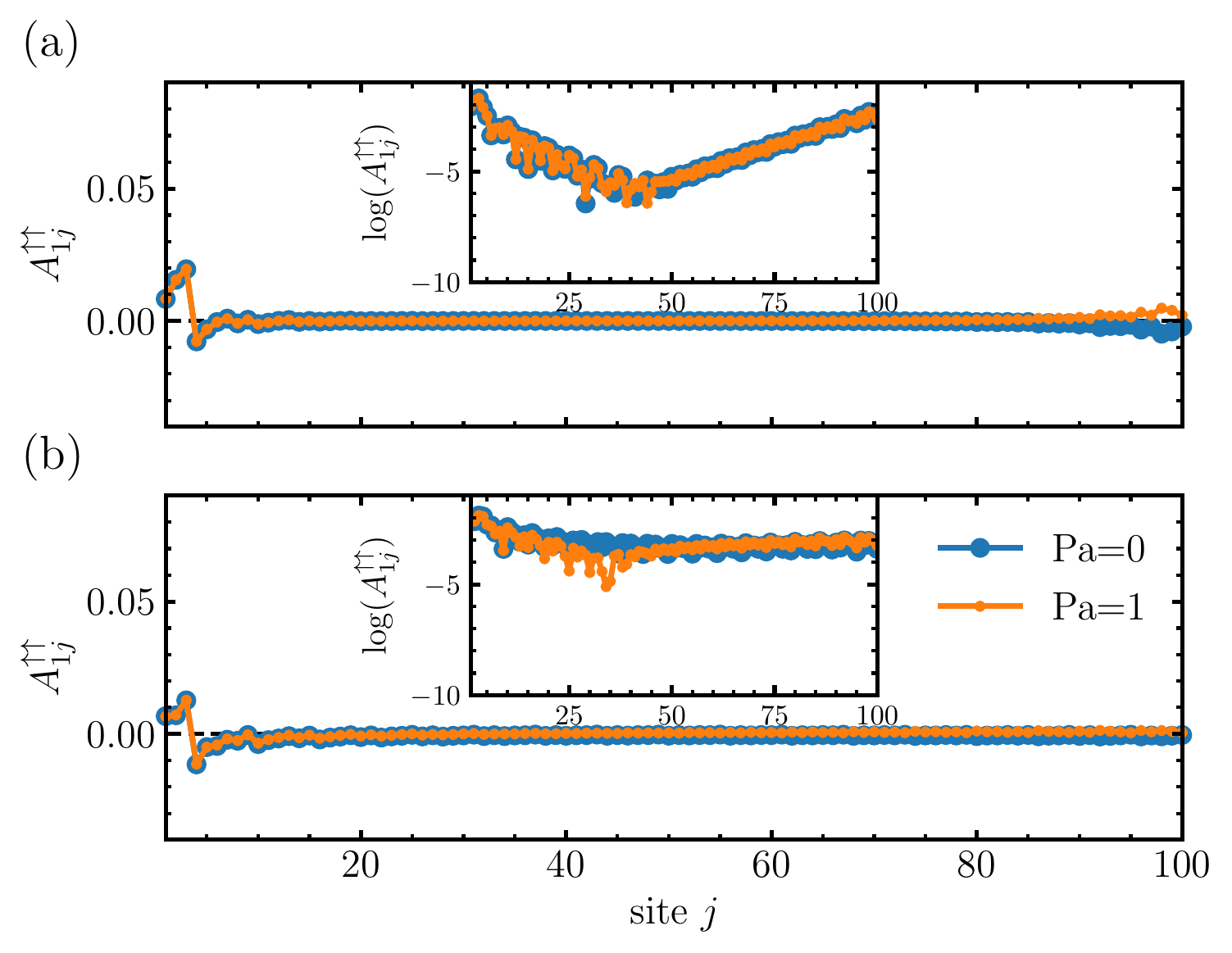}%
 \caption{ 
 Single particle correlations $ A^{\uparrow \uparrow}_{1j} $ for one leg of a ladder with $L=100$ rungs and for both parity sectors ($P_a = 0, 1$). 
 These results correspond to $t'=0$ for (a) $U=-0.4t$ and (b) $U=-0.6t$. 
 \label{FIG_CORR_T3}}
 \end{figure}

In order to quantify the revival of the single particle correlation function we are going to define the maximal revival MR as the difference between the maximum value of the single particle correlations in the last $10$ sites of the chain end of the chain and the maximum value in the $10$ central sites, i.e.~
\begin{equation}
\text{MR}
 = \max_{j\in \text{[L-10,L]}}\left(A^{\uparrow \uparrow}_{1j}\right)
 - \max_{j\in \text{[L/2-5,L/2+5]}}\left(A^{\uparrow \uparrow}_{1j}\right).
\end{equation}
We have calculated this quantity for lattices with $L=50$ rungs or more. 
We show cuts for $t'=0$ and $t'=0.2t$ of the maximal revival of the single particle correlation function in Fig.~\ref{FIG_REVMAXCUTS}. 
The shaded regions are the same that the ones depicted in Figs.~\ref{FIG_GAP}, \ref{FIG_ENTR_tp00} and \ref{FIG_ENTR_tp20}. 
A clear maximum is seen within the topologically non-trivial phase which decays towards the other phases. 
At $t'=0.2t$ a very steep behaviour is seen around $U=-t$ indicating a drastic change in the nature of the state. This is in agreement with the observations from the other observables. 
For large (attractive) values of the interaction $U$ the maximum of the revival decays to zero as $1/L$. In the topological phase the functional form is not that clear from our data and the behaviour is different depending if $L$ is even or odd-valued, but nonetheless the extrapolation to the thermodynamic limit goes to a finite value which we show in Fig.~\ref{FIG_REVMAXCUTS}b. 

In order to explore further the smooth behaviour at the upper edge, we show in Fig.~\ref{FIG_CORR_T1} the single particle correlations for one leg of a ladder with $L=100$ rungs and for both parity sectors ($P_a = 0, 1$) corresponding to $t'=0.2t$ and at $U=0$ and $U=-0.4t$. 
We can see that the Majorana edges states get broader and less confined towards the edges. The edge states delocalize into the center of the chain until they loose their edge character. At some point these states become so broad that they are able to interact with each other for the finite system sizes considered. Therefore, we find a strong system size dependence of the maximum value in this region.
When the Majoranas on opposite ends are able to interact then the topological phase is destroyed, provoking at this point the lift of the degeneracy on the entanglement spectrum, the opening of an energy difference between symmetry sectors, and the appearance of a trivial phase with non-vanishing correlations. 
As we see in panel (b) of Fig.~\ref{FIG_REVMAXCUTS}, the maximum of the revival in the opposite edge decreases and goes to negative values for larger values of $U$ even though the edge-edge correlations are finite. This is because we are in the trivial phase and the correlations decay without reviving. Hence, the edge-edge correlations are smaller than the correlations in the middle of the chain and the maximal revival is negative. 
In this case, the finite value of the correlations at opposite ends does not come from a revival and is not of topological nature. The maximum value therefore decays strongly with increasing system size. 

 This findings are further supported by the large-$U$ limit (Section \ref{SEC_LARGEU}) where we found that in the strongly repulsive limit the system should behave as two independent chains. 
Here we show that we can explain the main features of the correlations outside of the topologically phase for larger interaction already to a large extent by assuming two polarized chains along the $x$-direction (the direction of the magnetic field $h_x$). 
In Fig.~\ref{FIG_CORR_TFREE} we compare the correlations within this polarized state with $L=100$ sites and the ones obtained in our model for a ladder of $L=100$ rungs for $U=t$ and $t'=0.2t$. 
We see a good agreement between these models for these values of $U$. In particular, the correlations are still finite at the edge, but no revival is found. The small discrepancies between the original model and the free-fermion limit will vanish if we further increase the interaction and the magnetic field.

At the lower limit of the topological phase the behaviour is different. Here we concentrate on the horizontal transition line in the phase diagram at $U_C \simeq -0.9t$ present for $0.1t < t' < 0.3t$. 
If we look at the single particle correlations around this transition (see Fig.~\ref{FIG_CORR_T2}) we can see that even though the Majoranas do not spread into the bulk, as soon as the interaction goes below $U_C \simeq -0.9t$ we completely loose the correlations between different ends of the chain, and therefore the Majorana edge modes. In contrast to the previous case where the breakdown of the topological phase corresponded to a delocalization of the edge states, now is the amplitude of the edge to edge correlation function that goes down. 
As a consequence, in Fig.~\ref{FIG_SUM} and in particular in panel (b) of Fig.~\ref{FIG_REVMAXCUTS} we see that the maximal revival at the opposite end completely vanishes as we go to more attractive values of the interaction $U$. 
The single particle correlations are strongly suppressed in this phase below the topological region, this can be understood from the fact that the rungs are weakly coupled between them as this coupling arises from a 2nd order process in perturbation theory in contrast with the in-rung coupling that is of first order. 

Surprisingly enough, the lower limit of the topological phase seems to behave differently closer to $t'=0$ where the change in the observable are noticeable less abrupt. 
This behaviour can be seen on the oblique line on the lower part of the phase diagram and for small $t'$ (see Fig.~\ref{FIG_SUM}c). 
In Fig.~\ref{FIG_CORR_T3}, we show the behaviour of the single particle correlations at $t'=0$ for $U=-0.6t$ (topological) and $U=-t$ (trivial). 
In panel (a) of Fig.~\ref{FIG_REVMAXCUTS} we can see that maximal revival at the opposite end also vanishes for more attractive $U$, but in agreement with the previously mentioned observables this behaviour is also smoother than it was for larger $t'$ (see Fig.~\ref{FIG_REVMAXCUTS}b).

\subsection{Criterion (iv): Robustness against static noise}
\label{SEC_NOI}

 \begin{figure}
 \includegraphics[width=0.8\columnwidth]{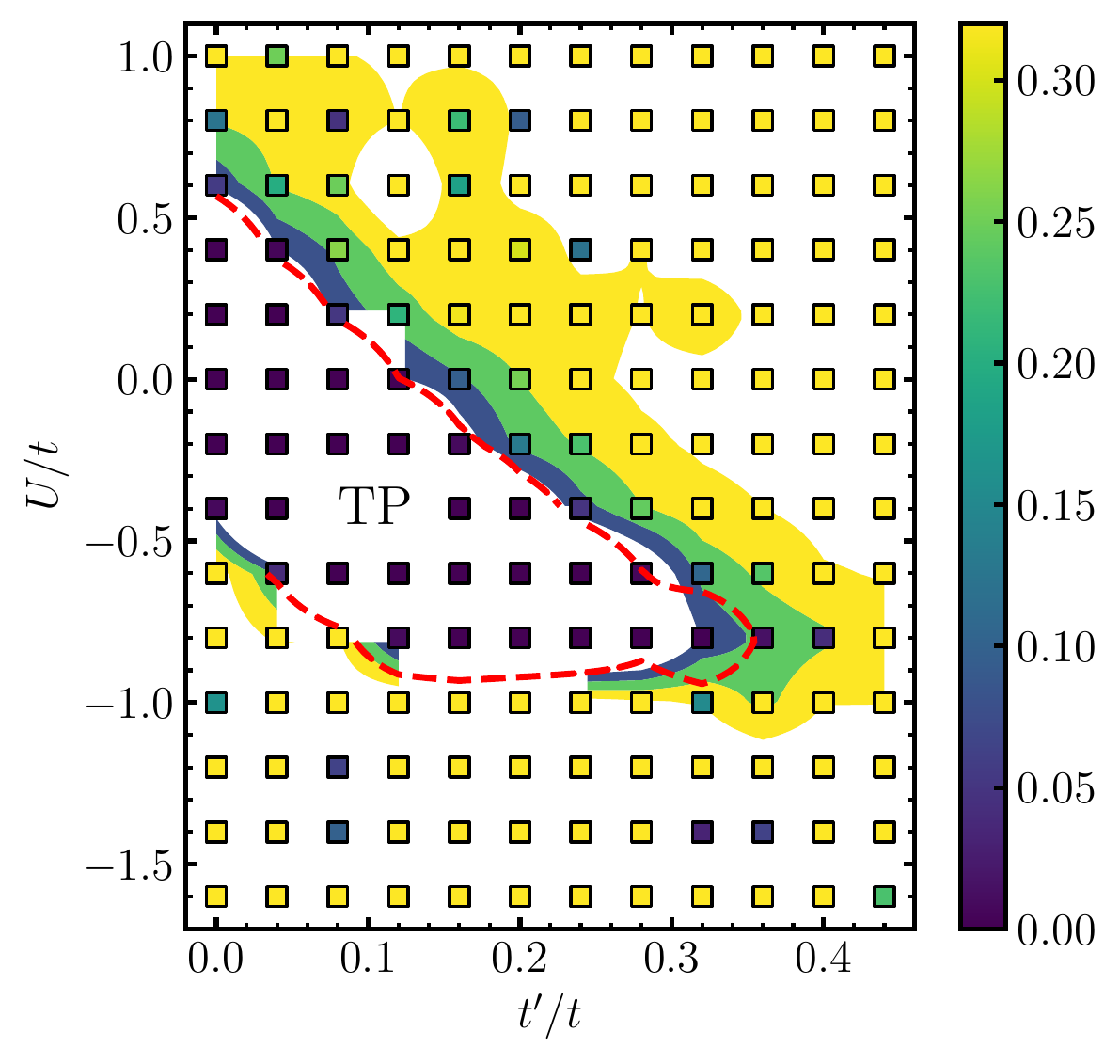}%
 \caption{ 
 Ground state phase diagram of the model $H$ in Eq.~\ref{EQ_H} for a system with $L=100$ rungs and on-site noise $V_r=0.1t$.
 The color plot in the background shows the logarithm of the energy difference between the two parity sectors [$ \log(\Delta_{Pa}) $]. The coloured symbols correspond to the difference of the first two eigenvalues $(\Lambda_0 - \Lambda_1)/(\Lambda_0 + \Lambda_1)$ of the reduced density matrix. 
The dashed red line shows the contour line for the gap at $\Delta_{Pa} = 10^{-5}$ for the case without static noise, i.e.~$V_r=0$ (already shown in Fig.~\ref{FIG_SUM}). 
 \label{FIG_Vr1}}
 \end{figure}
 
Another characteristic of a topological phase is its robustness against static local disorder. In this section we introduce two different kinds of disorder, a random onsite disorder potential and a small additional random pair-hopping. 

The random site-noise has a amplitude $V_{\alpha j} \in \left[ -V_r, V_r\right]$ which is randomly chosen for each site. Here $\alpha=a,b$ indicates the leg and $j$ is the rung index. For each parameter set a new random distribution is drawn. For the calculation of the gap we need to run twice each point in the phase diagram and use the same set of random terms both times. We present these results for $V_r$ in Fig.~\ref{FIG_Vr1}. 
As expected, even for moderated values of $V_r$, the noise does not have an strong influence on the phase diagram which is extracted using the energy gap between the two parity sectors and the degeneracy of the eigenvalues of the reduced density matrix. Only slight shifts in the phase boundary can be observed compared to the disorder free case which is indicated in red dashed lines. These mainly occur in the region of the tip of the lobe of the topological phase where finite size effects are still sizable. We additionally verified that the revival of the single particle correlations is also stable against this disorder. Thus, we verified the robustness of the Majorana modes to local noise. 

 \begin{figure}
 \includegraphics[width=0.8\columnwidth]{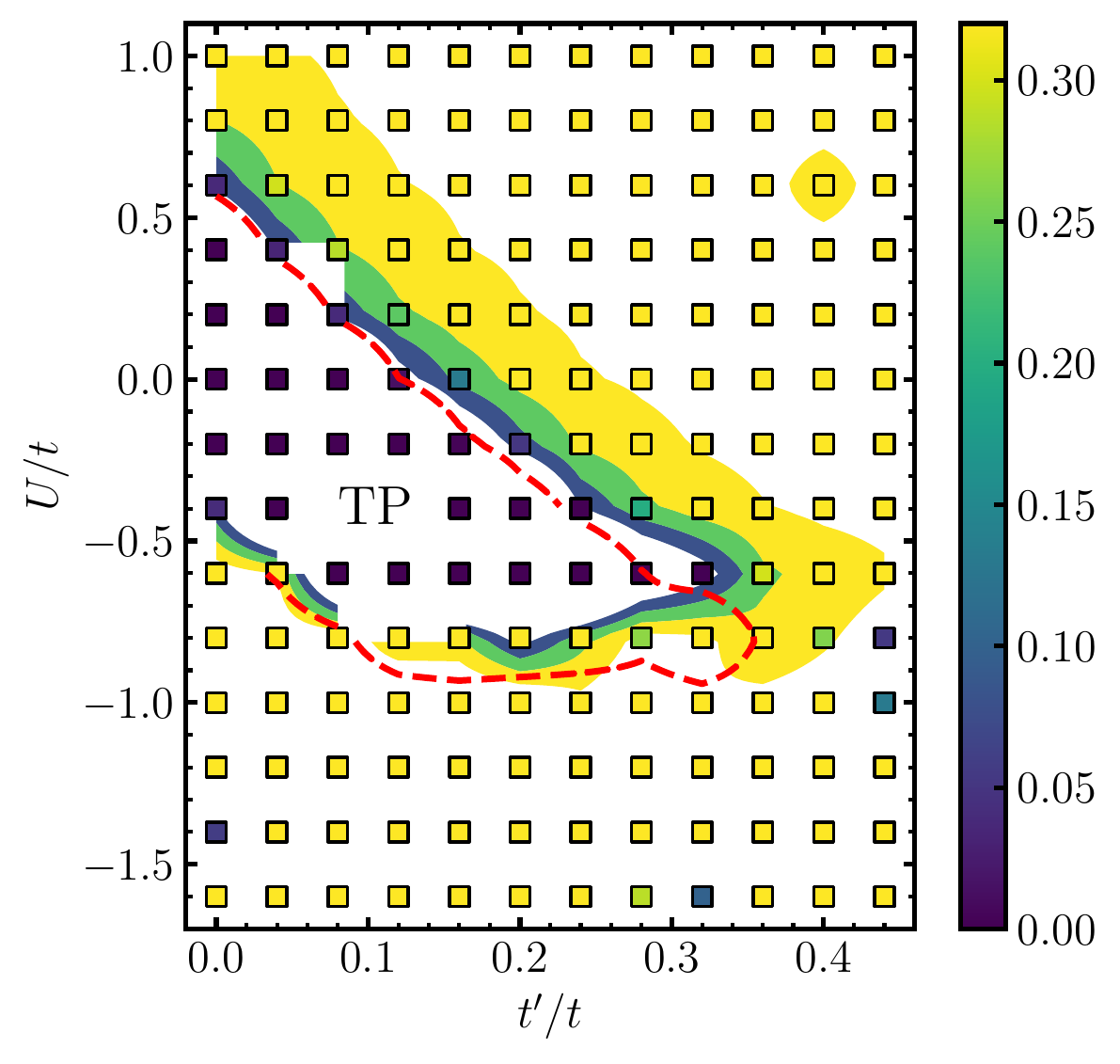}%
 \caption{ 
 Ground state phase diagram of the model $H$ in Eq.~\ref{EQ_H} for a system with $L=100$ rungs and pair-hopping noise $W_r=0.26t$.
 The color plot in the background shows the logarithm of the energy difference between the two parity sectors [$ \log(\Delta_{Pa}) $]. The coloured symbols correspond to the difference of the first two eigenvalues $(\Lambda_0 - \Lambda_1)/(\Lambda_0 + \Lambda_1)$ of the reduced density matrix. 
 The dashed red line shows the contour line for the gap at $\Delta_{Pa} = 10^{-5}$ for the case without static noise, i.e.~$W_r=0$ (already shown in Fig.~\ref{FIG_SUM}). 
 \label{FIG_Wr1}}
 \end{figure}
 
Additionally, we have studied the influence of disorder in the amplitude of the pair-hopping terms. The amplitude of the pair hopping terms is changed to $W_{\alpha j} = 2.6t + W_r$ with $ W_r \in \left[ -W_r, W_r\right]$. Again $\alpha$ and $j$ are the leg and rung index, respectively. 
Again for each point in the phase diagram we use a different set of random noise terms. In Fig.~\ref{FIG_Wr1} we show the result for $W_r=0.26t$ corresponding to a relative large disorder of 10 percent. The red dashed line again corresponds to the phase boundary of the noiseless case. 
Hence for this type of disorder, the influence on the phase diagram is also small. We further verified that the change on the revival of the single particle correlations is small. Thus, the phase shows it robustness against this new kind of disorder in the amplitude of the pair hopping.

\section{Stability of the topological phase to parameter changes}

Up to now we concentrated on one set of parameters for the pair hopping $W$ and the magnetic field $h_x$ in order to describe in details the properties of the topological phase. In this section we extend the previously presented result to smaller values of the pair-hopping $W$ and the magnetic field $h_x$ in order to show that the considered parameters are not special and we expect that topological phase and the Majorana modes to survive over a wide range of parameters.

\subsection{Change in the pairing hopping amplitude $W=0.1t$}
 \begin{figure}
 \includegraphics[width=0.8\columnwidth]{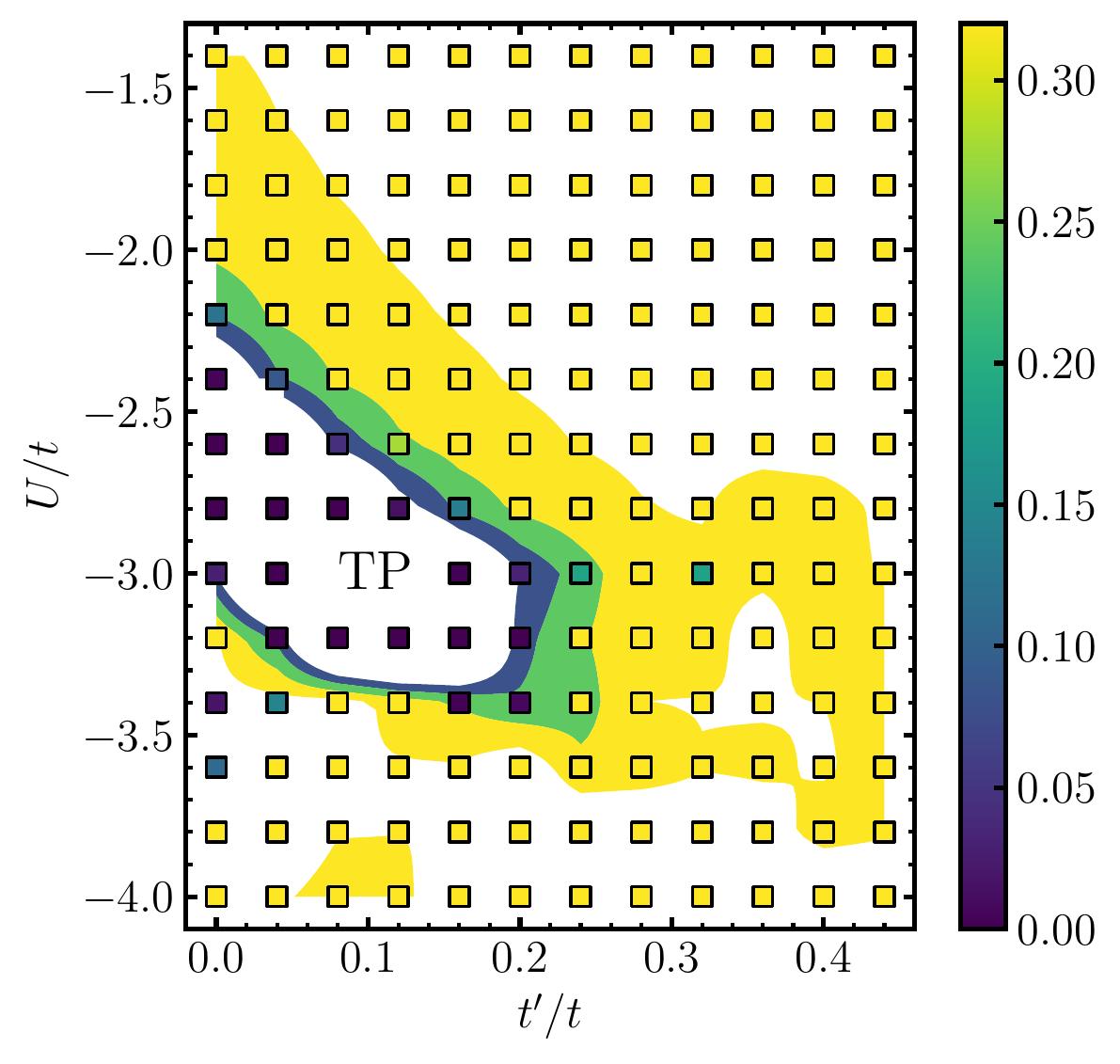}%
 \caption{ 
 Ground state phase diagram of the model $H$ in Eq.~\ref{EQ_H} for a system with $L=100$ rungs and $W=0.1t$.
 The color plot in the background shows the logarithm of the energy difference between the two parity sectors [$ \log(\Delta_{Pa}) $]. The coloured symbols correspond to the difference of the first two eigenvalues $(\Lambda_0 - \Lambda_1)/(\Lambda_0 + \Lambda_1)$ of the reduced density matrix. 
 \label{FIG_W01}}
 \end{figure}
 
In Fig.~\ref{FIG_W01} we show the results for a ladder with $L=100$ rungs and $W=0.1t$. All the other parameters are chosen as in Sec.~\ref{SEC_RES}. 
This plot is equivalent to Fig.~\ref{FIG_SUM}a with the only difference being the change of the value of $W$. The color plot in the background is the logarithm of the gap and the squares are the difference between the first two eigenvalues. We find that even for this small value of the pair hopping the topological phase is present. We verified that the criteria (i) to (iii) are fulfilled in the region identified as the topological phase. 
However, the phase boundaries have changed drastically. The topological phase occurs for this small pair hopping amplitude at large negative values of $U$. Its extension with the dimerized hopping $t' /t$ is smaller than for large values of the pair hopping, reaching $t'\simeq 0.2t$.

\subsection{Change of the value of the magnetic field}

 \begin{figure}
 \includegraphics[width=0.8\columnwidth]{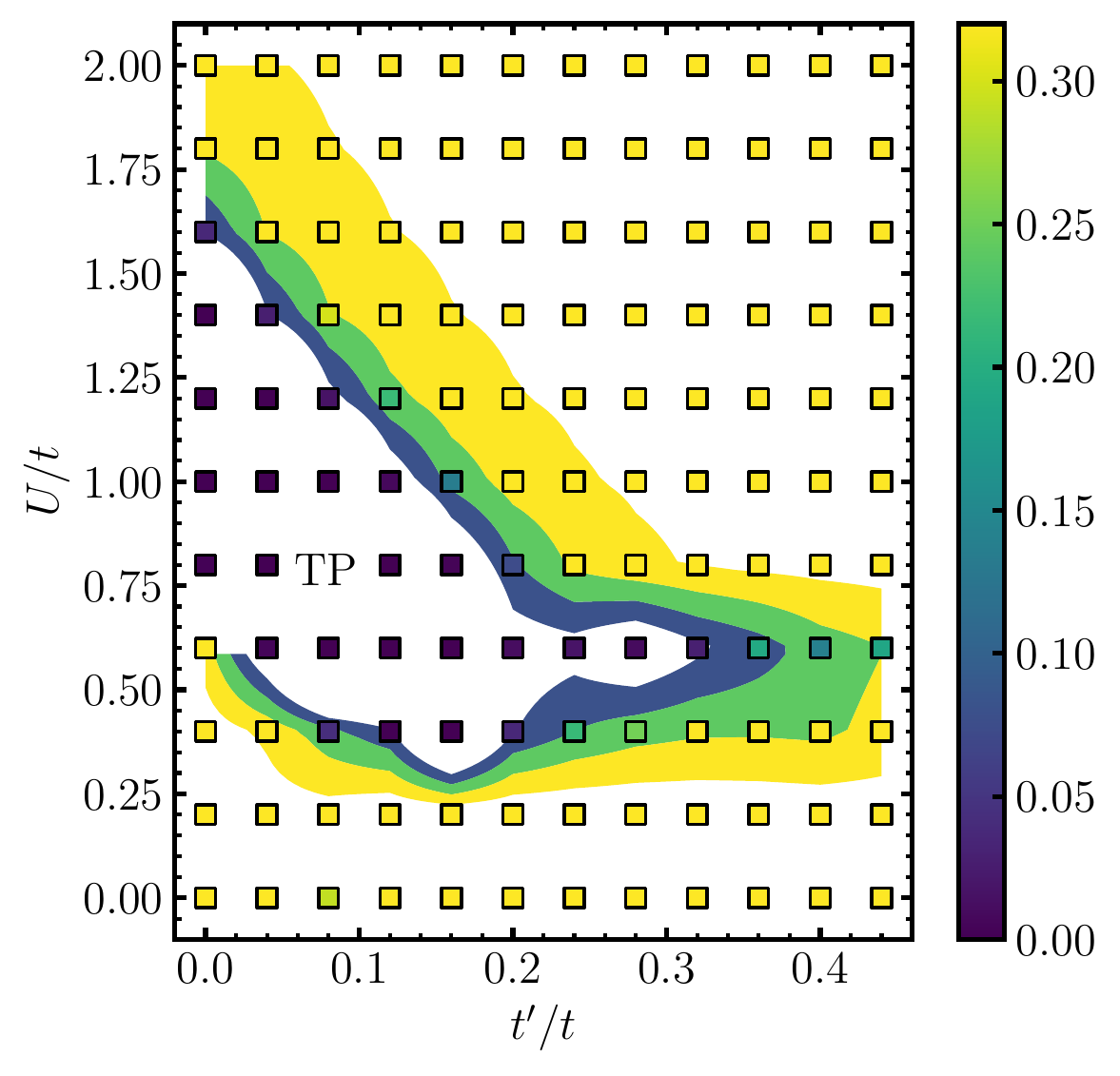}%
 \caption{ 
 Ground state phase diagram of the model $H$ in Eq.~\ref{EQ_H} for a system with $L=100$ rungs and $h_x=0.5t$.
 The color plot in the background shows the logarithm of the energy difference between the two parity sectors [$ \log(\Delta_{Pa}) $]. The coloured symbols correspond to the difference of the first two eigenvalues $(\Lambda_0 - \Lambda_1)/(\Lambda_0 + \Lambda_1)$ of the reduced density matrix. 
 \label{FIG_h05}}
 \end{figure}

The presence of a magnetic field in the considered model is crucial, since the magnetic field breaks the time reversal symmetry, a pre-requisite in order to be able to find a phase hosting Majorana zero modes. In the previous sections we have used the magnetic field $h_x=-t$. The goal of this section is to show the stability of the topological phase to a decrease of the value of the magnetic field.

In Fig~\ref{FIG_h05} we show the results for the ladder with $L=100$ rungs and $h_x=0.5t$. The remaining parameters are not changed compared to Sec.~\ref{SEC_RES}. 
Also for this much smaller value of the magnetic field we find the presence of the topological phase and the Majorana modes. We verified criteria (i) to (iii). 
The region in the phase diagram where the topological phase is present is smaller than for the larger magnetic field value. 
We note that for the points at $U=0.6t$ and $t' /t \in [0.24, 0.32]$ the revival of the correlations is not clear due to finite size effect that are still present. Calculations in larger systems would be needed if one wants to determine the precise boundary of this phase around those points. 
The whole region of the topological phase is displaced to positive values of $U$, i.e.~it lies entirely on the repulsive interaction side. 
The displacement of the topological region towards more repulsive interactions for smaller magnetic fields could have its origin in the fact that both the magnetic field term and the repulsion $U$ favours the breaking of pairs. Therefore if we reduce the magnetic field, the repulsion has to be increased to compensate. Its extension with the ratio of the dimerized hopping is slightly smaller than the one at $h_x=-t$. 

 \begin{figure}
 \includegraphics[width=0.8\columnwidth]{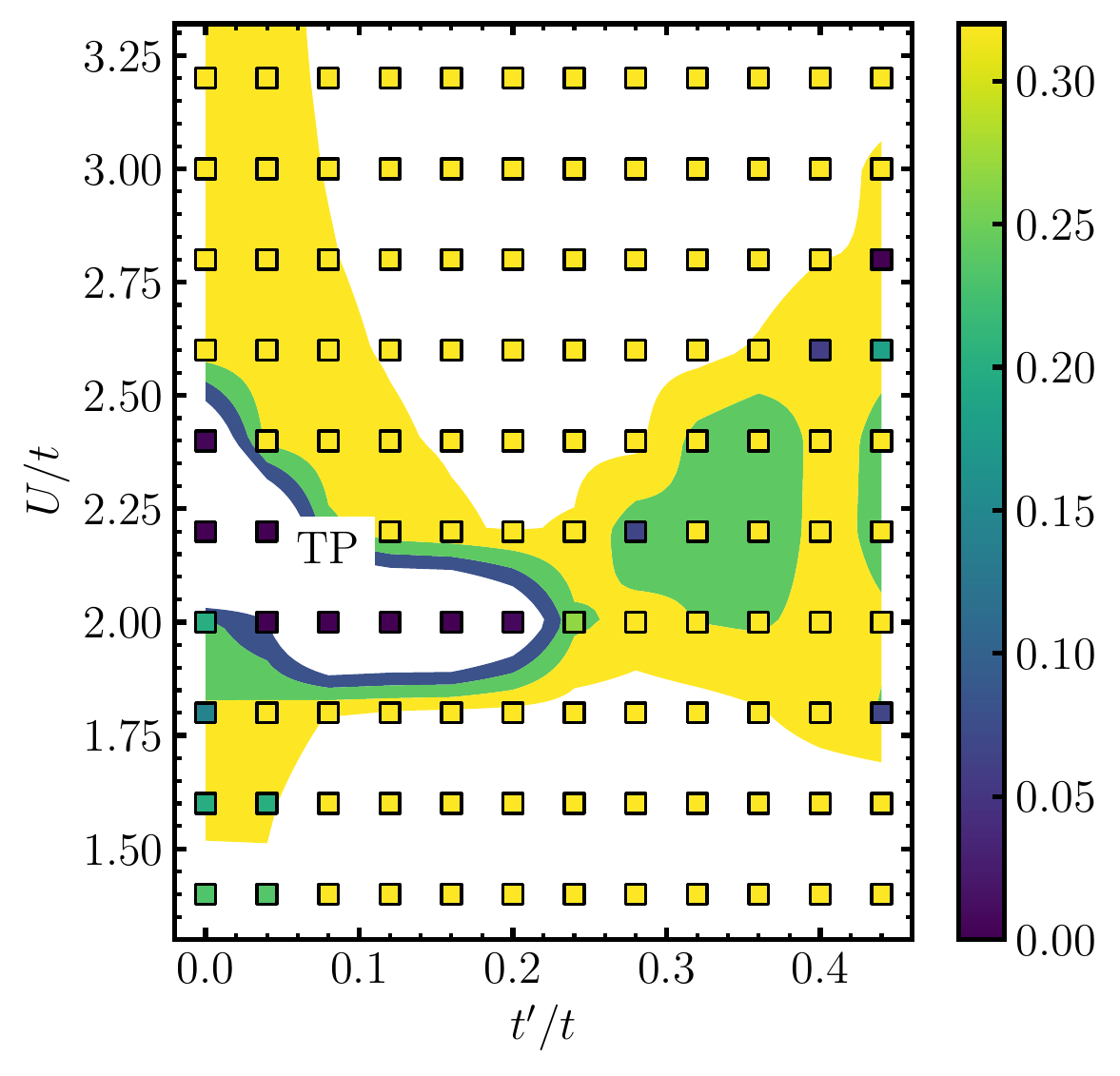}%
 \caption{ 
 Ground state phase diagram of the model $H$ in Eq.~\ref{EQ_H} for a system with $L=100$ rungs and $h_x=0.1t$.
 The color plot in the background shows the logarithm of the energy difference between the two parity sectors [$ \log(\Delta_{Pa}) $]. The coloured symbols correspond to the difference of the first two eigenvalues $(\Lambda_0 - \Lambda_1)/(\Lambda_0 + \Lambda_1)$ of the reduced density matrix. 
 \label{FIG_h01}}
 \end{figure}

Reducing even further the value of the magnetic field to $h_x=0.1t$ supports the stability of the topological phase. The results are shown in Fig.~\ref{FIG_h01}. 
For this small value of the magnetic field the topological phase is still present identified by criteria (i) to (iii) 
However, it seems considerably smaller than the one presented in Fig.~\ref{FIG_SUM}. 
On a further analysis of the points at $U=2.0t$ and $t' \geq 0.8t$ we detect that the revival of the correlations is replaced for a saturation in the decaying of the correlations. This is also due to the presence of finite size effects. To obtain a precise boundary here, calculations in larger systems would be needed. 
The phase boundaries are displaced to even more repulsive values of the interaction strength $U$ and the range of $U$ values is reduced. Thus, we expect that the size of the topological phase shrinks towards smaller values of the magnetic field. 
We verified that the results at vanishing magnetic field $h_x=0$ do not show a region compatible with this topological Majorana phase. 
 
\section{Conclusion}
\label{SEC_CONC}

We have studied the presence of Majorana edge-modes in a two-leg ladder of interacting spinful fermions where the total number of particles is conserved. 
By means of matrix product states algorithms we have obtained the phase diagram for the model. 
Specifically we have detected the topological phase by identifying its most characteristic features: (i) the vanishing difference in energy between the two parity sectors of a single chain, (ii) the even degeneracy on the entanglement spectrum, (iii) the finite end-end correlations with vanishing end-bulk correlations, and (iv) the robustness of this phase against static noise. 
We have encountered all these characteristics in the same finite region of the phase diagram clearly signalling the presence of the Majorana edge-modes. 
Additionally we have derived effective theories for the strongly interactive limits (strongly attractive and strongly repulsive interactions), allowing us to understand the topologically trivial phases around the non-trivial one. 

Due to the computationally expensive calculations, we performed the analysis at a few chosen sets of parameters of the magnetic field $h_x$ and the pair hopping amplitude $W$. However, since the topological phase occurred for all performed calculations we expect is to be quite robust to changes of the parameters $h_x$ and $W$. Preliminary calculations on smaller lattices also point towards the stability with respect to changes of the density.  
In particular, increasing $N$ but staying below half filling seems to broaden the parameter regime where we can find the topological phase. 

The model which we are considering has the potential to be realized in the field of cold atoms experiments by combining several known techniques with the Floquet driving of the lattice. Thus, it paves the way towards the experimental realizations of Majorana modes within the field of cold atoms.

\begin{acknowledgments}
We thank S. Diehl, M. Rizzi, and N. Tausendpfund for stimulating discussions, who were working on a related paper \cite{TausendpfundRizzi2022}. 
We acknowledge funding from the Deutsche Forschungsgemeinschaft (DFG, German Research Foundation) in particular under project number 277625399 - TRR 185 (B4) and project number 277146847 - CRC 1238 (C05) and under Germany’s Excellence Strategy – Cluster of Excellence Matter and Light for Quantum Computing (ML4Q) EXC 2004/1 – 390534769 and the European Research Council (ERC) under the Horizon 2020 research and innovation programme, grant agreement No.~648166 (Phonton). 
\end{acknowledgments}

\appendix

\section{Extraction of the central charge}
\label{SEC_CCFIT}

 \begin{figure}
 \includegraphics[width=1.0\columnwidth]{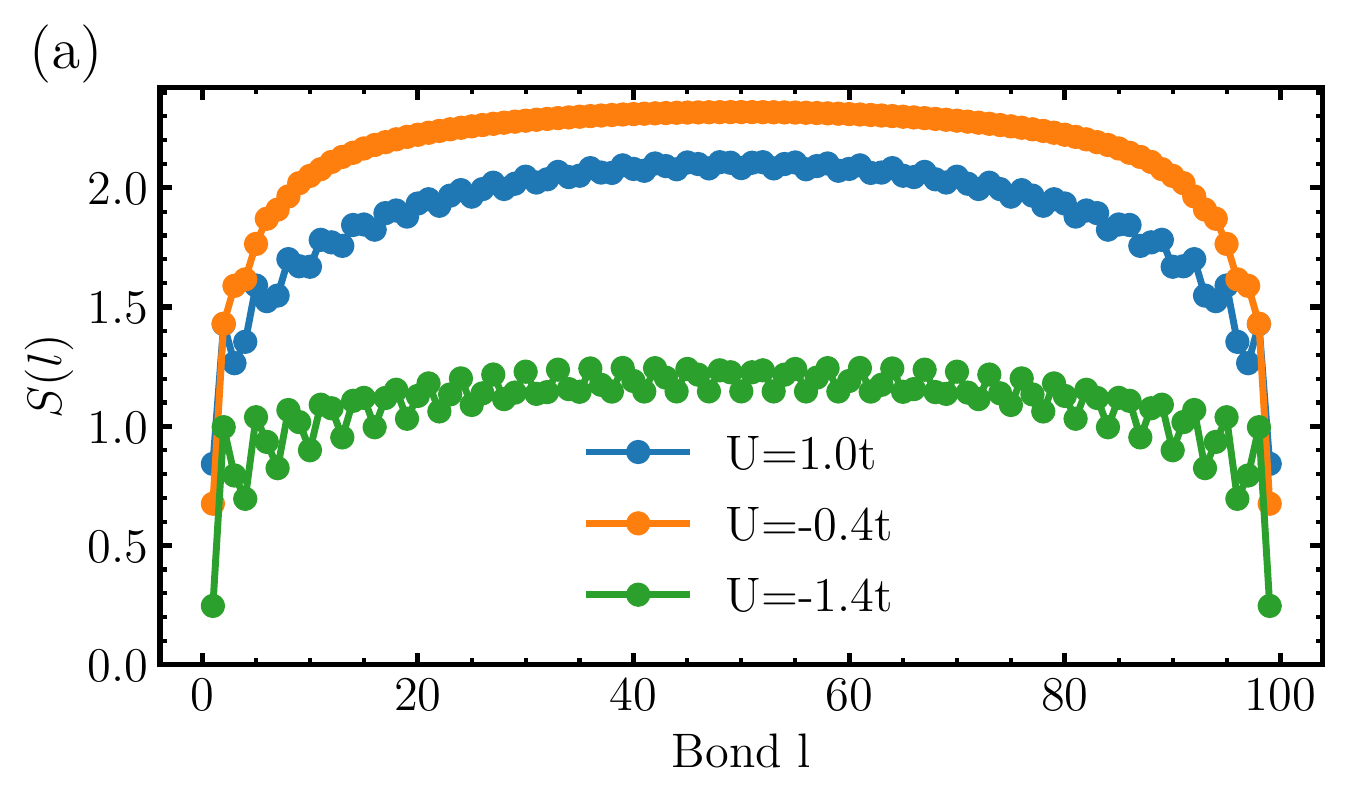}\\
 \includegraphics[width=1.0\columnwidth]{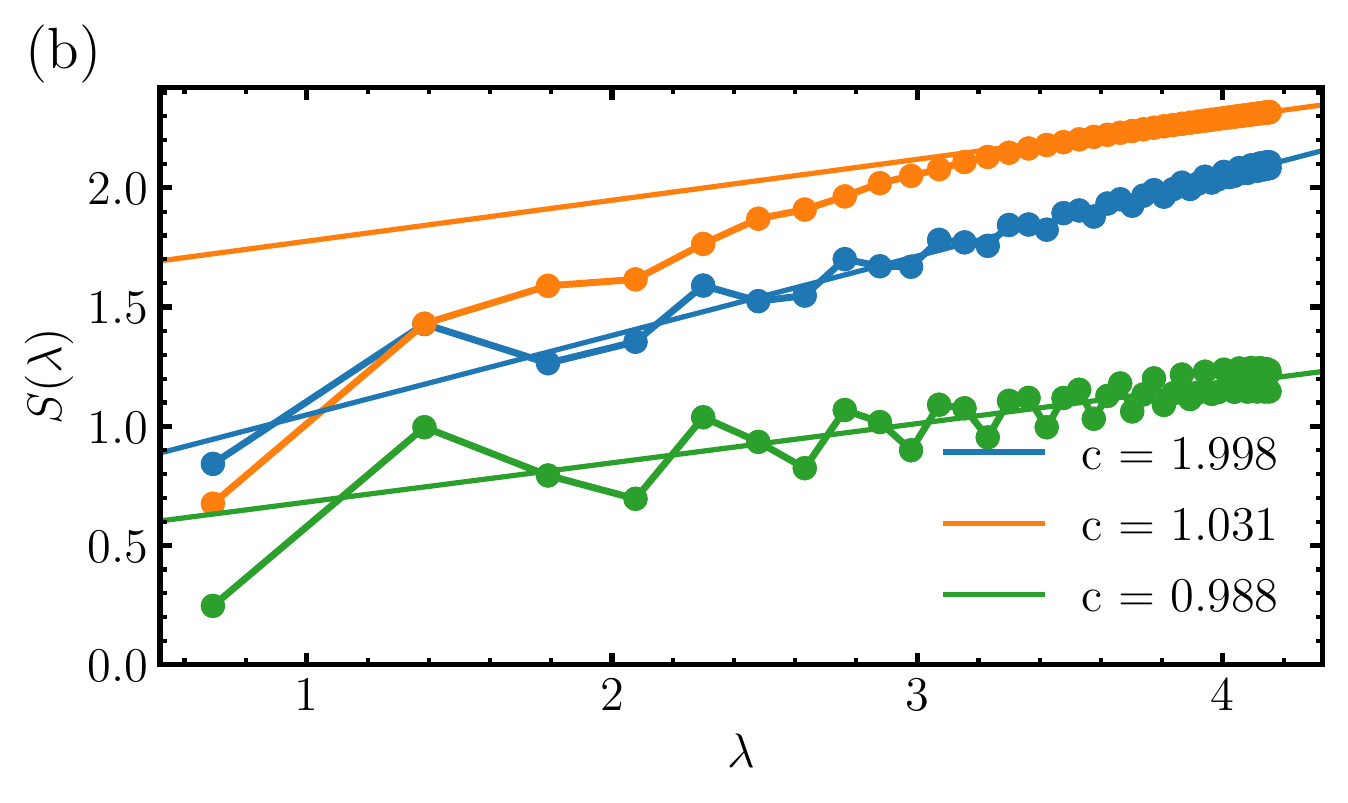}%
 \caption{ 
 von Neumann entropy for a ladder with $L=100$ rungs and $t'=0.2t$. The orange lines and points correspond to a point inside the topological region and the blue (green) ones to a point above (below) the topological region. 
 On panel (a) we plot the entropy as a function of the bond at which the system is bisected. On panel (b) we take this same result but now plotted as a function of the logarithmic conformal distance, $\lambda = \log\left[ \frac{2L}{\pi} \sin\left(\frac{\pi l}{L} \right) \right]$.
 The linear functions on panel (b) correspond to a linear fitting, and the slope is $c/6$. 
 \label{FIG_CCFIT}}
 \end{figure}

In this appendix we explain how we extract the central charge $c$ from the von Neumann entropy. The behaviour of the von Neumann entropy in one-dimensional critical systems depends on the size of the system and on the bond in which the system is bisected \cite{CalabreseCardy2004}. 
In a system with open boundary conditions of length $L$ bisected at the bond $l$ the von Neumann entropy is given by
\begin{equation}
S(l) = \frac{c}{6} \log\left[ \frac{2L}{\pi} 
\sin\left(\frac{\pi l}{L} \right) \right] +\log g+ c_1,
\end{equation}
where $\log g$ is the boundary entropy \cite{AffleckLudwig1991} and $c_1$ is a non-universal constant. An additional oscillating term beyond the conformal field theory prediction is present in many systems as for example in the critical phase of the XXZ spin model \cite{LaflorencieAffleck2006}.
When we perform the MPS simulations we map the ladder into a one-dimensional array. For that we choose a snake-like path (see Sec.~\ref{SEC_Methods}) to map the two sites on each rung, $a$ and $b$, to two consecutive matrices. 
When bisecting the system to measure the von Neumann entropy we can encounter two situations: either we cut a bond $a-b$ or a bond $b-a$. 
We note that if we cut the system at a bond $b-a$ both subsystems have an even number of sites and no rung is cut. 
If on the contrary we cut a bond $a-b$, then one would cut a rung and divide its two sites between the two subsystems. 
For the rest of the analysis we will consider only the entropy coming from cutting the system on a $b-a$ bond since we will be bisecting the system without cutting any rung. 

In panel (a) of Fig.~\ref{FIG_CCFIT} we show the results for the entropy as a function of the bisected bond $l$ for $t'=0.2t$ and different values of the interaction $U/t$. These results correspond to a ladder with $L=100$ rungs and open boundary conditions. 

To extract the central charge we re-write the von Neumann entropy as a function of the logarithmic conformal distance, which is defined as $\lambda = \log\left[ \frac{2L}{\pi} \sin\left(\frac{\pi l}{L} \right) \right]$. 
The von Neumann entropy can be expressed as linear function of the logarithmic conformal distance as $S(\lambda) = \frac{c}{6} \lambda + {\tilde c}_1$ neglecting the oscillating term and combining the boundary entropy and the constant $c_1$. In panel (b) of Fig.~\ref{FIG_CCFIT} we plot the von Neumann entropy as a function of the logarithmic conformal distance for a few examples. To extract the central charge we perform a linear fitting of the data. 

 \begin{figure}
 \includegraphics[width=1.0\columnwidth]{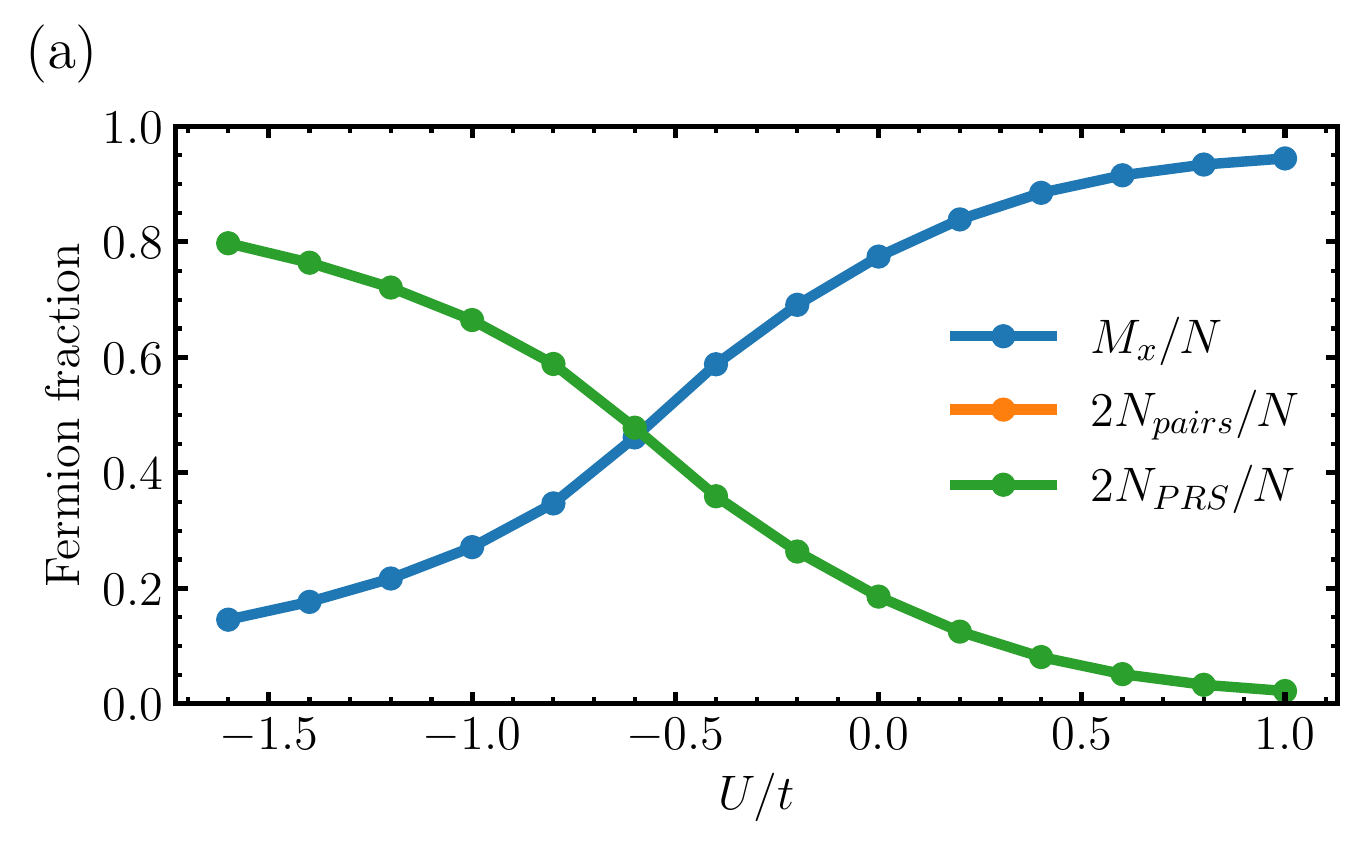}\\
 \includegraphics[width=1.0\columnwidth]{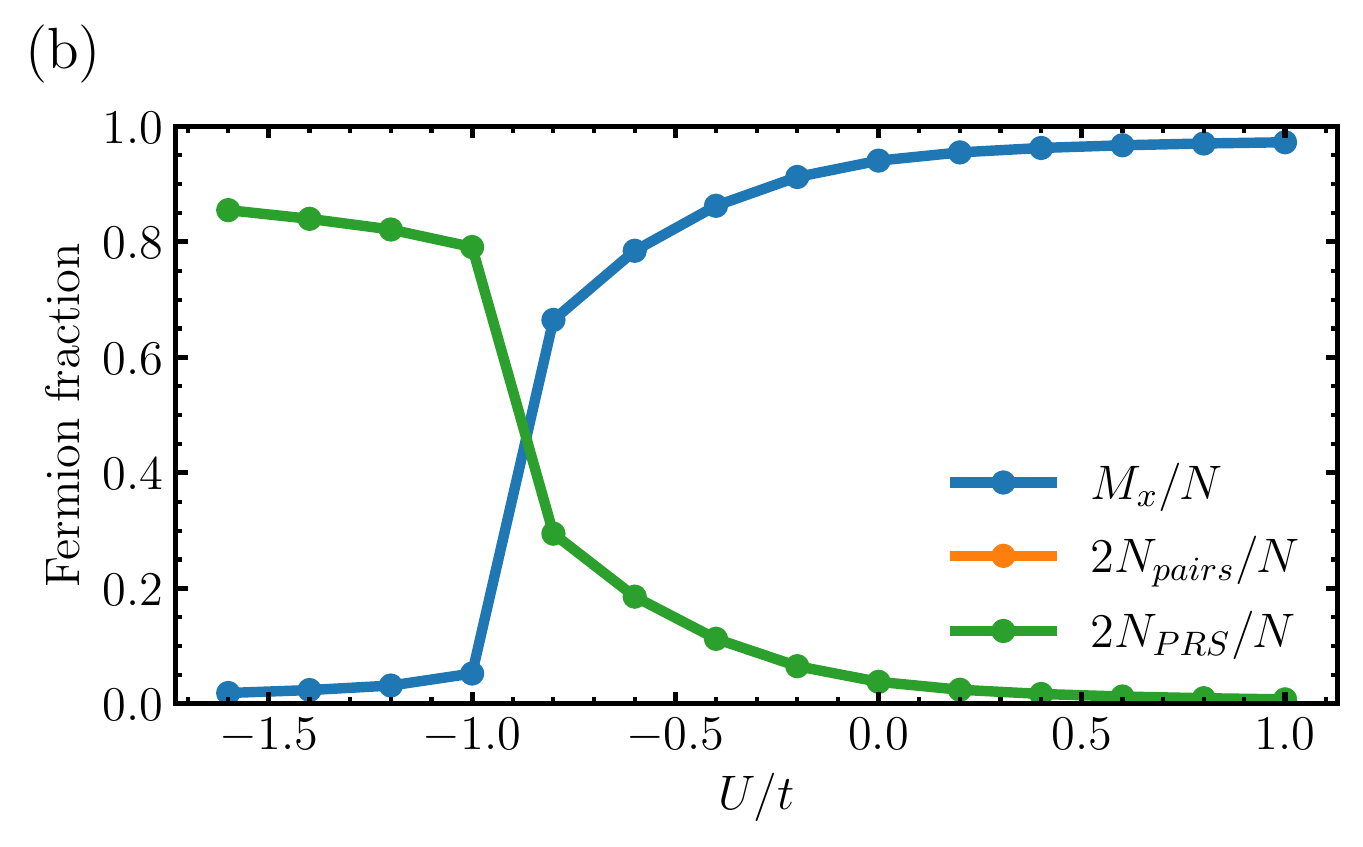}\\
 \includegraphics[width=1.0\columnwidth]{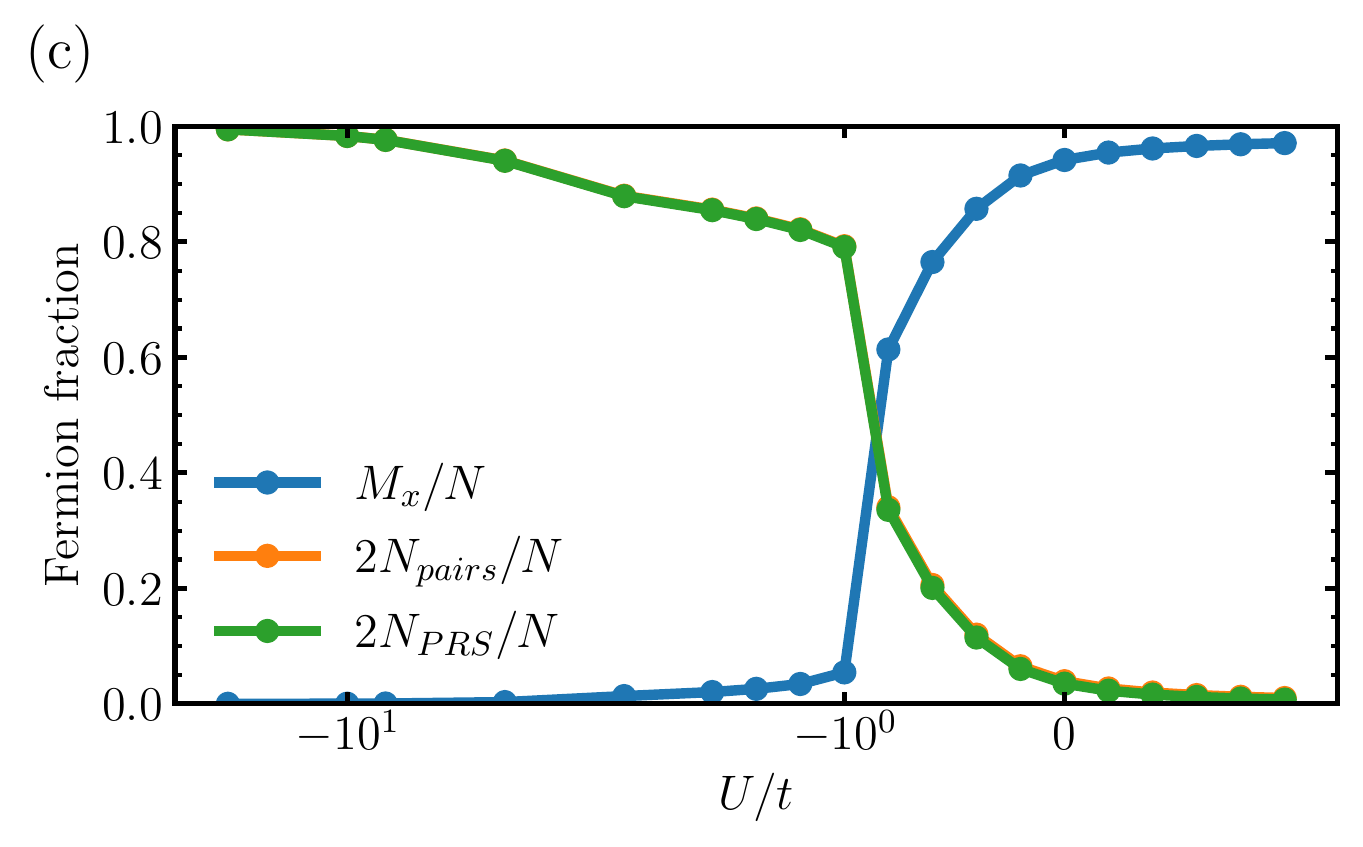}%
 \caption{ 
 Fraction of fermions forming pairs and fraction of fermions pointing to the $x$-direction against the Coulomb repulsion $U$. 
 The results in panel (a) correspond to a ladder with $L=100$ rungs at $t'=0$, in panel (b) $L=100$ and $t'=0.2t$, and in panel (c) $L=25$ and $t'=0.2t$.
 \label{FIG_PyPvsU}}
 \end{figure}

For small $l$ one can see deviations from the linear behavior of the entropy $S(\lambda)$. For a this reason we have chosen to fit only the points with large $\lambda$ near the center of the chain ($l \sim L/2$). In most cases we have fit only half of the points, discarding the ones coming from the first and last quarter of the chain (hence small $\lambda$). The linear functions plotted correspond to the linear fittings of $S(\lambda)$ performed on the data from the simulations. We see that the fit typically is well approximating the function. 
For the shown parameters we have extracted a central charge of approximately $c=1$ for the topological phase and the phase below it and a central charge of approximately $c=2$ in the phase above the topological phase.

\section{Validity of the effective models at large interaction strength}

In order to justify the validity of the effective models derived in Sections \ref{SEC_LARGEU} we compute the number of pairs, the polarization along the magnetic field $h_x$, and the opposite expectation value of the pair hopping operators given respectively by
\begin{align}
N_{pairs} &= \sum_{j \alpha} \expval{n_{\alpha \downarrow j} n_{\alpha \uparrow j} }, \\
M_x &= \sum_{j \alpha} \expval{\alpha^{\dagger}_{\uparrow j} \alpha^{}_{\downarrow j} + \text{h.c.}}, \\
N_{PRS} &= -\sum_{j} \expval{a^{\dagger}_{\uparrow j} a^{\dagger}_{\downarrow j} b^{\pdagger}_{\downarrow j} b^{\pdagger}_{\uparrow j} + \text{h.c.}}.
\end{align}
Note that $\expval{a^{\dagger}_{\uparrow j} a^{\dagger}_{\downarrow j} b^{\pdagger}_{\downarrow j} b^{\pdagger}_{\uparrow j} + \text{h.c.}}=-1$ if there is a pair and an empty state in the rung $j$ forming a singlet, if they form a triplet the expectation value is $+1$ and for any other state in the rung basis the result is zero. 
In that sense, $N_{PRS}$ counts the number of pairs forming a rung singlet with an empty state.
We show in Fig.~\ref{FIG_PyPvsU} these quantities normalized by its maximum possible value ($N$ for single-fermions and $N/2$ for pairs) for $t'=0$ and $t'=0.2t$. 
The fraction of fermions forming on-site pairs at attractive interaction $U = -1.6t$, lies above $\approx 0.8$ both for $t'=0$ and for $t'=0.2t$. 
For strongly attractive values of the interaction close to $U \simeq -10t$, one can see on panel (c) that the fraction of pairs almost reaches the unity. 
At repulsive interaction the fraction of pairs rapidly decreases due to the low filling, such that at $U=0$, its expectation value lies around $0.04$ for $t'=0.2t$ and $L=100$ rungs. 
Also the polarization along the magnetic field $h_x$ is rapidly growing towards an interaction strength of $U=t$, where the state is already almost fully polarized along the magnetic field. 
This justifies the use of our approximation of the fully polarized state at moderate to strong repulsive interactions. 
 
At the same time, the number of pair forming singlets behaves just like the number of pairs, is very large at attractive interactions and decreases strongly towards the repulsively interacting side. 
Remarkably, these two quantities are not only qualitatively similar but also one can see in Fig.~\ref{FIG_PyPvsU} that the symbols are larger than the difference between them. 
This indicates that almost all the pairs in the system are forming in-rung singlets with the empty state in the opposite leg for all the points studied in Fig.~\ref{FIG_PyPvsU}. This behaviour changes when going to lower values of $W$, where a discrepancy between $N_{pairs}$ and $N_{PRS}$ exists.

\section{Adiabatic connection to models in previous works}
\label{SEC_ADIAB}
 
In this appendix we show that there is an adiabatic connection between the topological phase studied in this work and the Majorana phase found in Ref.~\cite{KrausZoller2013}. 
For demonstrating this we study a trajectory in the phase diagram connecting the two models. We show that along this trajectory the properties of the system vary smoothly. 
In particular, the quantities considered are the entanglement spectrum, the energy difference between parity sectors and the energy difference between the lowest-lying state and the first excitation inside both sectors. 
We are going to require that the gap between parity sectors remains always negligible in comparison to the gap to the first excitation.

 \begin{figure}
 \includegraphics[width=1.0\columnwidth]{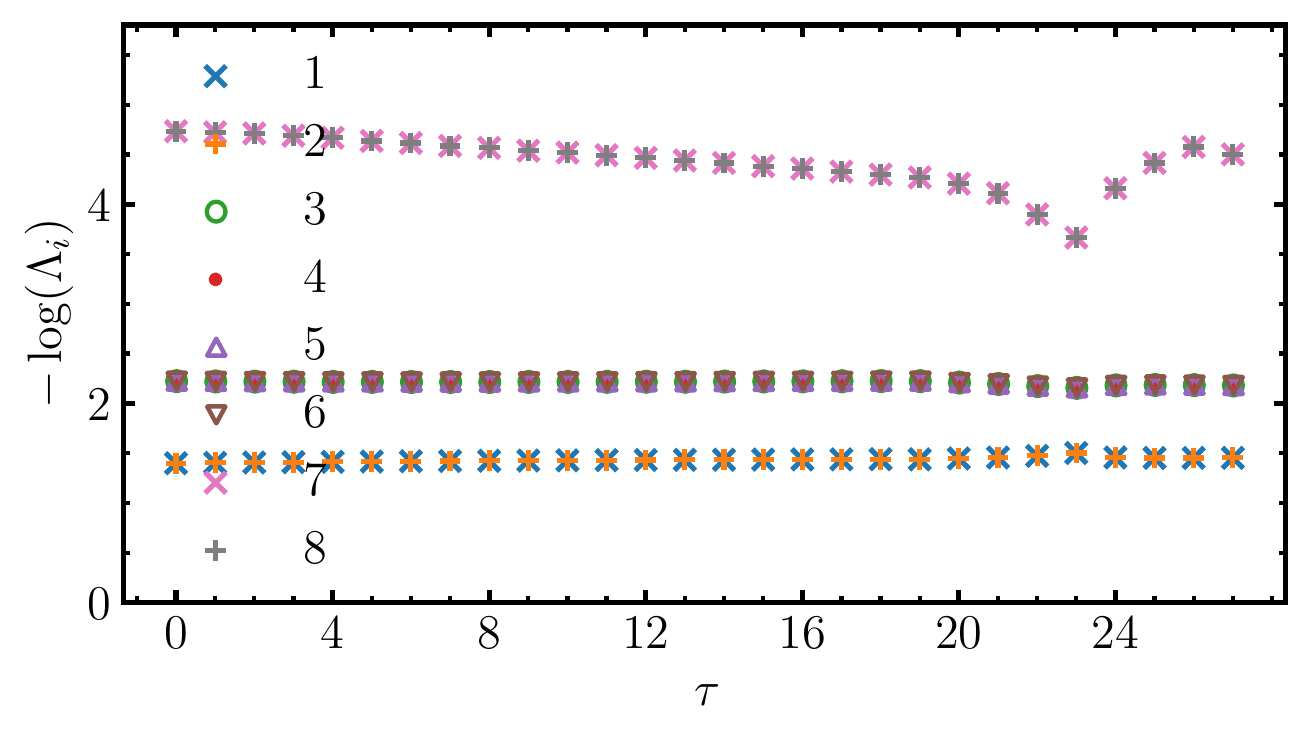}%
 \caption{ 
 The first eight eigenvalues of the entanglement spectrum for the spinless ladder with $L=100$ rungs. 
 For $\tau \in[0,18]$ it controls the pair-hopping amplitudes $W_{e,o}$ and the density is fixed to $n'=0.32$. 
 But for $\tau \in[19,27]$ the pair-hopping amplitudes are fixed to $W_e=2.6t$ and $W_o=0$ and the density varies from $n'=0.32$ to $n'=0.16$.
 \label{FIG_ADIABEIGS}}
 \end{figure} 

To find a path in the phase diagram that connects both models, we consider the $U=t'=0$-limit of our model and the exact mapping to spinless fermions in Eq.~\ref{EQ_KRAUS}. 
The following Hamiltonian includes in its parameter space our model at $U=t'=0$ and the model in Ref.~\cite{KrausZoller2013} 
\begin{align}
\label{EQ_KRAUS2}
&H = -t\sum_{j=1}^{L-1} \sum_{\alpha = a, b} 
( \tilde{\alpha}^{\dagger}_{j} \tilde{\alpha}^{\pdagger}_{j+1} + \text{h.c.} ) \\ \nonumber
&+ W_e\sum_{j=1}^{(L-1)/2} ( \tilde{a}^{\dagger}_{2j} \tilde{a}^{\dagger}_{2j+1} 
\tilde{b}^{\pdagger}_{2j+1} \tilde{b}^{\pdagger}_{2j} + \text{h.c.} )  \\ \nonumber
&+ W_o\sum_{j=1}^{L/2} ( \tilde{a}^{\dagger}_{2j-1} \tilde{a}^{\dagger}_{2j} 
\tilde{b}^{\pdagger}_{2j} \tilde{b}^{\pdagger}_{2j-1} + \text{h.c.} ) .
\end{align}
Here $W_e$ ($W_o$) is the pair-hopping amplitude on plaquettes which start with even (odd) sites. 
When $W_o = 0$, $W_e = 2.6t$, the system is equivalent to the $U=t'=0$ point in our phase diagram, where we clearly identify the topological phase. 
In contrast, when $W_e = W_o = 1.8t$, this corresponds to the model studied in Ref.~\cite{KrausZoller2013}, where the Majorana phase is found. 

Additionally, we need to connect a different density in the two models as the densities studied do not correspond to each other. We define in the spinless case the density as $n' = N/2L'$, with $N$ the number of fermions and $L'$ the number of rungs in the spinless model. This is the same definition used in Ref.~\cite{KrausZoller2013} where the density considered is $n'=1/3$.
Note, that the number of rungs after mapping to the spinless model has doubled with respect of the original number of rungs $L$ in the spinful case (since we divide each spinful site into two spinless sites), while the number of fermions $N$ remains the same. 
Therefore, the spinful density has to be divided by two to obtain the equivalent spinless density after the mapping. 
This is why the spinful model with density $n=0.32$ corresponds to an equivalent spinless model with density $n'=0.16$. 

We we will describe the pair-hopping amplitudes $W_{e,o}$ and the density $n'$ as a function of a parameter $\tau$. 
The trajectory in the parameter-space connecting both models for $\tau\in \left[ 0, 18 \right]$ is given by 
\begin{align}
\label{EQ_PARAM1}
&n' = 0.32, \\ \nonumber
&W_e/t=1.8+0.04\tau, \\ \nonumber
&W_o/t=1.8-0.01\tau.
\end{align}
And for $\tau\in \left[ 19, 27 \right]$, 
\begin{align}
\label{EQ_PARAM2}
&n' = 0.32 - 0.02 ( \tau - 19 ), \\ \nonumber
&W_e/t=2.6, \\ \nonumber
&W_o/t=0.0.
\end{align}
With this parametrization, $\tau=0$ corresponds to the model studied in Ref.~\cite{KrausZoller2013}. 
The only difference is that the density is $n'=0.32$ which is close to the value of $n'=1/3$, but not exactly. This choice had to be made since the system size does not allow us to choose a density value exactly at $n'=1/3$. 
At $\tau=19$ the pair-hoppings amplitudes are $W_e=2.6t$ and $W_o = 0$, as in Section \ref{SEC_RES}, but the density is $n'=0.32$ instead of $n'=0.16$.
When we reach $\tau=27$ the spinless density will correspond $n'=0.16$, equivalent to the spinful density $n=0.32$ we require. 

In Fig.~\ref{FIG_ADIABEIGS} the first eight eigenvalues of the entanglement spectrum for the spinless ladder with $L=100$ rungs are shown. 
As we mentioned, $\tau=0$ can be considered inside the Majorana phase studied in Ref.~\cite{KrausZoller2013}, and from Fig.~3 of that work one can see that the degeneracies on the first two branches of the entanglement spectrum coincide with the ones we show here. 
From there we vary the parameter $\tau$ and therefore $W_{e,o}$ and $n'$. 
We can see that the eigenvalues only change slightly their values without any crossing. Only in the highest shown value a considerably change occurs at $\tau = 23$ (which corresponds to a spinless density of $n'=0.24$). We remark that the double degeneracy of the whole spectrum is maintained during all this path. 
The smooth behaviour of the considered eigenvalues points strongly towards an adiabatic connection between $\tau=0$ and $\tau=27$.

 \begin{figure}
 \includegraphics[width=1.0\columnwidth]{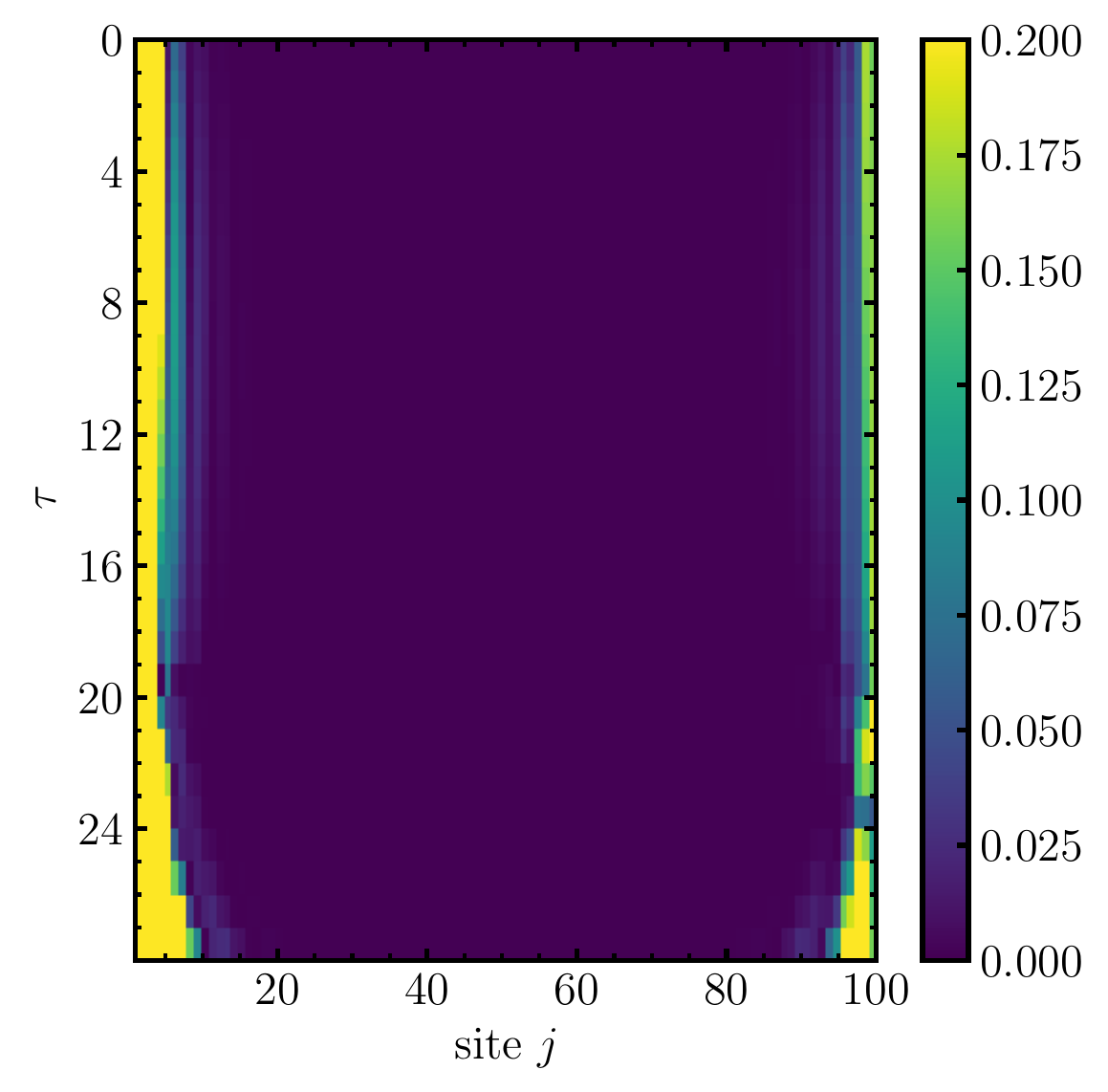}%
 \caption{ 
 Normalized absolute value of the single particle correlations $ A^{N}_{1j} = \left| \expval{ a^{\dagger}_{1} a^{\pdagger}_{j} } / \text{max}_j \expval{ a^{\dagger}_{1} a^{\pdagger}_{j} } \right| $ on a spinless ladder with $L=100$ rungs. 
 The horizontal axis corresponds to the site $j$ where the particle is annihilated and the vertical axis correspond to the trajectory parametrized by $\tau$. 
 The color bar saturates at $A^{N}_{1j} = 0.2$ for better visibility. 
 \label{FIG_ADIABCORR}}
 \end{figure} 

Furthermore, we followed the gap between energy sectors and the gap inside any parity sector along the path connecting the different models. 
The gap between the energy sectors remains always below $2.10^{-12}t$, order of magnitudes below the gap inside the energy sectors which remains above $0.02t$ for this system size. 
From the smooth behaviour of the considered quantities and the fact that the entanglement spectrum is qualitatively the same during all the trajectory (only smoothly varying between the extreme cases without showing any crossings) an adiabatic connection between the Majorana phase in Ref.~\cite{KrausZoller2013} and the phase studied in this work seems to exist. 
 
Additionally, we show in Fig.~\ref{FIG_ADIABCORR} the normalized absolute value of the single particle correlations 
$ A^{N}_{1j} = \left| \expval{ a^{\dagger}_{1} a^{\pdagger}_{j} } / \text{max}_j \expval{ a^{\dagger}_{1} a^{\pdagger}_{j} } \right| $ 
for every value of $\tau$. 
We have normalized this quantity and set the maximum of the colorscale to $0.2$ to better visualize the data. 
The general shape of the correlations remain almost unchanged for values of $\tau$ between $\tau=0$ and $\tau=20$. 
From $\tau=20$ to $\tau=27$ we see some changes in these correlations. The peak close to $j=1$ grows in intensity (not visible) and the edge states spread a little into the bulk. 
Despite the small quantitative changes we can appreciate that during the complete path connects both models the single particle correlations are zero in the bulk but have a revival on the opposite end leading to finite edge-edge correlations. 
Therefore, we take this as a signal that the nature of the edge modes does not change drastically and that the Majorana modes survives for every value of $\tau$.

\bibliography{}

\end{document}